\documentclass[iop,numberedappendix,appendixfloats]{emulateapj}
\usepackage{color}
\usepackage[hyperfootnotes, colorlinks, linkcolor=blue, citecolor=blue, urlcolor=blue]{hyperref}
\usepackage{amsmath}
\usepackage{enumerate}

\newcommand{\idrop}{$i$-dropout}

\newcommand{\ydrop}{$Y$-dropout}
\newcommand{\yjdrop}{$Y\!J$-dropout}

\newcommand{\bFilter}{$B_{435}$}
\newcommand{\vFilter}{$V_{606}$}
\newcommand{\iFilter}{$i_{814}$}

\newcommand{\yFilter}{$Y_{105}$}
\newcommand{\jFilter}{$J_{125}$}
\newcommand{\jhFilter}{$J\!H_{140}$}
\newcommand{\hFilter}{$H_{160}$}
\newcommand{\clone}{Abell 2744}
\newcommand{\cltwo}{MACS J0416.1$-$2403}
\newcommand{\clthree}{MACS J0717.5+3745}
\newcommand{\clfour}{MACS J1149.6+2223}
\newcommand{\clfive}{Abell S1063}
\newcommand{\clsix}{Abell 370}

\shorttitle{Size--luminosity relations and UV luminosity functions at $z\sim6-9$ from the {\it HFF} data}
\shortauthors{Kawamata et al.}

\slugcomment{Accepted for publication in The Astrophysical Journal}

\begin{document}

\title{Size--luminosity relations and UV luminosity functions at $z=6-9$\\simultaneously derived from the complete \textit{Hubble} Frontier Fields data}

\author{Ryota~Kawamata\altaffilmark{1},
Masafumi~Ishigaki\altaffilmark{2,3}, 
Kazuhiro~Shimasaku\altaffilmark{1,4},
Masamune~Oguri\altaffilmark{2,4,5}, 
Masami~Ouchi\altaffilmark{3,5},
and~Shingo~Tanigawa\altaffilmark{1,6}
}

\altaffiltext{}{Email: kawamata@astron.s.u-tokyo.ac.jp}
\altaffiltext{1}{Department of Astronomy, Graduate School of Science, The University of Tokyo, 7-3-1 Hongo, Bunkyo-ku, Tokyo 113-0033, Japan}
\altaffiltext{2}{Department of Physics, Graduate School of Science, The University of Tokyo, 7-3-1 Hongo, Bunkyo-ku, Tokyo 113-0033, Japan}
\altaffiltext{3}{Institute for Cosmic Ray Research, The University of Tokyo, 5-1-5 Kashiwanoha, Kashiwa, Chiba 277-8582, Japan}
\altaffiltext{4}{Research Center for the Early Universe, The University of Tokyo, 7-3-1 Hongo, Bunkyo-ku, Tokyo 113-0033, Japan}
\altaffiltext{5}{Kavli Institute for the Physics and Mathematics of the Universe (Kavli IPMU, WPI), The University of Tokyo, 5-1-5 Kashiwanoha, Kashiwa, Chiba 277-8583, Japan}
\altaffiltext{6}{Institute of Astronomy, University of Cambridge, Madingley Road, Cambridge, CB3 0HA, UK}

\begin{abstract}
We construct $z\sim6-7$, 8, and 9 faint Lyman break galaxy samples 
(334, 61, and 37 galaxies, respectively) with accurate 
size measurements with the software \texttt{glafic} from the complete \textit{Hubble} Frontier Fields cluster 
and parallel fields data. 
These are the largest samples hitherto and reach down to the faint 
ends of recently obtained deep luminosity functions.
At faint magnitudes, however, these samples are highly 
incomplete for galaxies with large sizes, implying that derivation of the luminosity 
function sensitively depends on the intrinsic size--luminosity relation.
We thus conduct simultaneous maximum-likelihood estimation of luminosity function and size--luminosity relation parameters 
from the observed distribution of galaxies on the size--luminosity plane 
with the help of a completeness map as a function of size and luminosity.
At $z\sim6-7$, we find that the intrinsic size--luminosity relation expressed as 
$r_\textrm{e} \propto L^\beta$ has a notably steeper slope of
$\beta=0.46^{+0.08}_{-0.09}$ than those at lower redshifts,
which in turn implies that the luminosity function has a relatively shallow faint-end slope of $\alpha=-1.86^{+0.17}_{-0.18}$.
This steep $\beta$ can be reproduced by a simple analytical model
in which smaller galaxies have lower specific angular momenta.
The $\beta$ and $\alpha$ values for the $z\sim8$ and 9 samples are consistent with 
those for $z\sim6-7$ but with larger errors.
For all three samples, there is a large, positive covariance between $\beta$ and $\alpha$,
implying that the simultaneous determination of these two parameters
is important.
We also provide new strong lens mass models of \clfive\ and \clsix, as well as 
updated mass models of \clone\ and \cltwo.
\end{abstract}


\section{Introduction}

Disk sizes of galaxies at very high redshifts are important in two
aspects.
One is that they provide information on the formation and early evolution
of galaxies.
The other is that they have a significant effect
on the determination of UV luminosity functions because the correction 
for detection incompleteness sensitively depends on size.

Concerning the first aspect, the size of galaxies is 
largely determined by their angular momentum \citep[e.g.,][]{fallefstathiou80, mmw98}
as is the case for disk galaxies, and angular momentum is 
one of the fundamental parameters of galaxies as argued by \citet{fall83}.
\citet{romanowskyfall12} and \citet{fall13} have discussed galaxy formation 
and evolution using the specific angular momentum--mass diagram.
Indeed, numerous simulations and analytical models of galaxy formation
suggest that the size of galaxies changes with a redistribution of 
the angular momentum in them due to stellar feedback 
such as galactic winds \citep[e.g.,][]{brooks11, wyithe11, brook12, danovich15, genel15}.
Recently, high-resolution cosmological simulations have succeeded in 
increasing sizes at a fixed luminosity or stellar mass
of simulated galaxies to reproduce observed sizes 
by incorporating stellar feedback
such as galactic winds of high mass-loading factors
\citep[e.g.,][]{brooks11, genel15}.
The luminosity dependence of the size is also affected by stellar feedback
as explained by simple analytical models.
For example, \citet{wyithe11} showed that the slope of 
the size--luminosity relation varies depending on the dominating
feedback such as energy-driven and momentum-driven feedback. 
Larger sizes indicate more efficient feedback, 
which suggests that the slope of the size--luminosity 
relation contains information on the dominant 
feedback process.

The second aspect concerning UV luminosity functions is also important
because luminosity functions are determined by correcting
for detection completeness, which depends on the intrinsic size distribution.
For a given magnitude, galaxies with larger sizes are less 
likely to be detected because of their lower surface brightness.
\citet{graz11}, based on the $z\sim7$ analysis, have 
pointed out that the assumed
size distribution critically alters the UV luminosity function,
especially the faint-end slope.

One of the main goals of recent observational projects targeting $z\gtrsim6$ 
galaxies \citep[e.g., HUDF09/12, CANDELS, XDF, GOLDRUSH;][]{oesch10a, grogin11, koekemoer11, ellis13, illingworth13, ono17} 
is to obtain the faint-end slope of luminosity functions, 
a key quantity for testing galaxy formation models. 
In addition, since $z\sim 6-10$ is the epoch of reionization
and faint galaxies are thought to be major sources of ionizing photons,
the abundance of faint galaxies, i.e., the faint-end slope, is important
for understanding the reionization of the universe. 

Recently, in order to derive luminosity functions at fainter
magnitudes,
deep observations combined with the power of the gravitational 
lensing by galaxy clusters have been conducted, such as the CLASH program
\citep[see][for more details]{postman12} and the \textit{Hubble} Frontier Fields
program \citep[HFF;][]{lotz17}.
Utilizing early-stage data from the HFF, the faint limits of
luminosity functions reach as faint as UV magnitudes ($M_{\mathrm UV}$)
of $M_{\mathrm UV} \sim -15.5$,
$-17$, and $-17.5$ at $z\sim 6-7$, 8, and 9, respectively 
\citep{atek14, atek15a, ishigaki15, mcleod15}.
More recently, very faint galaxies of $M_{\mathrm UV} \sim -13$ 
at $z\sim 6-7$ have been detected using one-third of the full
HFF data 
\citep{castellano16b, livermore16}, half of them \citep{laporte16}, 
two-thirds of them (\citealp{kawamata16}; hereafter K16, \citealp{yue17}), 
and all of them \citep{ishigaki17}.
However, the luminosity functions obtained in the previous studies, 
including those from the HFF, are still highly 
uncertain, especially at $M_{\mathrm UV} \gtrsim -18$ and $z\gtrsim6$,
because the size--luminosity relations are not determined well in that magnitude range 
\citep[see our Figure~\ref{fig:detected_fraction}, and Figure 2 of][]{bouw17size}
owing to an insufficient number of galaxies with size measurements.

There have been a number of studies that measure sizes of 
bright ($M_{\mathrm UV} \lesssim -18$) galaxies
\citep[e.g.,][]{ferg04, bouw04, curtislake16, laporte16, bowler17}.
At $z\sim 4$ and $5$, \citet{huang13} have carefully measured the size 
distributions of Lyman break galaxies (LBGs) with
$-22.5 \lesssim M_{\mathrm UV} \lesssim -17.5$
and find size--luminosity relations
of $L \propto r_{\mathrm e}^{0.22-0.25}$, where $L$ and $r_{\mathrm e}$
are the luminosity and effective half-light radius, respectively.
\citet{oesch10b} were among the first to measure the sizes of $z\sim7$ and 8 galaxies
with samples of 16 and five galaxies from HUDF09 \citep{oesch10a}
reporting that the decreasing trend of sizes with increasing redshifts 
continues to these redshifts.
This trend has been confirmed by \citet{ono13} by careful measurements
using the deeper imaging data from HUDF12 \citep{ellis13, koekemoer13}.
With a larger sample, \citet{graz12} have measured the sizes of $z\sim7$
LBGs of moderate magnitude ($M_{\mathrm UV} \lesssim -18.5$).
They have found that the size--luminosity 
relation is in the form of $L \propto r_{\mathrm e}^{0.5}$ at this
redshift, although their size measurements may suffer from 
systematic biases due to their measuring method.
More recently, \citet{shibuya15} have measured sizes for large
$z\sim6-10$ LBG samples with moderate magnitudes of 
$M_{\mathrm UV} \lesssim -18$.
However, since none of the above studies has reliably determined
the size--luminosity relation for $M_{\mathrm UV} \lesssim -18$
galaxies at $z\gtrsim6$,
a size--luminosity relation of $L \propto r_{\mathrm e}^{0.25}$
has been commonly adopted, given the results of 
\citet{huang13} obtained for $z\sim4-5$.
This relation is extrapolated and also applied to
fainter magnitudes down to $M_{\mathrm UV} \sim -13$,
beyond the magnitude range over which it is determined.

At faint magnitudes of $M_{\mathrm UV} \gtrsim -18$, 
\citet[][hereafter K15]{kawamata15} have used the 
first cluster and parallel fields data from the HFF
to find that the sizes of observed
faint galaxies ($-18.7 \lesssim M_{\mathrm UV} \lesssim -16.6$) are considerably
smaller than the sizes inferred from the extrapolated
size--luminosity relation of $L \propto r_{\mathrm e}^{0.25}$.
This result has subsequently been confirmed by
\citet{bouw17size}, \citet{laporte16}, and \citet{bouw17gc}, 
who have measured 
the sizes of faint galaxies using four, three, and four HFF cluster fields data, 
respectively. 
In addition, \citet{bouw17size} have indirectly indicated 
the absence of faint galaxies with large sizes using the dependence of 
the galaxy surface density on the lensing shear. 
They have concluded that the intrinsic sizes of the faintest galaxies 
are small, and the intrinsic size distribution assumed in the calculation 
of the luminosity function should be close to the observed one. 
This makes the faint-end slope of the luminosity function shallower. 
However, since none of \citet{kawamata15}, \citet{bouw17size, bouw17gc},
and \citet{laporte16} have considered an incompleteness correction 
due to galaxies with 
large sizes, the slope of the size--luminosity relation may be biased toward 
a steeper value. 
In addition, the indirect inference in \citet{bouw17size}
is subject to large uncertainties, which may result in weak constraints 
on the size distribution compared to inferences using direct size measurements.

In this paper, we provide direct size measurements of $z\sim6-7$,
8, and 9 
LBGs at $-21.6 \lesssim M_{\mathrm UV} \lesssim -12.3$ 
using all six HFF cluster and parallel fields data.
We show that the incompleteness effect is significant 
at $z\sim6-9$ for the first time.
We derive incompleteness-corrected intrinsic
size--luminosity relations simultaneously with luminosity functions,
which enables us to explore the correlation between 
these two functions.
We note that we do not discuss the UV luminosity density
and hence the contribution of galaxies to cosmic reionization,
because the normalization parameter of UV luminosity functions
is not determined in this paper.

The structure of this paper is as follows.
In Section \ref{sec:data}, we describe the data and 
samples, which are identical to those constructed in
\citet{ishigaki17} but with slight changes.
In Section \ref{sec:sizemeasurement}, we measure the 
sizes of the galaxies.
Our method to correct for systematic biases, 
which is updated from that in K15 in order to
deal with the increased number of galaxies,
is also described.
In Section~\ref{sec:RLrelations}, for each of the three
redshift ranges, we simultaneously estimate the intrinsic 
size--luminosity relation 
and the UV luminosity function from the observed 
distribution of galaxies on the size--luminosity 
plane, taking account of the incompleteness effect.
The correlations between the size--luminosity and luminosity function
parameters are also obtained.
We discuss our findings in Section~\ref{sec:discussion} and 
give a summary in Section~\ref{sec:conclusion}.

Throughout this paper, 
we adopt a cosmology with $\Omega_{M} = 0.3$, $\Omega_{\Lambda} = 0.7$, 
and $H_{0} = 70\>\mathrm{km\>s^{-1}\>Mpc^{-1}}$.
Magnitudes are given in the AB system \citep{okeg83}.
Galaxy sizes are measured in the physical scale.

\section{Data and Sample Selection}\label{sec:data}
Here we describe the data, sample selection, and 
obtained samples.
The data and the criteria for the sample selection are the same 
as those in \citet{ishigaki17}, but we remove two galaxies from
their samples.
Only a brief description is given in this section, and readers are referred 
to the above paper for further details.

\subsection{HFF Mosaic Data}
We use the reduced image mosaics obtained in the HFF program, which are made 
publicly available through the STScI 
website\footnote{\url{http://www.stsci.edu/hst/campaigns/frontier-fields/}}. 
This program targets six cluster fields, \clone, \cltwo, \clthree, \clfour, 
\clfive, and \clsix, and their accompanying six parallel fields.
Those fields have been observed
deeply with the \textit{Hubble Space Telescope} using three bands of
the Advanced Camera for Surveys (ACS) and four bands of the IR channel of 
the Wide Field Camera 3 (WFC3/IR).
We utilize the v1.0 standard calibrated (i.e., without `self-calibration') mosaics for 
the three ACS bands F435W (\bFilter), F606W (\vFilter), and F814W (\iFilter).
For the four WFC3/IR bands F105W (\yFilter), F125W (\jFilter), F140W 
(\jhFilter), and F160W (\hFilter), we use the v1.0 standard calibrated mosaics 
for the \clone\ parallel and \cltwo\ cluster fields and v1.0 mosaics 
corrected for `time-variable sky emission' for the other ten fields. 
The $5 \sigma$ limiting magnitudes of the mosaics are $\sim 29$ mag
on a $0\farcs35$ diameter aperture.
All the images have a pixel scale of $0\farcs03$.

\subsection{Sample Selection}\label{subsec:sampleselection}

\begin{figure}[t]
  \centering
      \includegraphics[width=\linewidth]{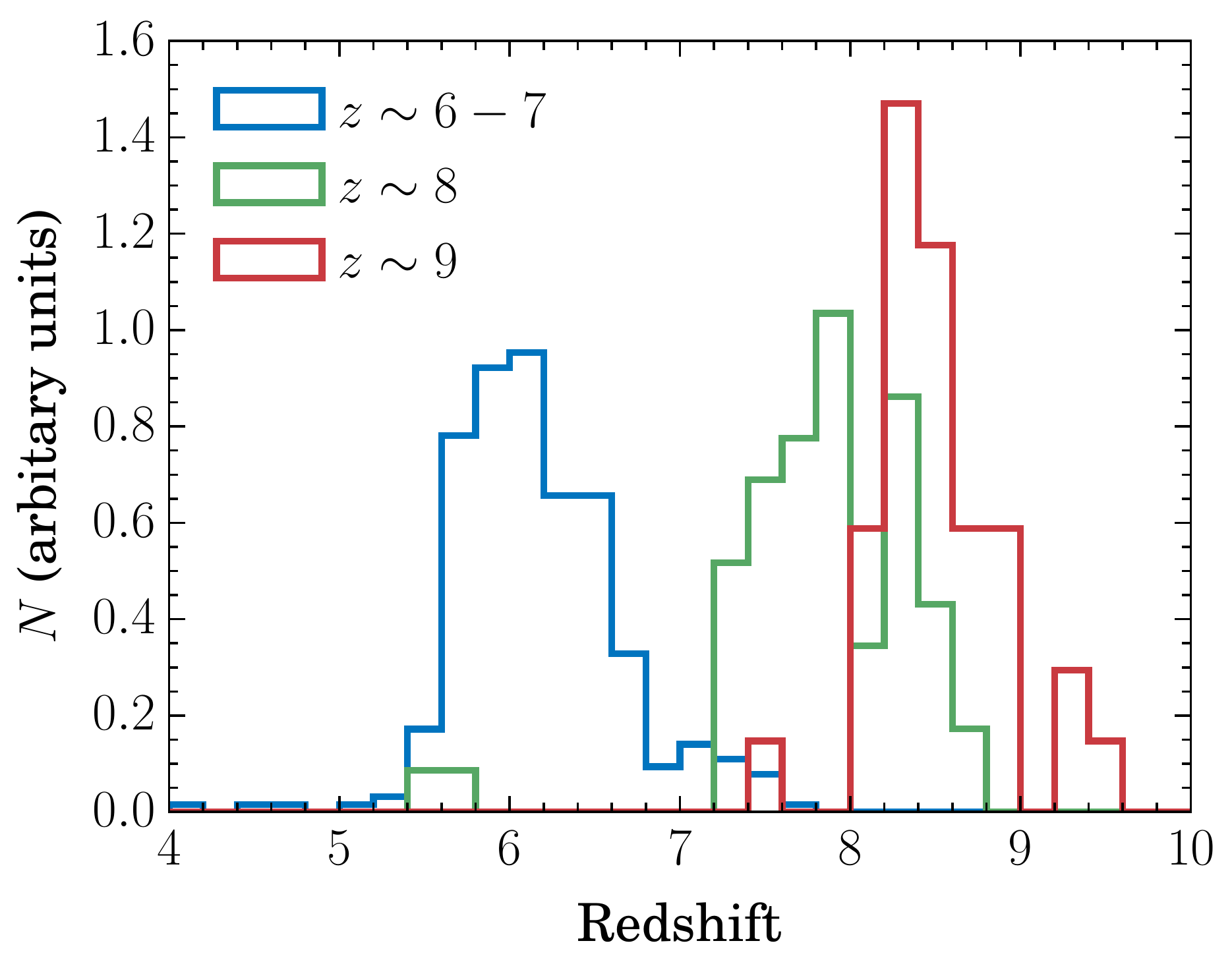}
  \caption{Normalized distributions of the photometric redshifts of our samples
  for $z\sim6-7$ (blue), 8 (green), and 9 (red).
  }
  \label{fig:redshiftdist}
\end{figure}

We make two catalogs with different detection images, 
which are referred to as the $J\!J\!H\!H$ and $J\!H\!H$ catalogs.
The detection image for the former is a
\jFilter, \jhFilter, and \hFilter\ combined image,
and for the latter it is a \jhFilter\ and \hFilter\ combined image;
these are created using \texttt{SWarp} v2.38.0 \citep{bertin02}
together with their weight maps.
To make the catalogs, we run \texttt{SExtractor} v2.8.6 \citep{bertin96} on the seven 
bands' images using the detection images.
The photometric redshifts of galaxies in these catalogs are 
estimated using \texttt{BPZ} v1.99.3 \citep{benitez00}.
From the catalogs, we 
select $i$-, $Y$-, and $Y\!J$-dropout galaxies using the Lyman break 
technique. 
For $i$- and $Y$-dropout selections, we use the $J\!J\!H\!H$ catalog
and for $Y\!J$-dropout selection, we use the $J\!H\!H$ catalog.
For \idrop s or $z\sim6-7$ galaxies, we use the criteria of 
\begin{align}
i_{814} - Y_{105} &> 0.8,\\
Y_{105} - J_{125} &< 0.8,\\
i_{814} - Y_{105} &> 2(Y_{105} - J_{125}) + 0.6,
\end{align}
for \ydrop s or $z \sim 8$ galaxies, 
\begin{align}
Y_{105} - J_{125} &> 0.5,\\
J_{125} - J\!H_{140} &< 0.5,\\
Y_{105} - J_{125} &> 0.4 + 1.6(J_{125} - J\!H_{140}),
\end{align}
and for \yjdrop s or $z \sim 9$ galaxies, 
\begin{align}
(Y_{105} + J_{125})/2 - J\!H_{140} &> 0.75,\\
(Y_{105} + J_{125})/2 - J\!H_{140} \nonumber\\
 > 0.75 &+ 0.8 (J\!H_{140} - H_{160}),\\
J_{125} - H_{160} &< 1.15,\\
J\!H_{140} - H_{160} &< 0.6.
\end{align}

For \idrop s, we use additional signal-to-noise ratio constraints 
that require objects not to be detected at $>2 \sigma$ levels 
in both the \bFilter- and \vFilter-band images or in a \bFilter\ + \vFilter\ 
stacked image.
Detections at $>5 \sigma$ levels are also required
in both the \yFilter- and \jFilter-band images.
For a conservative selection, \iFilter\ magnitudes
are replaced by the \iFilter\ $2 \sigma$ limiting magnitude 
if the signal is below that level.
For \ydrop s, objects are required to be detected at 
$>2 \sigma$ levels in none of the \bFilter-, \vFilter-, and 
\iFilter-band images.
In addition, detections at $>5 \sigma$ levels are required
in all of the \jFilter-, \jhFilter-, and \hFilter-band images.
For \yjdrop s, objects are required to be detected at 
$>2 \sigma$ levels in none of the \bFilter-, \vFilter-, and 
\iFilter-band images.
In addition, detections at $>3 \sigma$ levels are required
in all of the \jhFilter- and \hFilter-band images and at $>3.5 \sigma$ levels 
in at least one of these band images.
Magnitudes of \yFilter\ and \jFilter\ are replaced 
by their $0.9 \sigma$ limiting magnitudes
if the signal is below that level.
Finally, we remove objects whose pseudo-$\chi^2$ is larger than 2.8,
with $\chi^2 = \sum_i \mathrm{SGN} (f_i)(f_i / \sigma_i)^2$,
where the summation runs over all the ACS bands.
Here $f_i$ and $\sigma_i$ are the flux density and its uncertainty 
in the $i$-th band image,
respectively, and $\mathrm{SGN} (f_i)$ is the sign function,
whose definition is $\mathrm{SGN} (x) = 1$ if $x>0$, 
$\mathrm{SGN} (x) = 0$ if $x=0$, and 
$\mathrm{SGN} (x) = -1$ if $x<0$.
The selected dropout galaxies are presented in Tables~4--6 
in \citet{ishigaki17}. 

From the \citet{ishigaki17} samples,
we remove a \ydrop\ galaxy, HFF6P-1733-6559,
and a \yjdrop\ galaxy, HFF6P-1732-6562, in the \clsix\ parallel field,
because they appear to be spurious sources by visual inspection.
These are indeed the same object meeting both the $Y$- and 
$Y\!J$-dropout selections.
As a result, the total numbers of the selected galaxies are 350, 64, and 39
for $i$-, $Y$-, and $Y\!J$-dropouts, respectively.
Their photometric redshift distributions are shown
in Figure~\ref{fig:redshiftdist}.
The averages of the reliable ($z>4$) photometric redshifts of 
the $i$-, $Y$-, and $Y\!J$-dropouts are 
$z=6.2$, $7.8$, and $8.5$, respectively.
Therefore, we use $z=6$, $8$, and $9$ in the calculation of 
the sizes, magnitudes, and magnification factors
for $i$-, $Y$-, and $Y\!J$-dropouts, respectively.
Fixing the redshift to these values does not cause any systematic
errors in the following results.

\section{Size and Magnitude Measurements}\label{sec:sizemeasurement}

\subsection{Two-dimensional Profile Fitting}
In this subsection, we estimate lensing-corrected, i.e., intrinsic, 
sizes and magnitudes of the dropout galaxies.

The lensing effects are calculated using the software 
\texttt{glafic} v1.2.7 \citep{oguri10}.
For the mass distributions of \clone\ and \cltwo, 
we use our version 4 mass models updated to reflect 
the latest MUSE observations 
by \citet{mahler17} and \citet{caminha16}, respectively.
For \clthree\ and \clfour, we use our version 3 mass models 
constructed in K16.
For \clfive\ and \clsix, we newly construct version 4 mass models following 
the method established in K16.
Modeling details about the four version 4 mass models are described in Appendix~\ref{sec:appendix_models}.
All of the mass models are available on 
the Space Telescope Science Institute website\footnote
{\url{https://archive.stsci.edu/prepds/frontier/lensmodels/}}. 
The uncertainty in each magnification factor is calculated 
from ten-thousand models sampled from
a Markov chain Monte Carlo (MCMC) chain
(see Section~\ref{subsec:errorestimate}). 
This uncertainty is smaller than the scatter in
magnification factors among all modeling teams' models.
The typical scatters are $30\%$ at $\mu\sim2$ and 
$70\%$ at $\mu\sim40$ as reported in \citet{priewe17},
who have conducted a thorough comparison between the mass 
maps of \clone\ and \cltwo\ by all
modeling teams \citep[see also][]{meneghetti17}.
The smaller uncertainties in our models are
due to limited flexibilities inherent in parametric modeling
methods, while  the predicted magnification factors are consistent
with those by the other teams
\citep[see Figures 10--11 and 12--13 in][]{priewe17}.

The method to measure intrinsic sizes and magnitudes 
is identical to that in K15.
However, while the measurements in K15
were conducted only for bright galaxies,
here we deal with all the galaxies in the samples.
We fit a S\'ersic profile to a galaxy image in an $8\farcs4 \times 8\farcs4$ cutout image 
using a two-dimensional fitting algorithm conducted
by the command \texttt{optimize} in \texttt{glafic}, which simultaneously 
corrects for the lensing and point-spread function (PSF) effects. 
In order to correct for the lensing effects, an ellipsoidal S\'ersic profile on 
the source plane is lensed onto the image plane, and 
the galaxy image is fitted with the lensing-distorted S\'ersic profile.
In order to correct for the PSF effects, the lensing-distorted S\'ersic profile
is convolved with an average stellar image on the image plane, 
which is generated by stacking
$5$--$20$ stars found in each field.
The S\'ersic profile is defined as
\begin{align}
\Sigma(r) = \Sigma_{0} \exp\left[-b_{n} \left( \frac{r}{r_{\rm e}} \right)^{1/n} \right ],
\end{align}
where $\Sigma(r)$, $\Sigma_{0}$, $b_{n}$, $r_{\rm e}$, and $n$ represent 
the surface brightness profile, surface brightness at $r=0$, parameter 
to convert the scale radius to the half-light radius,
half-light radius, and S\'ersic index, respectively.
The ellipticity $e$ and position angle are introduced 
by a simple variable transformation 
\citep[see][for details]{oguri10}.
In what follows, $r_{\rm e}$ means the circularized half-light radius, 
$r^{\rm maj}_{\rm e} \sqrt{1-e}$, where $r^{\rm maj}_{\rm e}$ 
is the radius along the major axis.
The magnitude is calculated from $r_{\rm e}$ and $\Sigma_{0}$. 
During the fitting, the S\'ersic index is fixed to $n=1$ and 
the maximum ellipticity is set to 0.9.
A uniform sky background is assumed, and the normalization
is optimized at the same time.
When nearby objects may introduce any bias to the fitting result, we mask 
these objects or add additional profiles to fit the nearby objects simultaneously.
The fittings are conducted using the $Y\!J\!J\!H$, $J\!J\!H\!H$, 
and $J\!H\!H$ combined 
images at $z\sim6-7$, 8, and 9, respectively.
Although we have already constructed size 
samples in K15 from the \clone\ cluster and parallel fields,
we conduct the fittings again because there are updates 
on the mass map of the cluster.
The obtained morphological properties and magnitudes are presented 
in Tables~\ref{tab:i-results}--\ref{tab:yj-results} in 
Appendix~\ref{sec:fitting_results}.
The fitting results for galaxies fainter than $-18$ mag
are also graphically shown in 
Figures~\ref{fig:faceimages_z7_1} and \ref{fig:faceimages_z89}
in Appendix~\ref{sec:fitting_results}.

\subsection{Error Estimations}\label{subsec:errorestimate}

\begin{figure}[t]
  \centering
      \includegraphics[width=\linewidth]{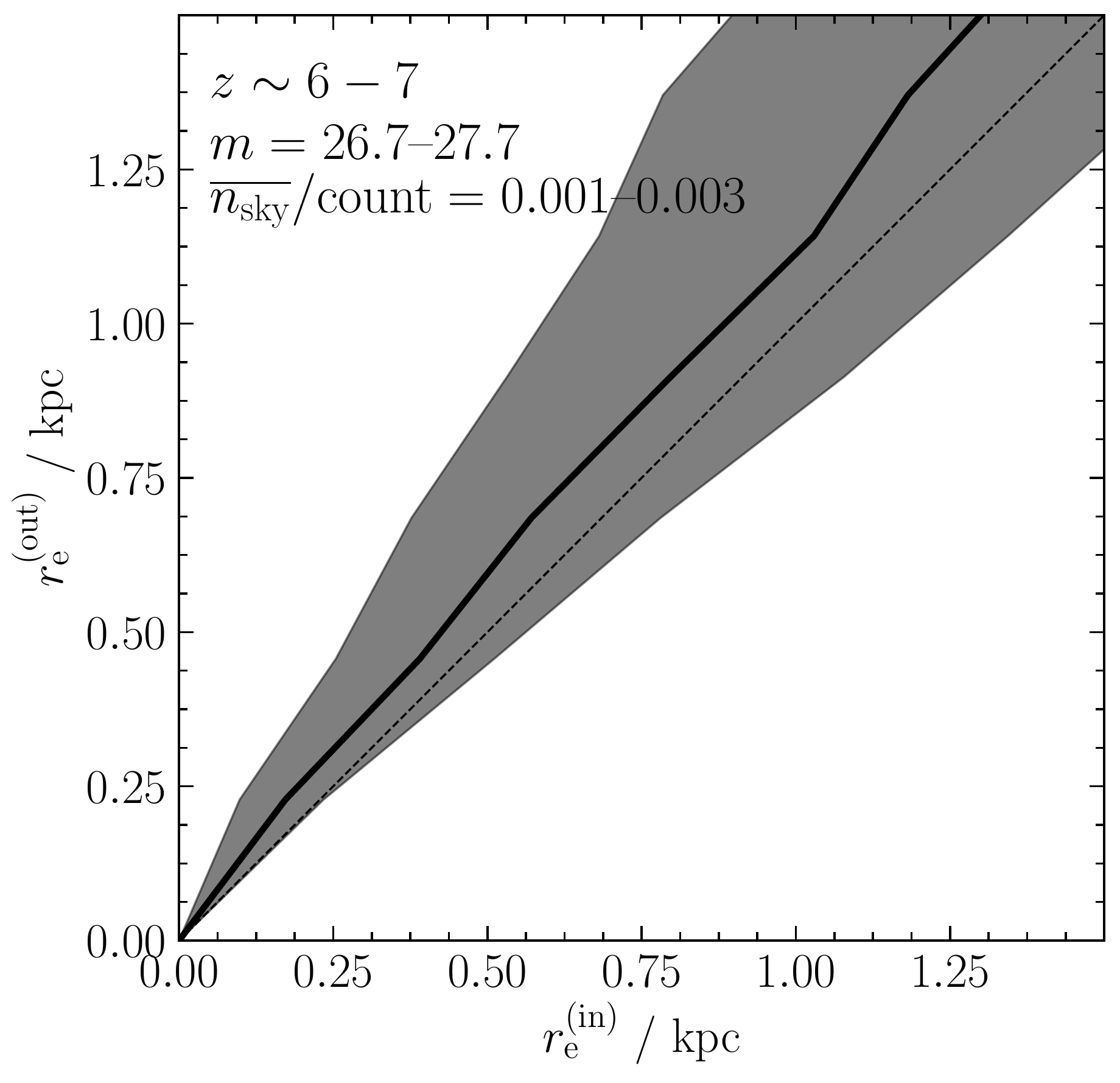}
      \includegraphics[width=\linewidth]{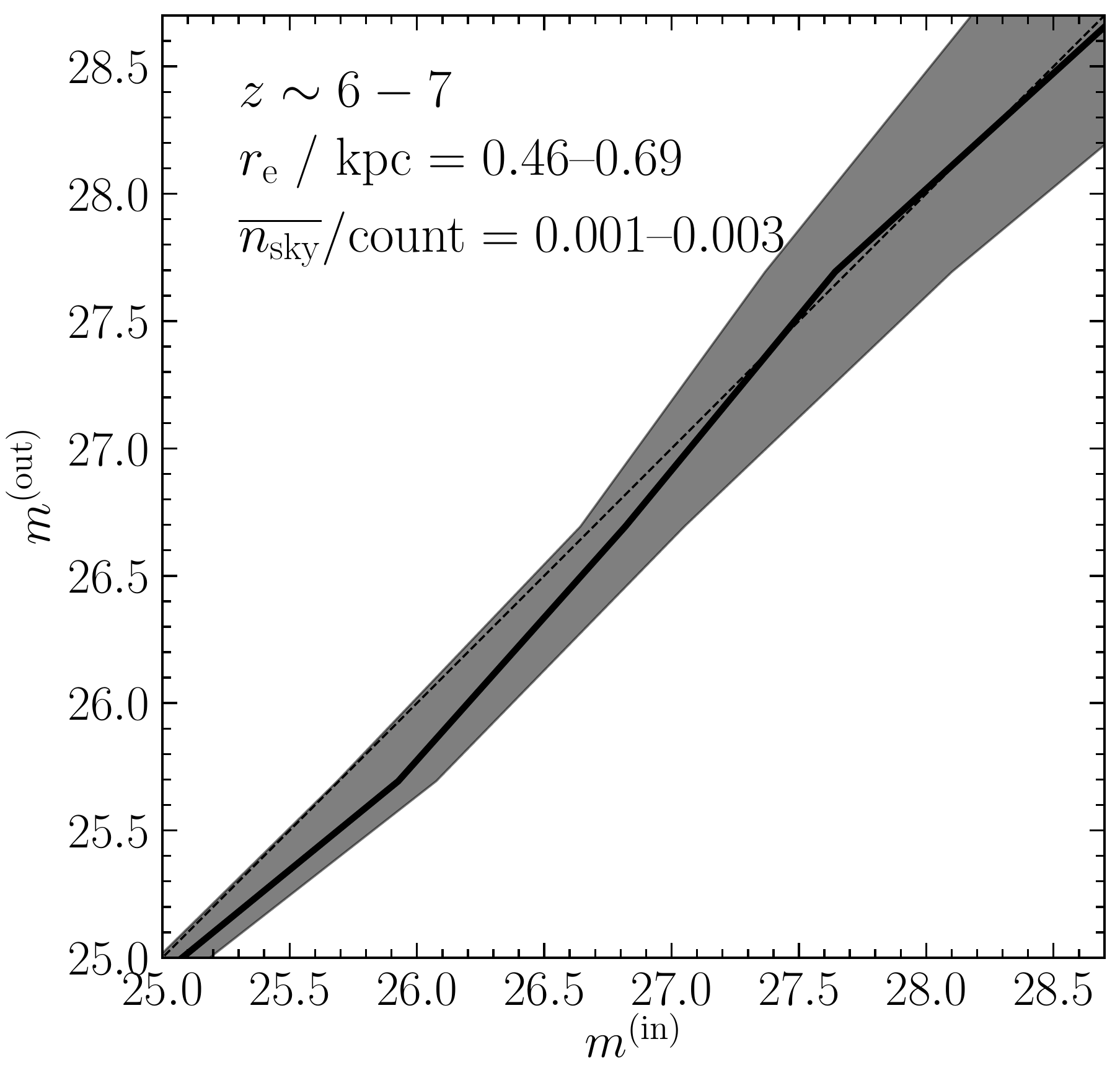}
  \caption{Examples of the Monte Carlo simulations
  to estimate the systematic and random errors in size and magnitude measurements.
  The top panel shows the median and $1\sigma$ distribution of 
  output radii as a function of input radius for galaxies with an
  apparent magnitude of $m/\mathrm{mag} = 26.7$--$27.7$ and for a sky value of
  $\overline{n_\mathrm{sky}} / \mathrm{count} = 0.001$--$0.003$.
  The bottom panel shows the median and $1\sigma$ distribution of 
  output magnitudes as a function of input magnitude for galaxies with 
  $r_\mathrm{e} / \mathrm{kpc} = 0.46$--$0.69$ and for a sky value of 
  $\overline{n_\mathrm{sky}} / \mathrm{count} = 0.001$--$0.003$.
}
  \label{fig:simulation}
\end{figure}

In this subsection, we evaluate errors in the measured sizes and 
magnitudes following the method in K15, but in a more efficient way.
We consider two sources of errors: errors in the fitting procedure
and errors in the mass map.

There are two types of errors in the fitting procedure.
One is a systematic bias, 
by which the sizes and magnitudes of larger (smaller) 
galaxies are underestimated (overestimated).
The other is a random error, which arises from random sky noise that 
disperses the estimated size and magnitude.
In order to estimate these errors, we conduct 
Monte Carlo simulations, 
in which we bury simulated galaxies in a real image and 
perform the same fitting 
procedure as for real dropout galaxies.
Since these systematic and random errors are primarily
dependent on the galaxy apparent magnitude, apparent radius, and
sky value in the vicinity, we estimate the two errors
as a function of the three parameters.
We use the \clone\ cluster field image for this derivation
and apply the relation to all twelve fields.
In detail, first, we select a random position in the image and 
bury an $n=1$ S\'ersic profile, whose 
magnitude, radius, ellipticity, and position angle are chosen randomly.
Second, we conduct the same procedure on this pseudo-galaxy as for 
real galaxies.
We repeat these two processes until we obtain a sufficient 
number of measurements in each parameter bin.
Third, for each real dropout galaxy, we choose a set of simulated galaxies 
whose apparent magnitudes, apparent radii, and sky values in the vicinity
are close to those of the dropout galaxy. Using the intrinsic magnitudes and radii
of the simulated galaxies in this set, we estimate the random errors 
and correct for the systematic errors in size and magnitude.
Examples of the Monte Carlo simulations are presented in Figure~\ref{fig:simulation}.

Systematic errors in mass maps also affect measurement results.
Since the apparent magnitudes and sizes of lensed galaxies are converted into
intrinsic values
using mass maps, an overestimate of the magnification factor results in 
an underestimate of the intrinsic sizes and magnitudes, and vice versa.
In order to estimate the errors in magnification, we generate an MCMC chain
of the mass model parameters using the 
command \texttt{mcmc} in \texttt{glafic}.
From ten-thousand samples in the chain, we estimate the error in magnification 
factor at the positions 
of each dropout galaxy with the \texttt{mcmc\_calcim} command.
For each cluster, one hundred mass maps generated from randomly selected 
MCMC samples are available on the Space Telescope Science 
Institute website.

The random errors in size and magnitude
due to the fitting procedure and random errors in 
magnification factor
are presented in
Tables~\ref{tab:i-results}--\ref{tab:yj-results}.

\section{Size--luminosity Distributions at $z\sim 6-9$}\label{sec:RLrelations}
In this section, we first present the distribution of our galaxies
on the size--luminosity plane.
Then, detection incompleteness is calculated as a function of 
absolute magnitude and size for each field and redshift range.
Finally, we use these incompleteness maps on the size--luminosity plane to 
simultaneously derive intrinsic size--luminosity relations and 
luminosity functions for the first time at these redshift ranges.

\subsection{Galaxy Distribution on the Size--luminosity Plane}

\begin{figure*}[t]
  \centering
      \includegraphics[width=0.8\linewidth]{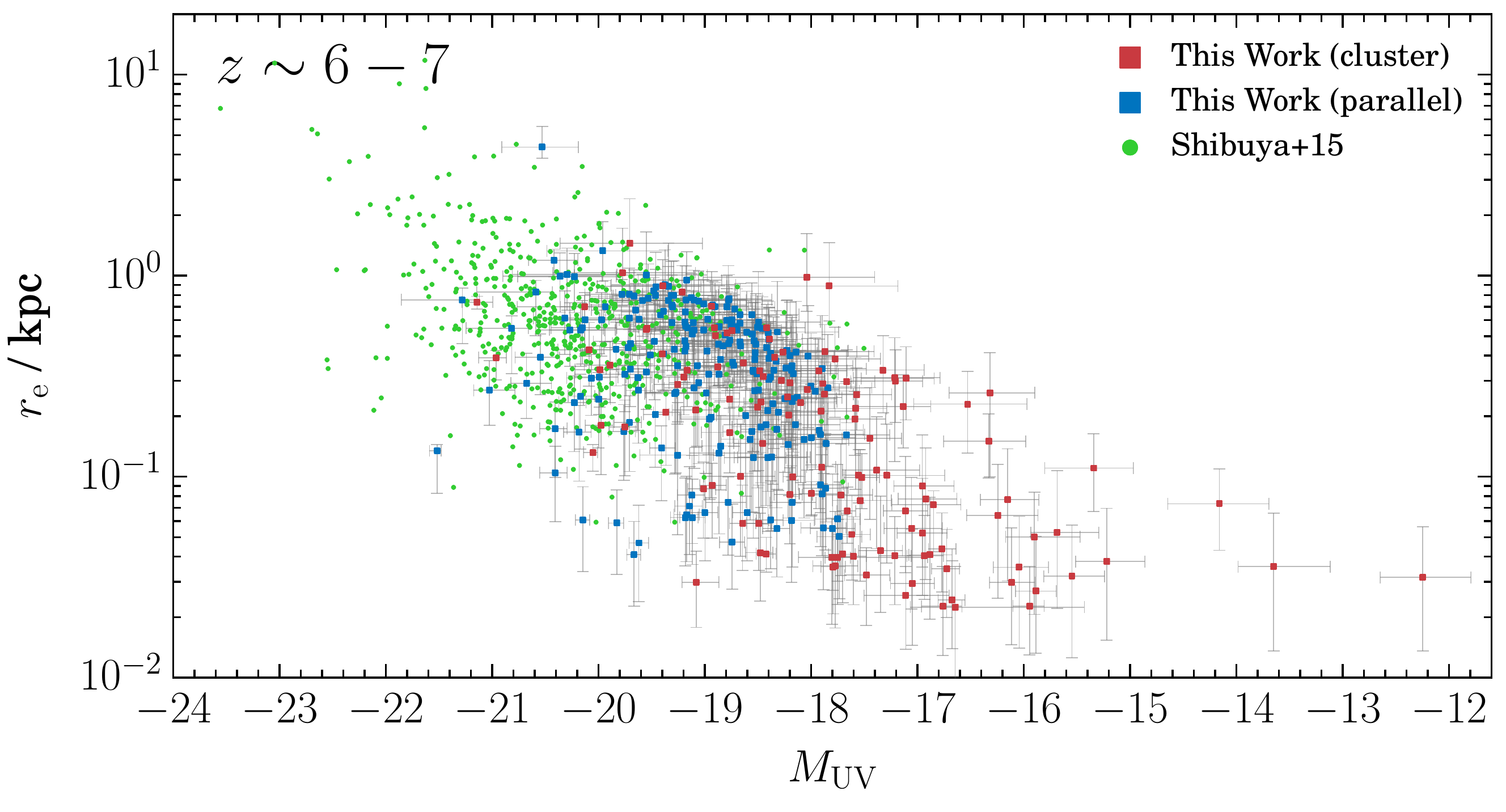}
      \includegraphics[width=0.8\linewidth]{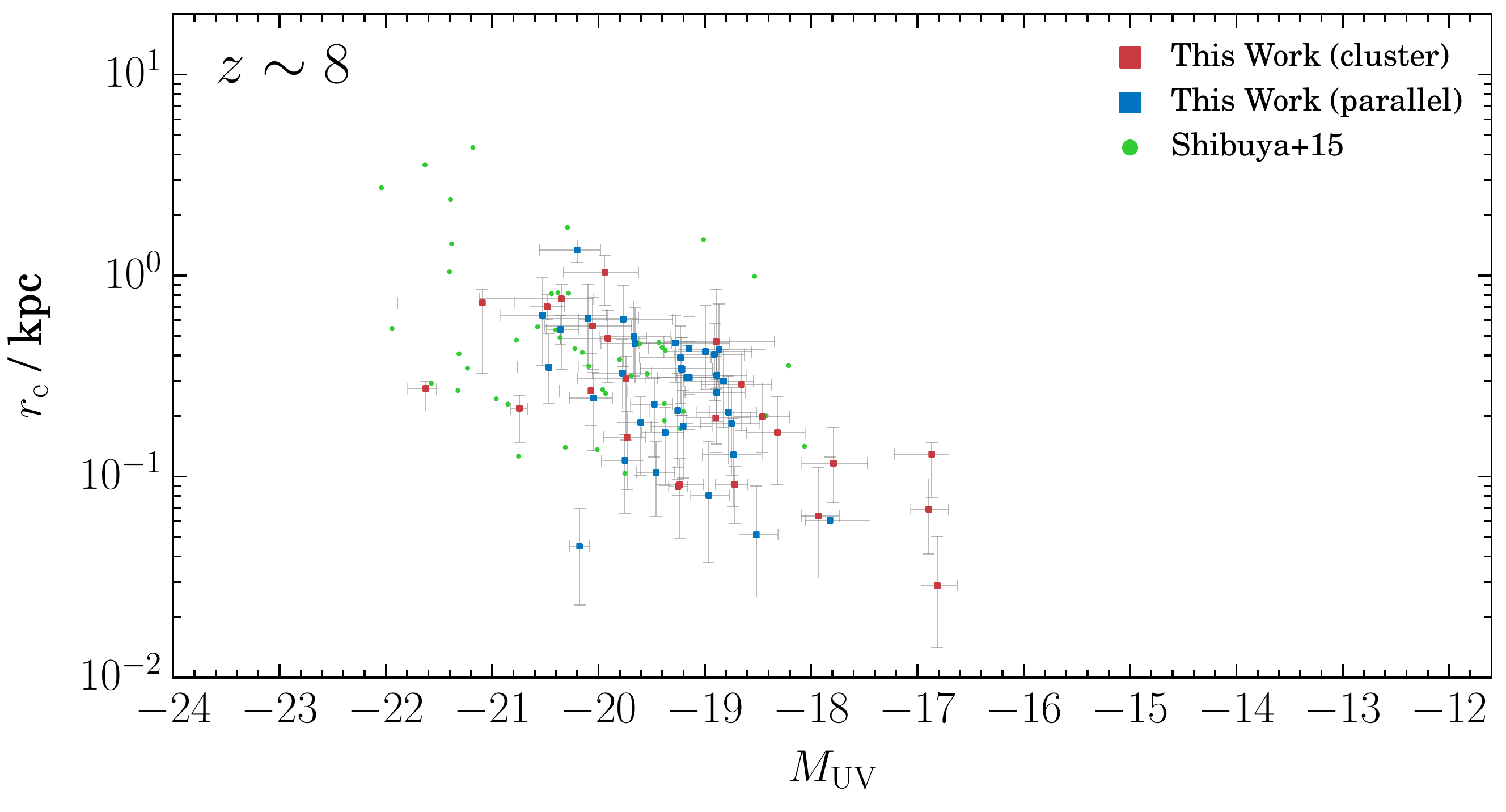}
      \includegraphics[width=0.8\linewidth]{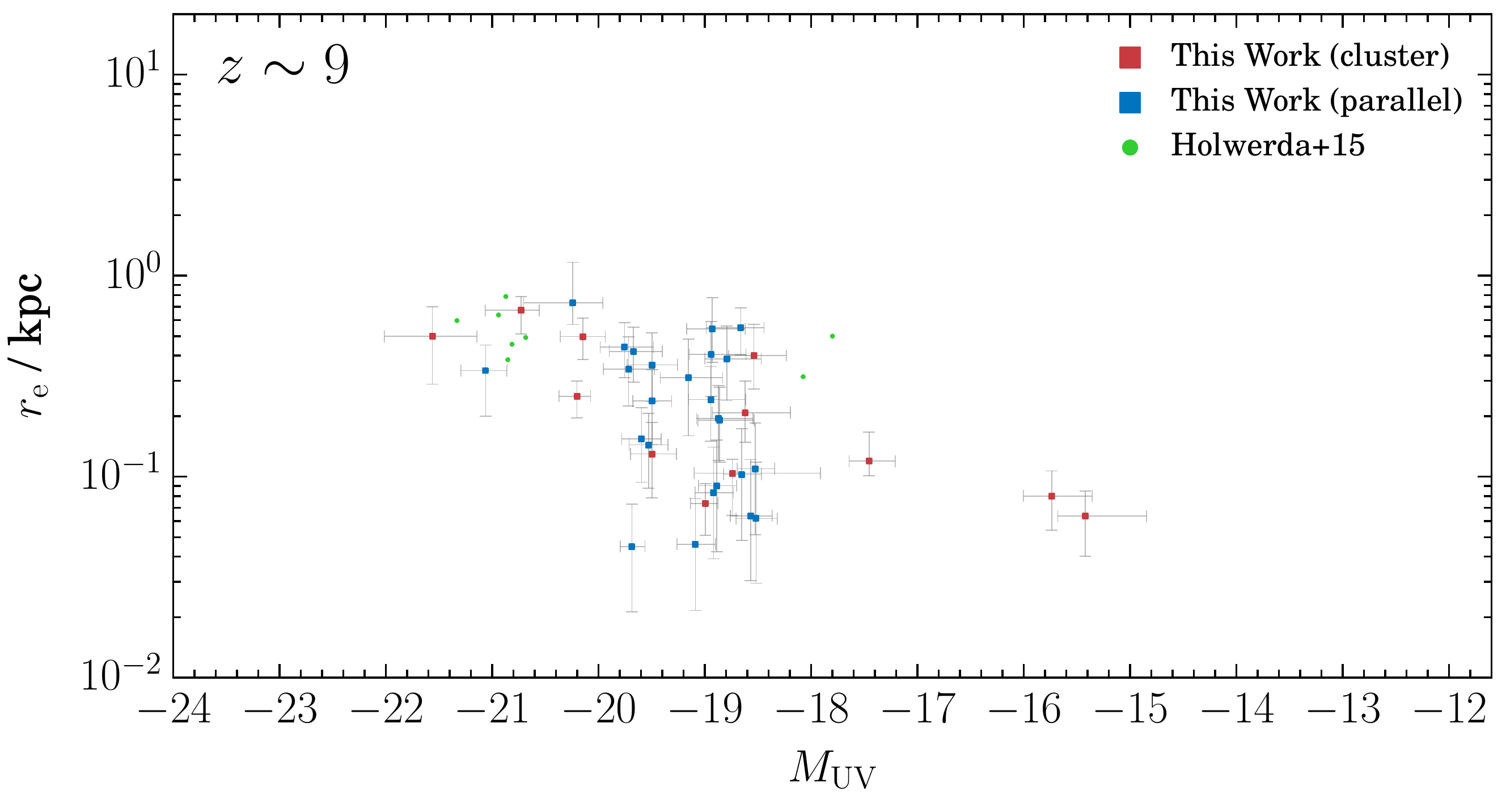}
  \caption{Galaxy distributions on the size--luminosity plane 
  at $z\sim6-7$ ({\it top}), 
  $8$ ({\it middle}), and $9$ ({\it bottom}).
  The red and blue points represent our galaxies from
  the cluster fields and parallel fields, respectively.
  The green points represent galaxies at $z\sim6-7$ and $z\sim8$
  in \citet{shibuya15} 
  ({\it top} and {\it middle}, respectively) and at $z\sim9$ in \citet{holwerda15} ({\it bottom}).
  }
  \label{fig:rlrelations}
\end{figure*}

\begin{deluxetable*}{lllll}
\tablecolumns{5}
\tabletypesize{\scriptsize}
\tablecaption{Number of $M_{\mathrm{UV}}\gtrsim-18$ galaxies in the present and previous samples
\label{tab:numberofgalaxies}}
\tablewidth{0pt}
\tablehead{
\colhead{References} &
\colhead{$z\sim6-7$} &
\colhead{$z\sim8$} &
\colhead{$z\sim9$} &
\colhead{Data}
}
\startdata
This work & 91 (350) & 7 (64) & 3 (39) & Six HFF cluster and parallel fields\\
\citet{ono13} & 0 (9) & 0 (6) & --- & HUDF12 \\
\citet{holwerda15} & ---  & --- & 1 (8) & XDF and CANDELS\\
\citet{kawamata15} & 4 (31) & 0 (8) & --- & First HFF cluster and parallel fields\\
\citet{shibuya15} & $7+1$ $(422+173) \tablenotemark{a}$ & 0 (46) & --- & CANDELS, HUDF09/12, and first two HFF parallel fields \\
\citet{bouw17size} & 47 (76)  & --- & --- & First two HFF cluster fields
\enddata
\tablecomments{The number of galaxies in the full sample is shown in parentheses.}
\tablenotetext{a}{Numbers at $z\sim6$ and $7$ are presented.}
\end{deluxetable*}

Figure~\ref{fig:rlrelations} shows the size--luminosity distributions 
of our galaxies at 
$z\sim6-7$, $8$, and $9$, together with those from previous studies
that adopt
two-dimensional profile fittings in size measurements. 
The error bars include the errors in the fitting 
process and our mass maps. 
Our samples occupy either the same regions as the previous samples
or their reasonable extrapolations toward much fainter magnitudes.

As summarized in Tables~\ref{tab:i-results}--\ref{tab:yj-results}, 
some galaxies are multiply imaged on the image plane. 
The physical parameters of these galaxies are calculated 
by averaging over the multiple images.
The numbers of independent galaxies with size measurements are 
thus reduced to 334, 61, and 37 at $z\sim6-7$, 8, and 9, respectively.
Among them, the numbers of faint ($M_{\mathrm UV} \gtrsim -18$) galaxies 
are 83, six, and three, respectively.
These numbers should be compared only with those from previous
studies that adopt parametric size measurements such as \texttt{GALFIT} 
\citep{peng02, peng10}, not with those based on  
nonparametric methods such as 
``curve-of-growth.''
This is because these two methods rely on 
different assumptions, which may introduce different biases
and therefore make comparisons of the results difficult.
At faint magnitude ranges, as investigated in this work,
previous studies that adopt parametric size measurements 
are \citet{ono13}, K15, \citet{holwerda15}, \citet{shibuya15}, 
and \citet{bouw17size} \citep[see also][]{oesch10b}.
The numbers of galaxies in our samples and in the previous studies 
are presented in Table~\ref{tab:numberofgalaxies}.
For $z\sim6-7$ and $9$, the addition of our samples increases the numbers of 
faint ($M_{\mathrm UV} \gtrsim -18$) galaxies with size measurements 
about $2.5$ and $4$ times, respectively.
For $z\sim8$, our sample is the first that contains faint galaxies
with size measurements.
The faintest objects among the previous samples have
$M_\mathrm{UV} \simeq-14.48$ \citep{bouw17size}, 
$-18.1$ \citep{shibuya15}, and $-17.8$ \citep{holwerda15}
at $z\sim6-7$, 8, and 9, respectively.
We push the faint limits down to $M_\mathrm{UV} \simeq-12.3$, $-16.8$, and
$-15.4$ at $z\sim6-7$, $8$, and 9, respectively.

\subsection{Completeness Estimation}
\label{subsec:completeness}

\begin{figure}[t]
  \centering
      \includegraphics[width=\linewidth]{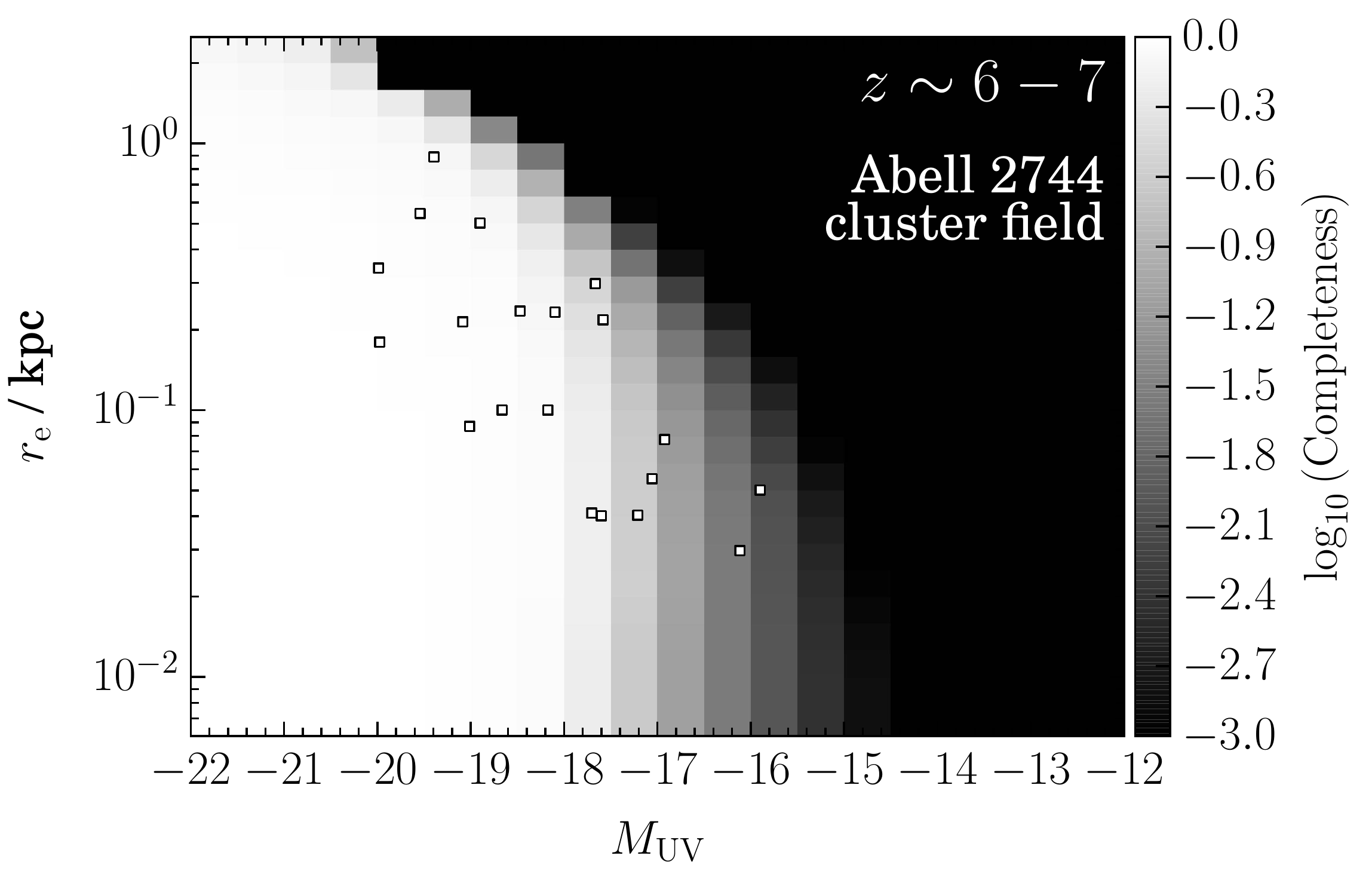}
      \includegraphics[width=\linewidth]{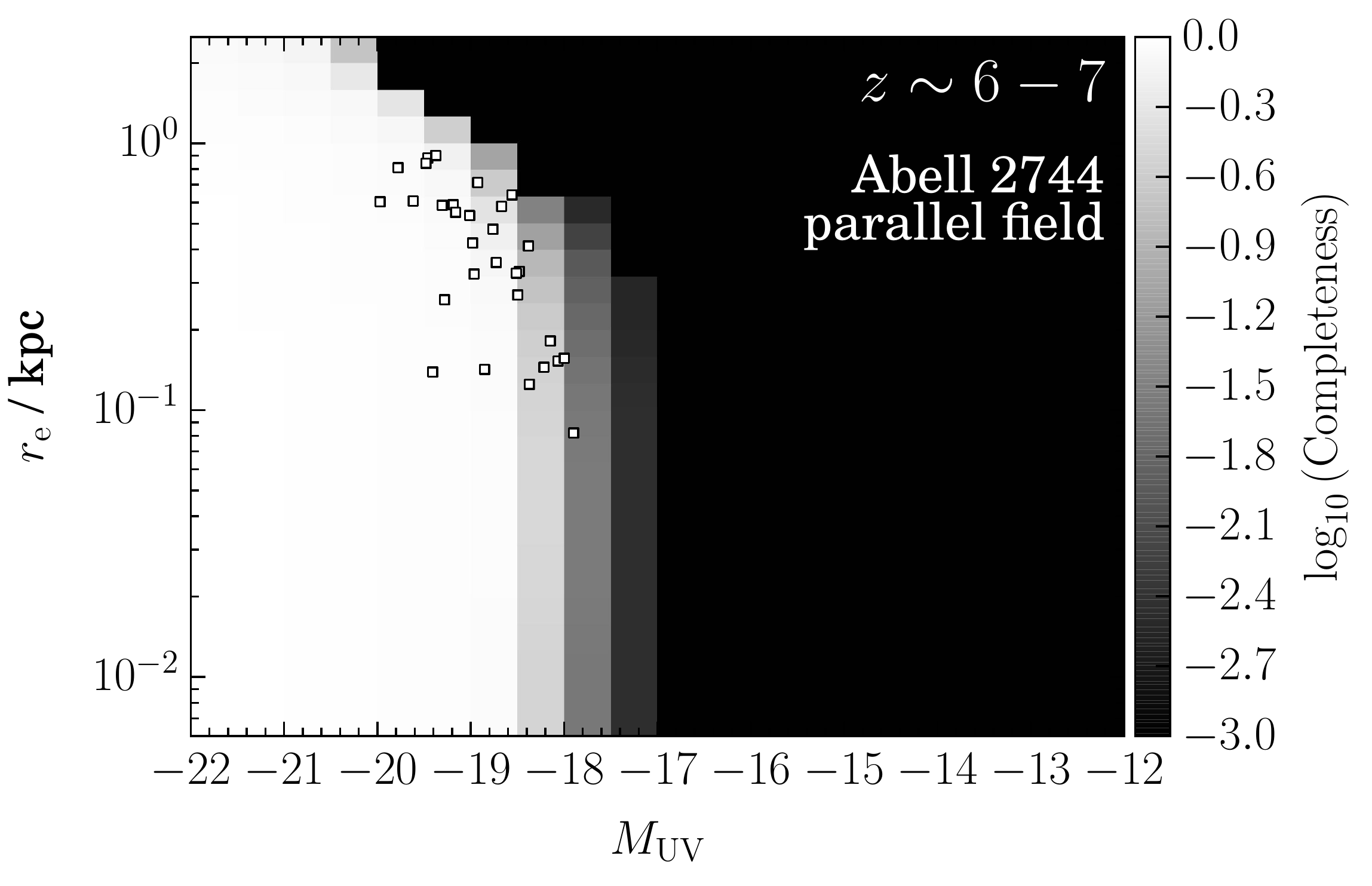}
  \caption{Detection completeness at $z\sim 6-7$ as a 
  function of absolute magnitude 
  and size for the \clone\ 
  cluster (\textit{top}) and parallel (\textit{bottom}) fields, shown on a logarithmic scale.
  Galaxies detected in each field are plotted 
  with squares.
  }
  \label{fig:completeness}
\end{figure}

For a given total magnitude, galaxies with larger sizes
are less likely to be detected in observations because
of their low surface brightnesses.
Since this effect is more prominent for
fainter objects, observed size--luminosity relations can 
become significantly steeper than intrinsic ones.
We conduct the following Monte Carlo simulations to 
calculate detection completeness as a function of absolute 
magnitude and size.
The detection completeness is defined
as the fraction of galaxies that are detected and pass 
the dropout selection described in Section~\ref{subsec:sampleselection}.
(1) We select random positions uniformly on the source plane.
(2) For each position, we generate an artificial galaxy with a certain size 
and magnitude and place it, taking the lensing and PSF effects into account,
into the combined image, which is used as the detection image in 
the catalog construction.
The galaxy is modeled with a S\'ersic profile of the index $n=1$.
The ellipticity is randomly chosen from a uniform distribution
between 0 and 0.9.
(3) We run \texttt{SExtractor} on the image with artificial galaxies 
and calculate the fraction of artificial galaxies that are detected 
by \texttt{SExtractor} and bright enough to meet the criteria of dropout selection. 
(4) We repeat steps (1)--(3), changing the size and magnitude of 
artificial galaxies.
It should be noted that we do not assume any specific 
spectral energy distribution (SED) shape.
This is because, primarily, the completeness is not dependent on the 
SED shape but only on size and magnitude.
As an example, the obtained completeness maps at $z\sim 6-7$ in 
the \clone\ cluster and parallel fields are shown in 
Figure~\ref{fig:completeness}.
Note that although faint galaxies are bright enough to be detected
if highly magnified, their completeness is significantly low
because they rarely fall onto highly magnified regions.

As seen in Figure~\ref{fig:2Ddistributions}, the observed size--luminosity distributions
can be significantly deformed by incompleteness, which depends on 
size and luminosity.
We discuss the impact of incompleteness on the estimation of 
the intrinsic size--luminosity relations in Section~\ref{subsec:67relations}.
In the cluster fields, even galaxies fainter than $\sim-18$ mag 
are detected, but with low completeness.
For example, at $M_\mathrm{UV} = -16$, only those with $r_\mathrm{e}<0.1$ kpc are 
included in the samples.
This means that while the HFF has opened a window to faint galaxies, 
it is open only to very small objects.
On the other hand, galaxies detected in the parallel fields 
are limited to $\sim -18$ mag, but with a relatively high completeness over 
a wide size range because completeness drops sharply at $M_\mathrm{UV} \sim -18$.
Therefore, the cluster fields require a more careful consideration 
of incompleteness effects.

\subsection{Maximum-likelihood Estimation\\of the Intrinsic Size--luminosity Distribution}
\label{subsec:mle}

\begin{figure}[t]
  \centering
      \includegraphics[width=\linewidth]{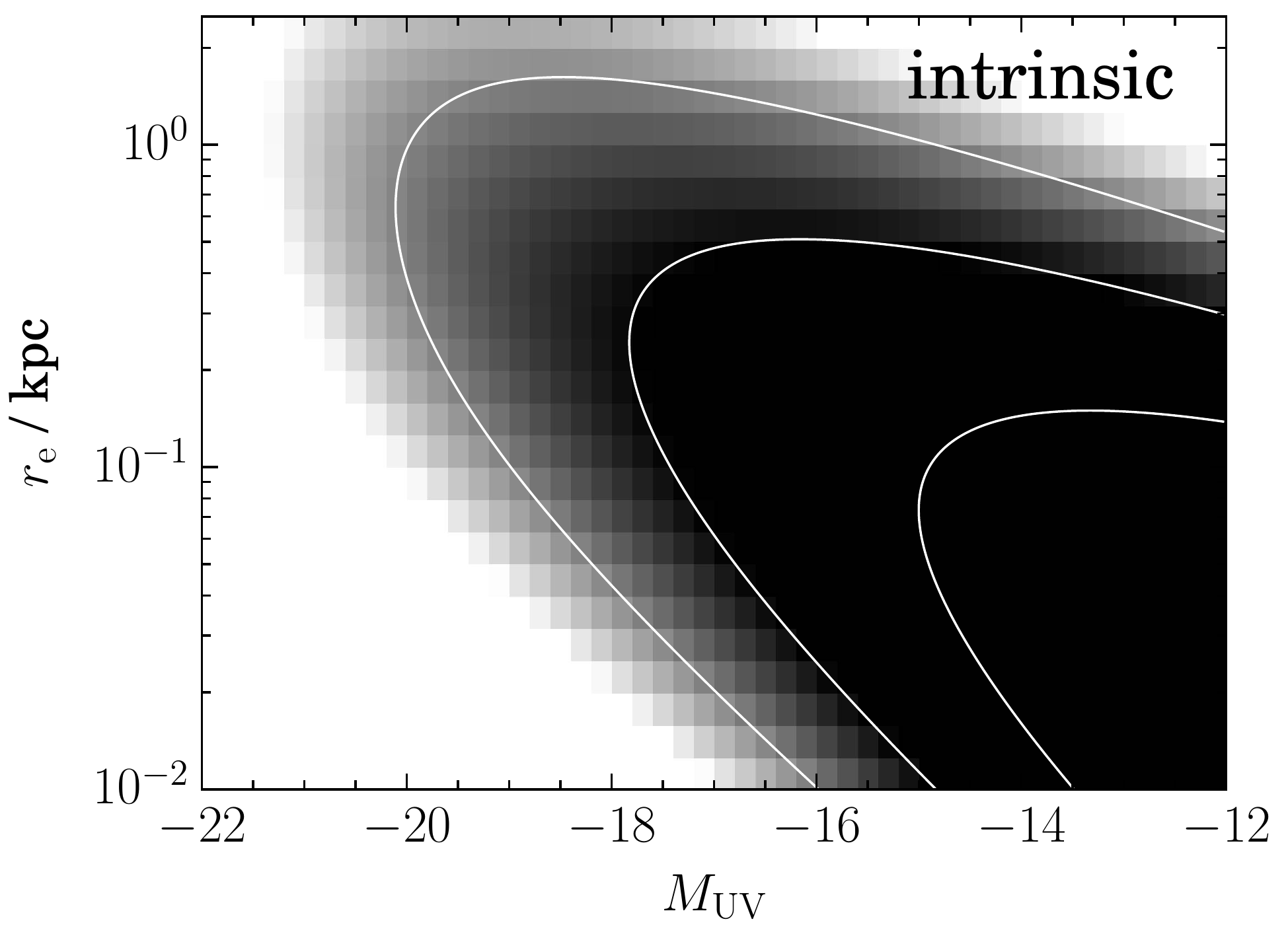}
      \includegraphics[width=\linewidth]{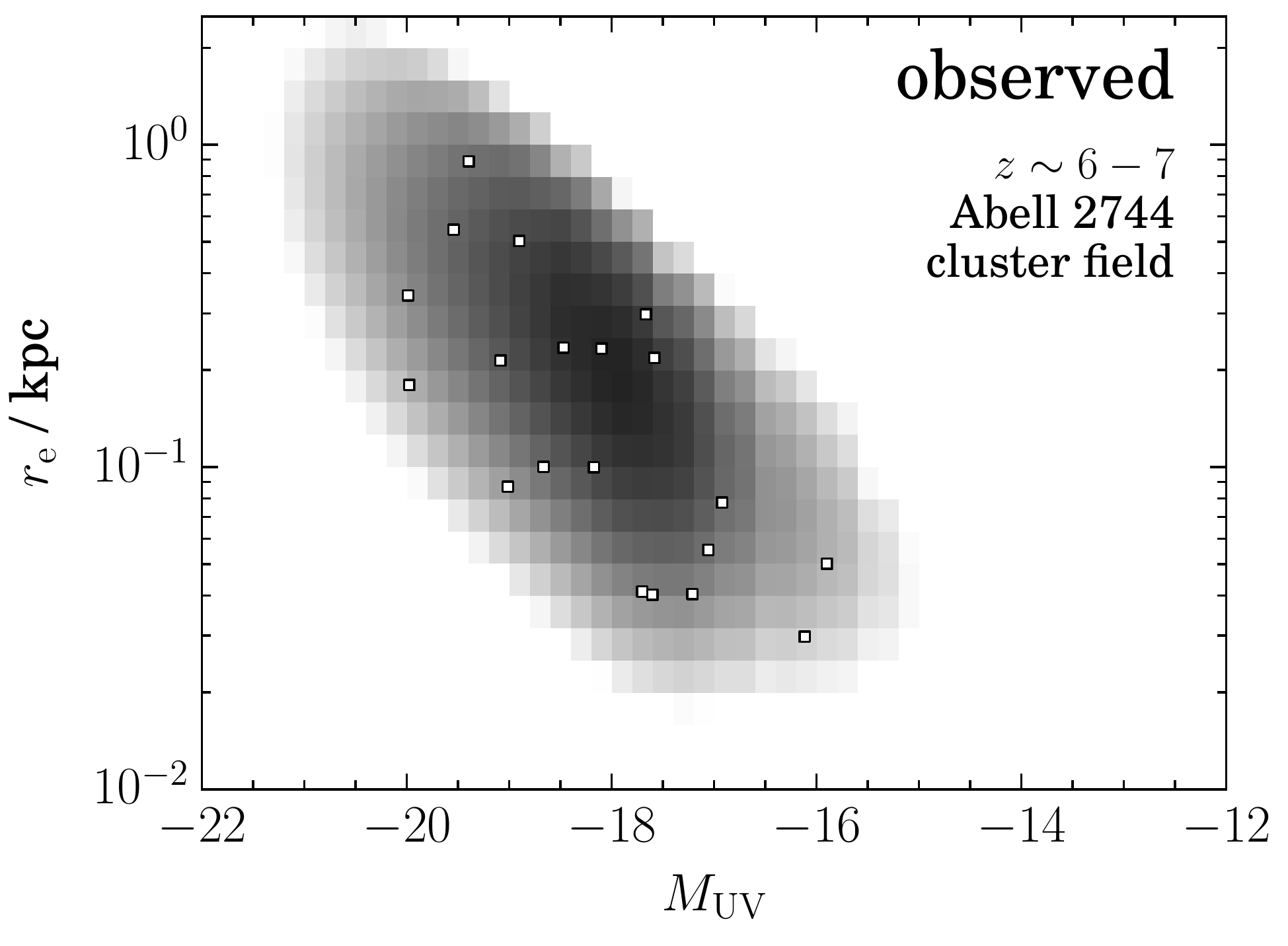}
      \includegraphics[width=\linewidth]{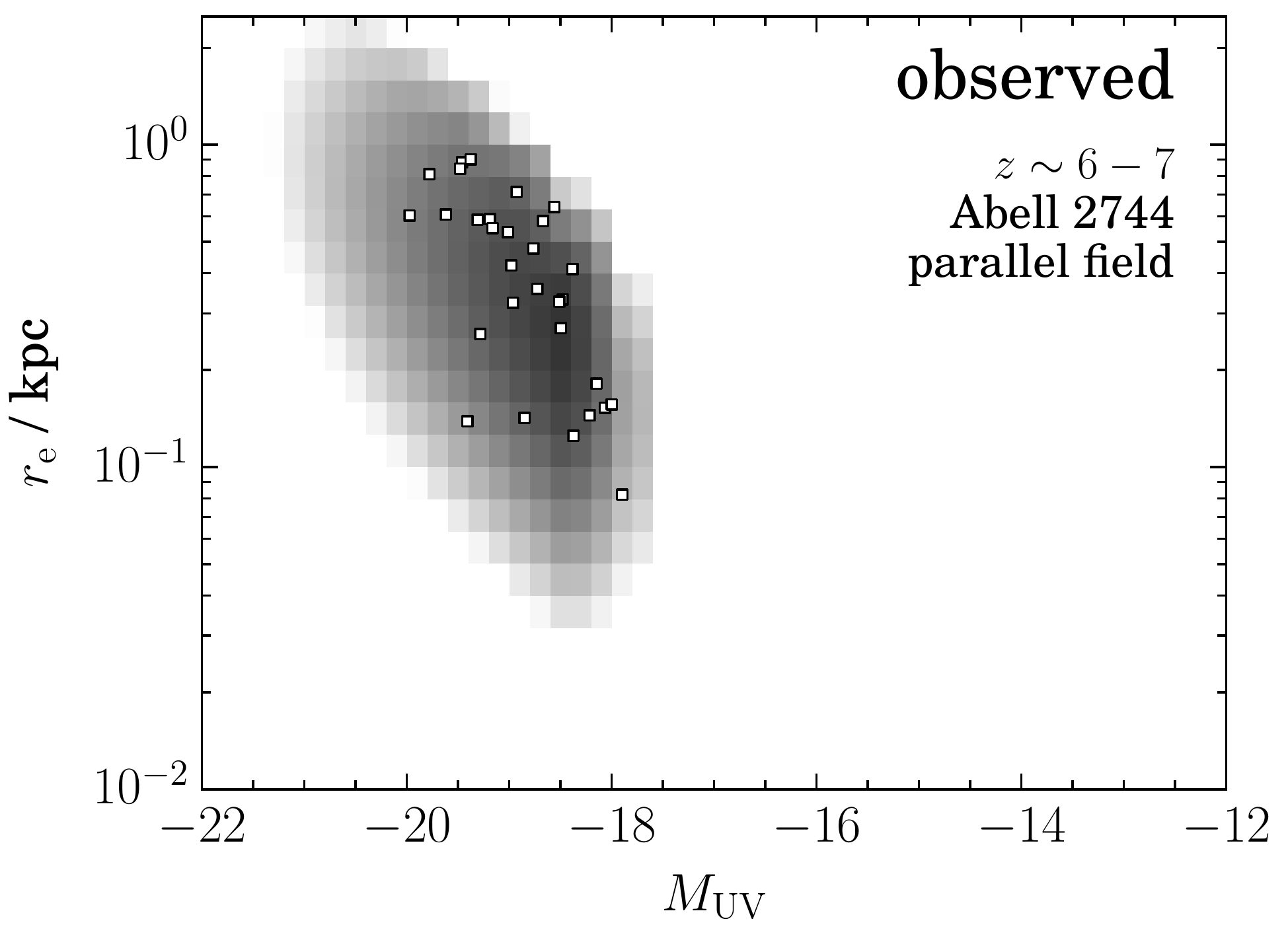}
  \caption{Bivariate probability distributions of $z\sim6-7$ galaxies
  on the size--luminosity plane shown on a logarithmic scale.
  The top panel shows the intrinsic distribution with an arbitrary normalization.
  The contour levels are logarithmically equidistant with 1 dex steps.
  The middle and bottom panels are for the observed distributions 
  in the \clone\ cluster and parallel fields, respectively, 
  calculated by multiplying the intrinsic distribution by the completeness 
  map for each field. 
  Galaxies detected in each field are shown with squares in the 
  lower two panels.
  The parameters of the intrinsic bivariate distribution presented
  here are the best-fit parameters obtained in Section~\ref{subsec:mle}.}
  \label{fig:2Ddistributions}
\end{figure}

In this subsection, we obtain for each of the three redshift ranges
the incompleteness-corrected or intrinsic bivariate size--luminosity 
distribution of galaxies, which is a product of the intrinsic 
size--luminosity relation and the luminosity function.
We model the size--luminosity relation by a log-normal distribution with three free
parameters while modeling the luminosity function by a Schechter function with two 
free parameters; the total number of free parameters is thus five.
Then, by multiplying the intrinsic distribution by the incompleteness map,
we model the observed size--luminosity distribution of galaxies.
Maximum-likelihood estimation (MLE) is used to obtain the best-fit 
values of these parameters that best reproduce the observed bivariate 
distribution.

This bivariate method has been exploited in \citet{dejonglacey00} and \citet{huang13} 
to simultaneously derive the size--luminosity relation and 
UV luminosity function for
local spiral galaxies and LBGs at $z\sim4-5$, respectively.
A similar method has also been adopted in \citet{schmidt14lf}.
This method has two advantages over binning methods conventionally adopted
as described in \citet{schmidt14lf}; one is that no information is lost 
because data are not binned, and the other is that photometric errors 
in magnitude are also considered.
In addition, by determining the size--luminosity relation and luminosity
function simultaneously, we are able to evaluate the degeneracy between
those two relations.
Furthermore, in most previous studies, size--luminosity relations 
have been determined to minimize the residuals in size,
which is equivalent to MLE that assumes observed galaxies have
a flat distribution in luminosity. 
On the other hand, our method correctly derives
the size--luminosity relation and, consequently, the 
luminosity function because
the luminosity distribution is also modeled using luminosity functions.

The probability density function (PDF) of the intrinsic galaxy distribution on
the size--luminosity plane 
$\Psi(r_{\rm e}, M_\mathrm{UV})$ is modeled as 
\begin{align}
&\Psi(r_{\rm e}, M_\mathrm{UV}; r_{0}, \sigma, \beta, M^{*}, \alpha) \nonumber\\
&= P(r_{\rm e}, M_\mathrm{UV}; r_{0}, \sigma, \beta) \, \phi(M_\mathrm{UV}; M^{*}, \alpha),
\label{eqn:intrinsicmodel}\end{align}
where $P(r_{\rm e}, M_\mathrm{UV})$ is the PDF of size
and $\phi(M_\mathrm{UV})$ is that of luminosity.
As $P(r_{\rm e}, M_\mathrm{UV})$, we adopt a log-normal distribution described as 
\begin{align}
P(r_{\rm e}, M_\mathrm{UV}; r_{0}, \sigma, \beta) = \frac{1}{\sigma r_{\rm e} \sqrt{2\pi}} \exp\left[-\frac{\ln^{2}(r_{\rm e}/\overline{r_{\rm e}})}{2\sigma^{2}}\right], \label{eqn:sizedistribution}
\end{align}
where 
\begin{align}
\overline{r_{\rm e}}(L) = r_{0} \left(\frac{L}{L_{0}}\right)^{\beta},
\end{align}
and $r_{0}$, $\sigma$, $\beta$, and $L_{0}$ are the modal radius at $M_\mathrm{UV} = -21$, 
width of the log-normal distribution, slope of the size--luminosity relation, 
and luminosity corresponding to $M_\mathrm{UV} = -21$, respectively.
As $\phi(M_\mathrm{UV})$, we adopt a Schechter function described as 
\begin{align}
&\phi(M_\mathrm{UV}; M^{*}, \alpha) \nonumber\\
&= 10^{-0.4(\alpha+1)(M_\mathrm{UV} - M^{*})} \exp{\left[-10^{-0.4(M_\mathrm{UV} - M^{*})}\right]},
\end{align}
where $M^{*}$ and $\alpha$ are the characteristic magnitude
and power-law slope at the faint end.
Note that we do not determine the normalization parameter $\phi_{*}$
of the Schechter function because we are interested not in 
the absolute number of galaxies but only in their relative distribution 
on the size--luminosity plane.

\begin{figure}[t]
  \centering
      \includegraphics[width=\linewidth]{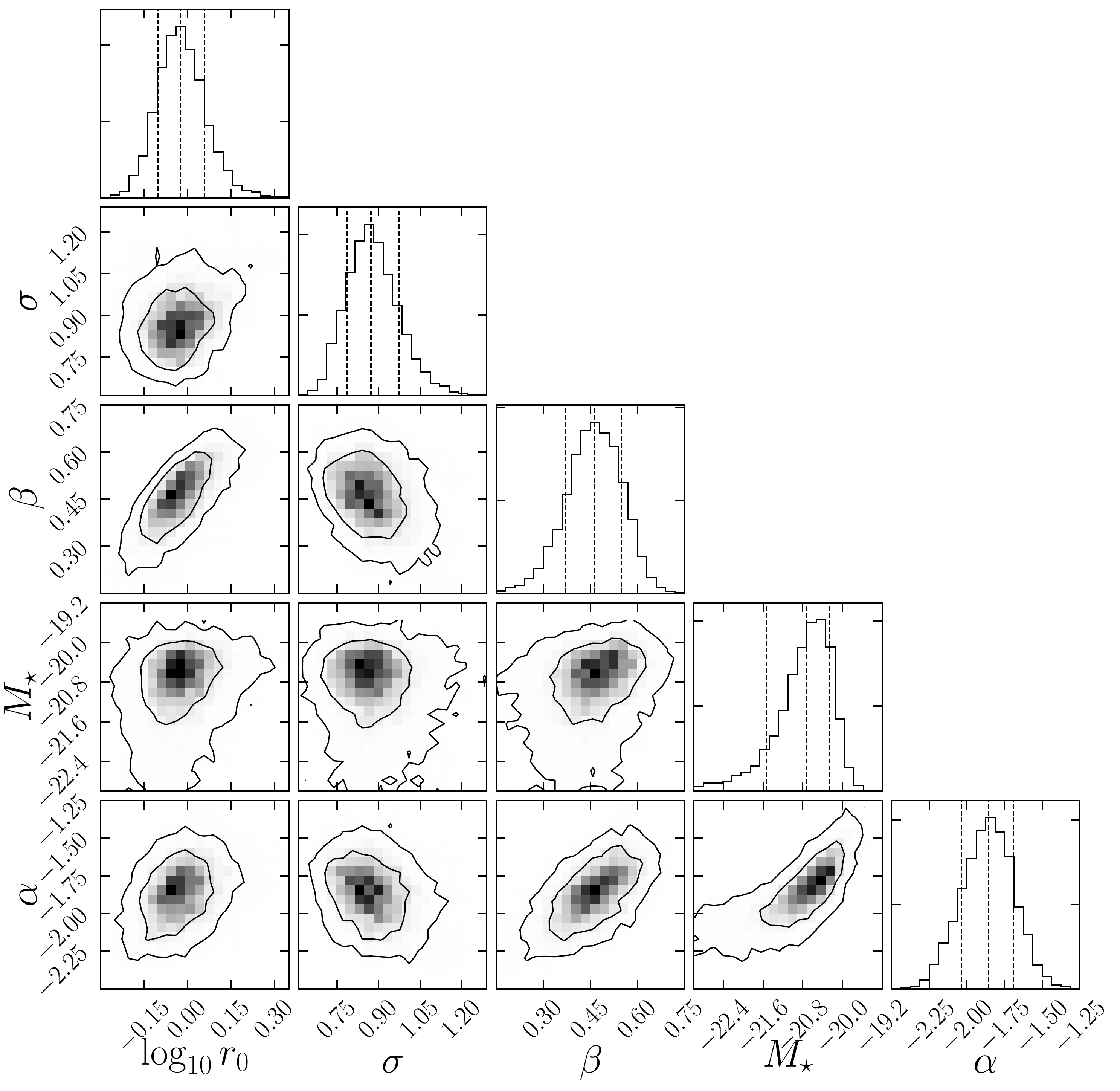}
  \caption{Two-dimensional projections of the MCMC samples at $z\sim6-7$.
  The inner and outer contours represent $68\%$ and $95\%$
  confidence intervals. The three vertical dashed lines on the histograms show the
 16th, 50th, and 84th percentiles.
  Plotted using the \texttt{corner.py} module \citep{foremanmackey16}.
  }
  \label{fig:mcmcz7}
\end{figure}

\begin{figure}[t]
  \centering
      \includegraphics[width=\linewidth]{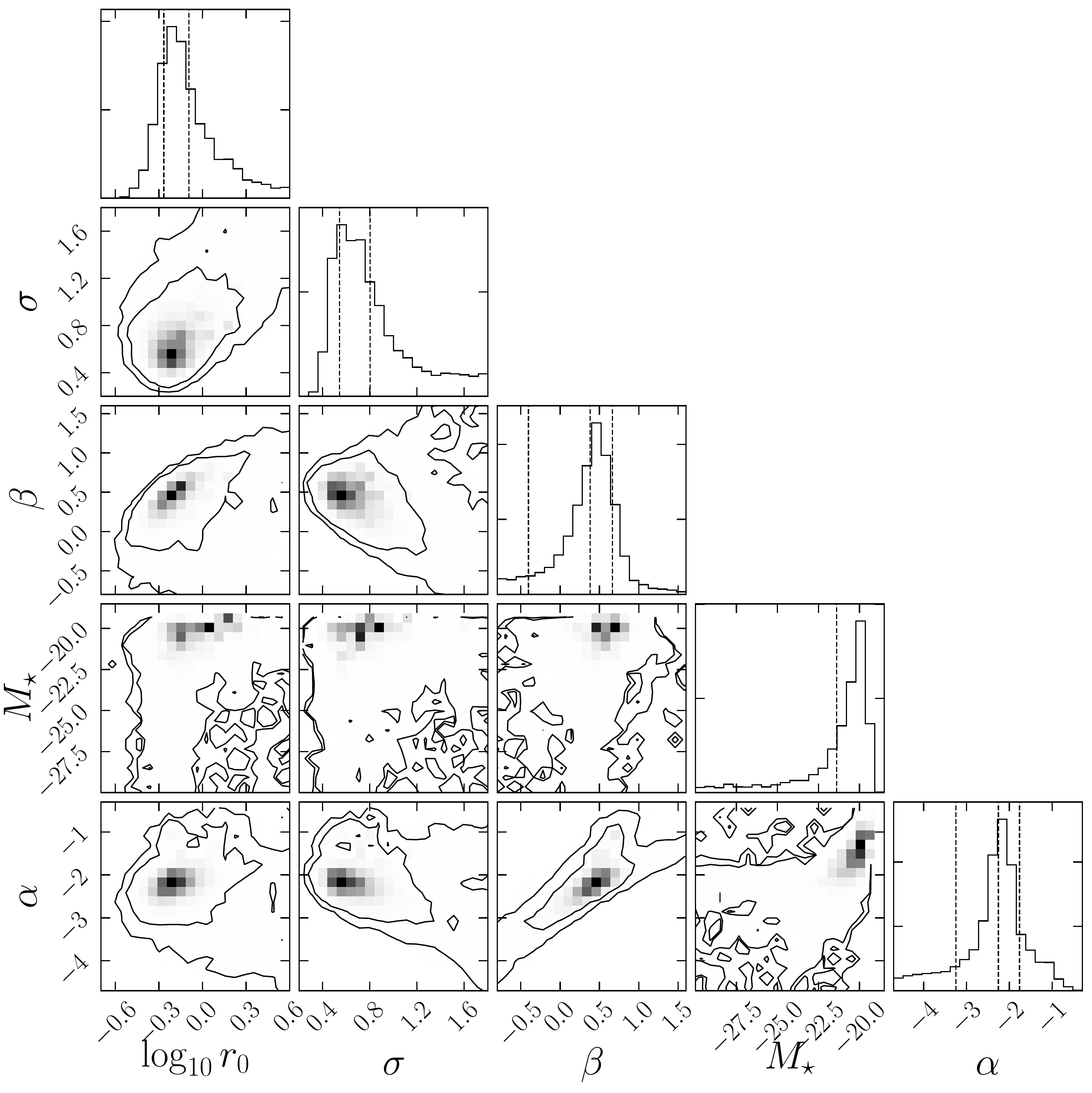}
  \caption{Same as Figure~\ref{fig:mcmcz7} but for $z\sim8$.
  }
  \label{fig:mcmcz8}
\end{figure}

\begin{figure}[t]
  \centering
      \includegraphics[width=\linewidth]{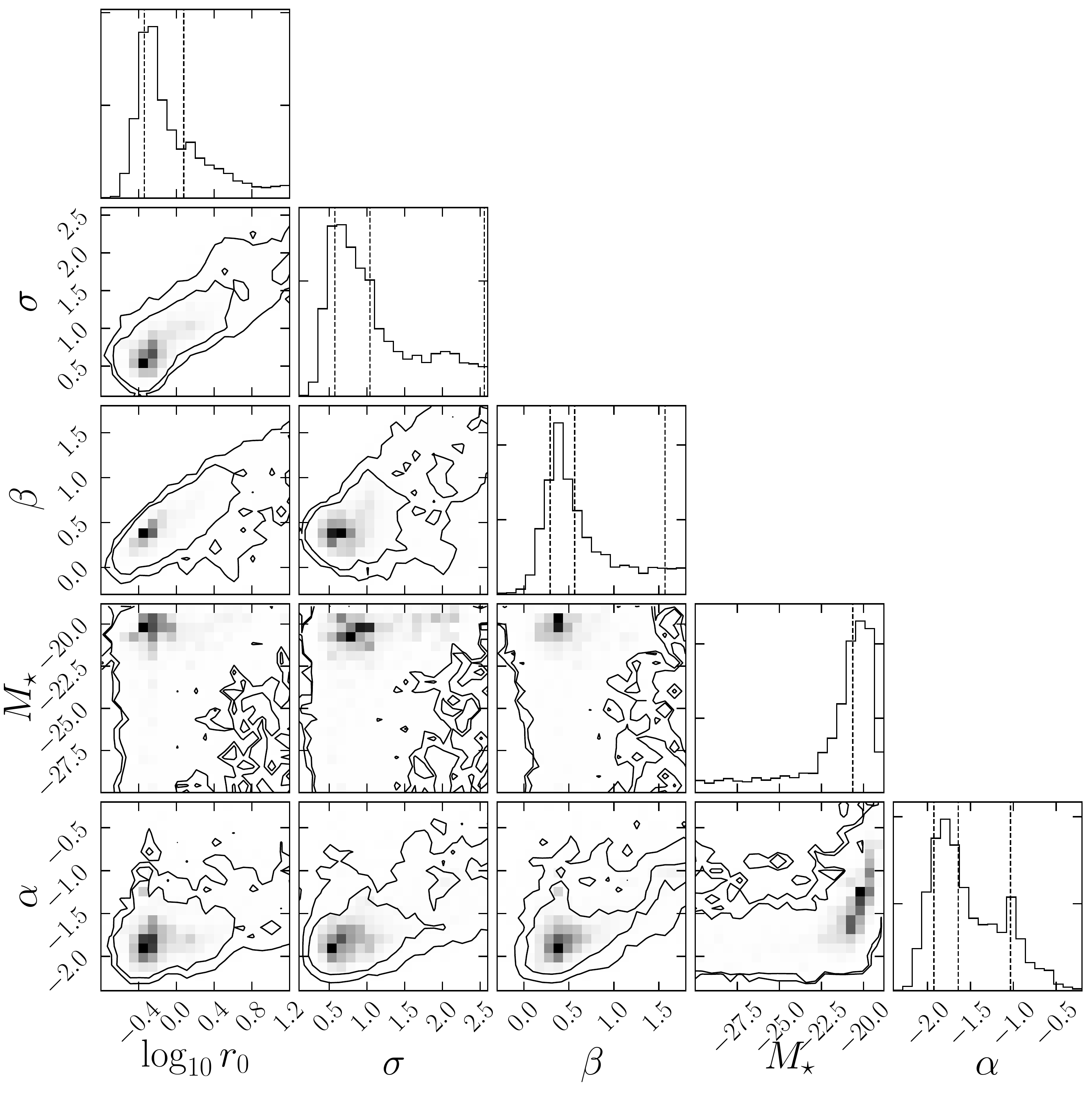}
  \caption{Same as Figure~\ref{fig:mcmcz7} but for $z\sim9$.
  }
  \label{fig:mcmcz9}
\end{figure}

\begin{deluxetable*}{llllll}
\tablecolumns{6}
\tabletypesize{\scriptsize}
\tablecaption{Best-fit parameters of size--luminosity relations and luminosity functions
\label{tab:bestparams}}
\tablewidth{0pt}
\tablehead{
\colhead{References} &
\colhead{$r_{0} / \mathrm{kpc}$} &
\colhead{$\sigma$} &
\colhead{$\beta$} &
\colhead{$M^{*}$} &
\colhead{$\alpha$} 
}
\startdata
$z\sim6-7$ &  &  &  &  & \\
This work & $0.94^{+0.20}_{-0.15}$ & $0.87^{+0.10}_{-0.09}$ & $0.46^{+0.08}_{-0.09}$ & $-20.73^{+0.46}_{-0.81}$ & $-1.86^{+0.17}_{-0.18}$\\
This work (mode) & $0.94$ & $0.86$ & $0.44$ & $-20.56$ & $-1.86$\\
This work (LF fixed) & $0.95^{+0.18}_{-0.14}$ & $0.86^{+0.09}_{-0.07}$ & $0.47^{+0.06}_{-0.06}$ & $[-20.73]$ & $[-1.86]$\\
This work (apparent) & $0.75$ & $0.66$ & $0.52$ & \multicolumn{1}{c}{---} & \multicolumn{1}{c}{---} \\[6pt]
\citet{atek15b} & $[0.81]$ & $[0.90]$ & $[0.25]$ & $-20.89^{+0.60}_{-0.72}$ & $-2.04^{+0.17}_{-0.13}$\\
\citet{bouw15} & \multicolumn{1}{c}{---\tablenotemark{a}} & \multicolumn{1}{c}{---\tablenotemark{a}} & $[\sim 0.25]$\tablenotemark{a,b} & $-20.94^{+0.20}_{-0.20}$ & $-1.87^{+0.10}_{-0.10}$\\
\citet{laporte16} & $[0.81]$ & $[0.90]$ & $[0]$ & $-20.33^{+0.37}_{-0.47}$ & $-1.91^{+0.26}_{-0.27}$\\
\citet{livermore16} & $[0.5]$ & $[0]$ & $[0]$ & $-20.819^{+0.044}_{-0.034}\ {}^{+0.001}_{-0.031}$ & $-2.10^{+0.03}_{-0.03}\ {}^{+0.07}_{-0.01}$\\
\citet{ishigaki17} & \multicolumn{1}{c}{---\tablenotemark{c}} & \multicolumn{1}{c}{---\tablenotemark{c}} & $[\sim 0.25]$\tablenotemark{b,c} & $-20.89^{+0.17}_{-0.13}$ & $-2.15^{+0.08}_{-0.06}$\\
\citet{bouw17magnif} & $[0.80]$ & $[0.69]$ & $[0.27]$ & $[-20.94]$ & $-1.92^{+0.04}_{-0.04}$\\
\tableline 
$z\sim8$ &  &  &  &  &  \\
This work & $0.81^{+5.28}_{-0.26}$ & $0.80^{+1.07}_{-0.26}$ & $0.38^{+0.28}_{-0.78}$ & $-151.98^{+130.60}_{-314.19}$ & $-2.26^{+0.49}_{-0.99}$\\
This work (mode) & $0.58$ & $0.56$ & $0.44$ & $-19.95$ & $-2.14$\\
This work ($M_{*}$ fixed) & $0.75^{+0.53}_{-0.16}$ & $0.65^{+0.35}_{-0.14}$ & $0.50^{+0.16}_{-0.21}$ & $[-20.73]$ & $-1.80^{+0.22}_{-0.30}$\\
This work (LF fixed) & $0.69^{+0.24}_{-0.14}$ & $0.62^{+0.18}_{-0.12}$ & $0.49^{+0.13}_{-0.14}$ & $[-20.73]$ & $[-1.86]$\\
This work (apparent) & $0.57$ & $0.48$ & $0.52$ & \multicolumn{1}{c}{---} & \multicolumn{1}{c}{---} \\[6pt]
\citet{bouw15} & \multicolumn{1}{c}{---\tablenotemark{a}} & \multicolumn{1}{c}{---\tablenotemark{a}} &  $[\sim 0.25]$\tablenotemark{a,b} & $-20.63^{+0.36}_{-0.36}$ & $-2.02^{+0.23}_{-0.23}$\\
\citet{laporte16} & $[0.81]$ & $[0.90]$ & $[0]$ & $-20.32^{+0.49}_{-0.26}$ & $-1.95^{+0.43}_{-0.40}$\\
\citet{livermore16} & $[0.5]$ & $[0]$ & $[0]$ & $-20.742^{+0.195}_{-0.152}\ {}^{+0.006}_{-0.014}$ & $-2.02^{+0.08}_{-0.07}\ {}^{+0.01}_{-0.03}$\\
\citet{ishigaki17} & \multicolumn{1}{c}{---\tablenotemark{c}} & \multicolumn{1}{c}{---\tablenotemark{c}} & $[\sim 0.25]$\tablenotemark{b,c} & $-20.35^{+0.20}_{-0.30}$ & $-1.96^{+0.18}_{-0.15}$\\
\tableline 
$z\sim9$ &  &  &  &  & \\
This work & $1.20^{+367.64}_{-0.74}$ & $1.04^{+1.52}_{-0.46}$ & $0.56^{+1.01}_{-0.27}$ & $-82.74^{+62.10}_{-763.40}$ & $-1.64^{+0.61}_{-0.28}$\\
This work (mode) & $0.42$ & $0.54$ & $0.40$ & $-19.80$ & $-1.82$\\
This work ($M_{*}$ fixed) & $0.59^{+0.61}_{-0.16}$ & $0.69^{+0.40}_{-0.20}$ & $0.42^{+0.17}_{-0.15}$ & $[-20.73]$ & $-1.59^{+0.19}_{-0.18}$\\
This work (LF fixed) & $0.53^{+0.27}_{-0.13}$ & $0.68^{+0.27}_{-0.18}$ & $0.34^{+0.13}_{-0.14}$ & $[-20.73]$ & $[-1.86]$\\
This work (apparent) & $0.43$ & $0.47$ & $0.39$ & \multicolumn{1}{c}{---} & \multicolumn{1}{c}{---}\\[6pt]
\citet{oesch13} & \multicolumn{1}{c}{---} & \multicolumn{1}{c}{---} & \multicolumn{1}{c}{---} & $-18.8^{+0.3}_{-0.3}$ & $[-1.73]$\\
\citet{laporte16} & $[0.81]$ & $[0.90]$ & $[0]$ & $[-20.45]$ & $-2.17^{+0.41}_{-0.43}$\\
\citet{ishigaki17} & \multicolumn{1}{c}{---\tablenotemark{c}} & \multicolumn{1}{c}{---\tablenotemark{c}} & $[\sim 0.25]$\tablenotemark{b,c} & $-51.39^{+18.51}_{-44.73}$ & $-2.22^{+0.26}_{-0.17}$
\enddata
\tablecomments{Numbers in square brackets are fixed during the fitting.}
\tablenotetext{a}{Size--luminosity relation is presented in their Appendix D.}
\tablenotetext{b}{Effective slope of the size--luminosity relation, although its parameterization is different from ours.}
\tablenotetext{c}{Size--luminosity relation is presented in their paper and the bottom panel of our Figure~\ref{fig:detected_fraction}.}
\end{deluxetable*}

The observed size--luminosity distribution $\Psi'$ in the $i$-th field 
is modeled by multiplying
the parameterized intrinsic size--luminosity distribution and 
the completeness map in that field $\mathcal{C}_{i}$ obtained 
in Section~\ref{subsec:completeness},
\begin{align}
\Psi'_{i} (r_{\mathrm{e}}, M_{\mathrm{UV}}; r_{0}, \sigma, \beta, M^{*}, \alpha) \equiv \mathcal{N}_{i} \, \Psi (r_{\mathrm{e}}, M_{\mathrm{UV}}) \, \mathcal{C}_{i} (r_{\mathrm{e}}, M_{\mathrm{UV}}),
\label{eqn:psiobserved}\end{align}
where $\mathcal{N}_{i}$ is 
the normalization parameter to make the volume unity.
The probability that a galaxy with $(r_\mathrm{e}, r_\mathrm{e}+d r_\mathrm{e})$ 
and $(M_\mathrm{UV}, M_\mathrm{UV} + d M_\mathrm{UV})$
is found is $\Psi'(r_\mathrm{e}, M_\mathrm{UV})\ dr_\mathrm{e}\ dM_\mathrm{UV}$.
In order to calculate the probability of the $j$-th galaxy in the $i$-th
field $f_{i,j}$ considering 
the observed errors in size and magnitude,
we convolve the modeled observed size--luminosity distribution $\Psi'$
with a two-dimensional gaussian centered on the observed size and 
magnitude, whose variances are equal to their observed errors,
\begin{align}
&f_{i,j} =\int \, dr'_{\mathrm{e}} \, dM'_{\mathrm{UV}} \nonumber\\
&\times \Psi'_{i} (r'_{\mathrm{e}}, M'_{\mathrm{UV}}) \, g(r'_{\mathrm{e}}, M'_{\mathrm{UV}}; r_{\mathrm{e}, j}, M_{\mathrm{UV},j}, \delta r_{\mathrm{e},j}, \delta M_{\mathrm{UV},j}),
\label{eqn:gaussian}\end{align}
where $g(r'_{\mathrm{e}}, M'_{\mathrm{UV}}; r_{\mathrm{e}, j}, M_{\mathrm{UV},j}, \delta r_{\mathrm{e},j}, \delta M_{\mathrm{UV},j})$
is a gaussian function whose peak is at the observed size and magnitude
($r_{\mathrm{e}, j}, M_{\mathrm{UV},j}$)
and the variances are equal to their observed errors 
$(\delta r_{\mathrm{e},j}, \delta M_{\mathrm{UV},j})$.
The likelihood in the $i$-th field $\mathcal{L}_{i}$ is given by
\begin{align}
\mathcal{L}_{i} (r_{0}, \sigma, \beta, M^{*}, \alpha) = \prod_{j} f_{i,j}(r_{0}, \sigma, \beta, M^{*}, \alpha).
\end{align}
The total likelihood $\mathcal{L}$ is 
the product of the likelihood in each field,
\begin{align}
\mathcal{L} \equiv \prod_{i} \mathcal{L}_{i}.
\end{align}

We use the MCMC procedure
to estimate the best-fit values and uncertainties for the five parameters
and the degeneracy between them.
We assume flat priors on all five parameters.
Note that we do not use the galaxies HFF5P-1940-3315 at $z\sim6-7$ 
and HFF5P-2129-2064 at $z\sim8$ in the \clfive\ parallel field
because they are outliers.
For the MCMC sampling, we use the public software \texttt{emcee}
\citep{foremanmackey13}. 
The MCMC results are shown in Table~\ref{tab:bestparams}
and Figures~\ref{fig:mcmcz7}--\ref{fig:mcmcz9}.
As an example, the obtained intrinsic bivariate size--luminosity 
distribution at $z\sim6-7$ is presented in the top panel of 
Figure~\ref{fig:2Ddistributions}. 

\section{Discussion}\label{sec:discussion}
In this section, we first discuss the intrinsic size--luminosity relations 
and luminosity functions at $z\sim6-9$.
Second, we construct a model to reproduce
the steep size--luminosity relation at $z\sim6-7$ using the result of the 
abundance matching in \citet{behroozi13}.
Third, we show that there are large uncertainties in the $z>6$ luminosity functions 
derived in previous studies because of a large variance in the 
assumed size--luminosity relations and that those uncertainties 
are greatly reduced at least for $z\sim6-7$
by using the size--luminosity relation obtained in this work.
Finally, we discuss the redshift evolution of size.

\subsection{The Intrinsic Size--luminosity Relation \\and Luminosity Function at $z\sim6-7$}\label{subsec:67relations}

\begin{figure*}[t]
  \centering
      \includegraphics[width=0.8\linewidth]{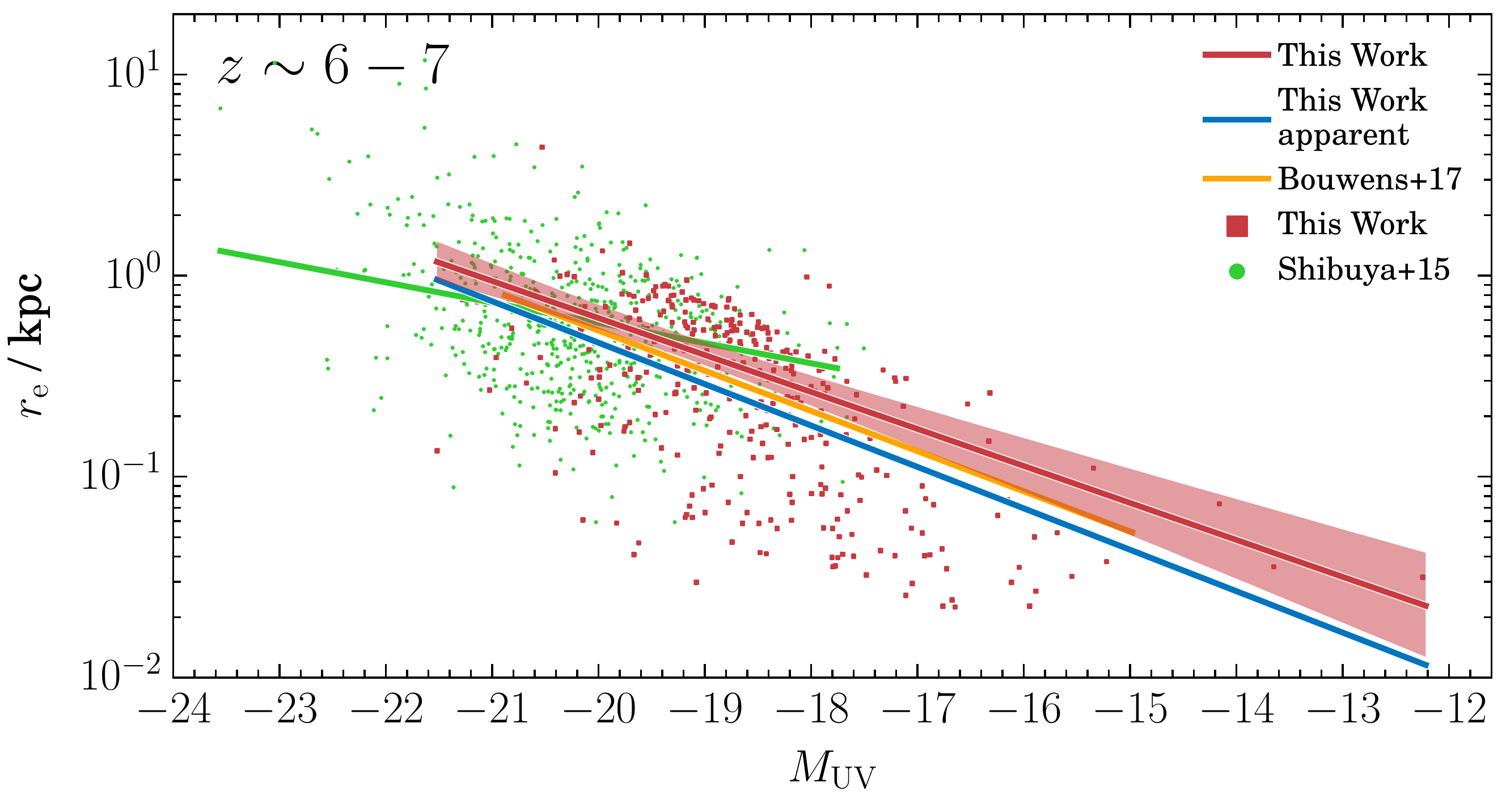}
      \includegraphics[width=0.8\linewidth]{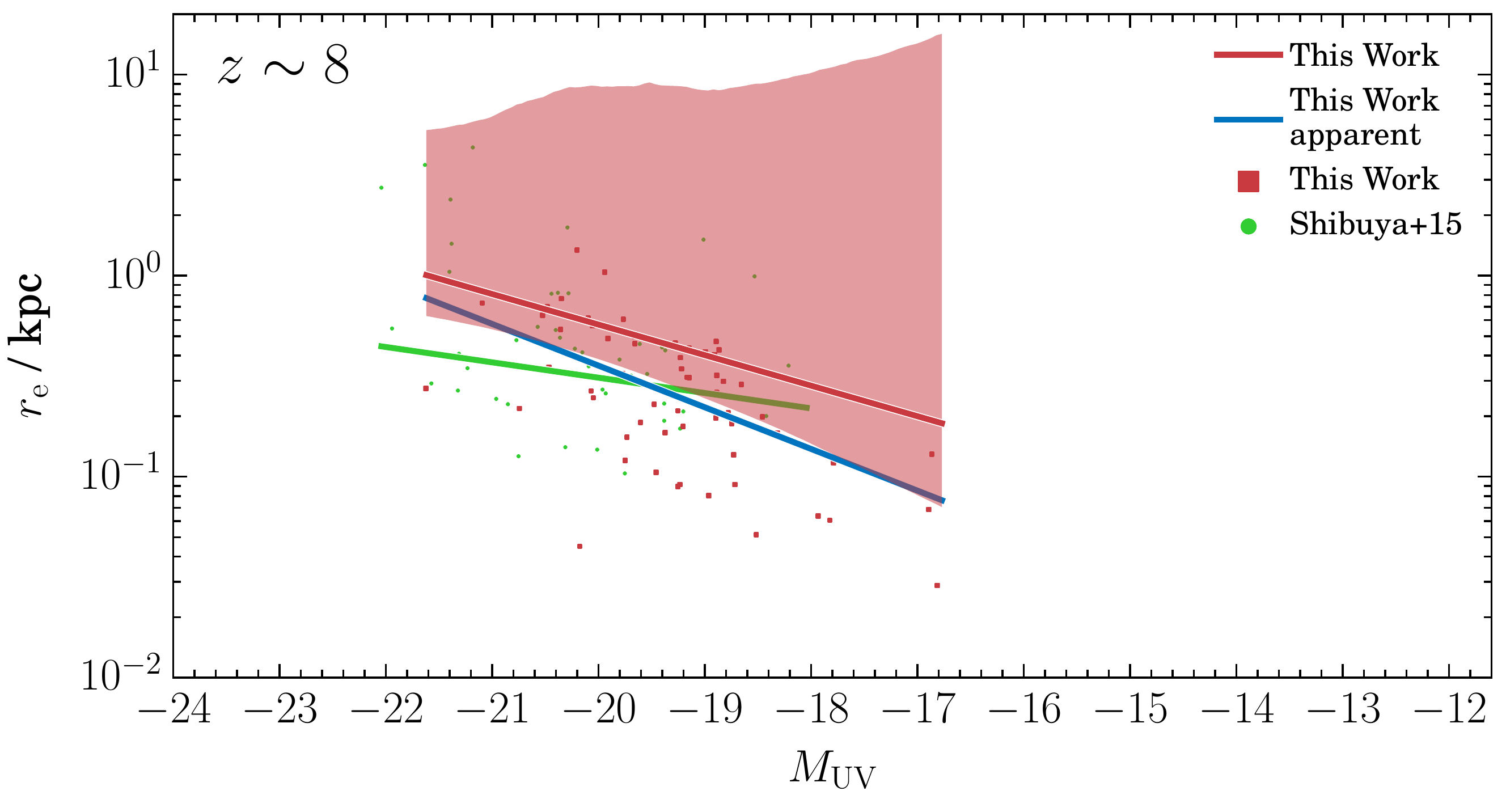}
      \includegraphics[width=0.8\linewidth]{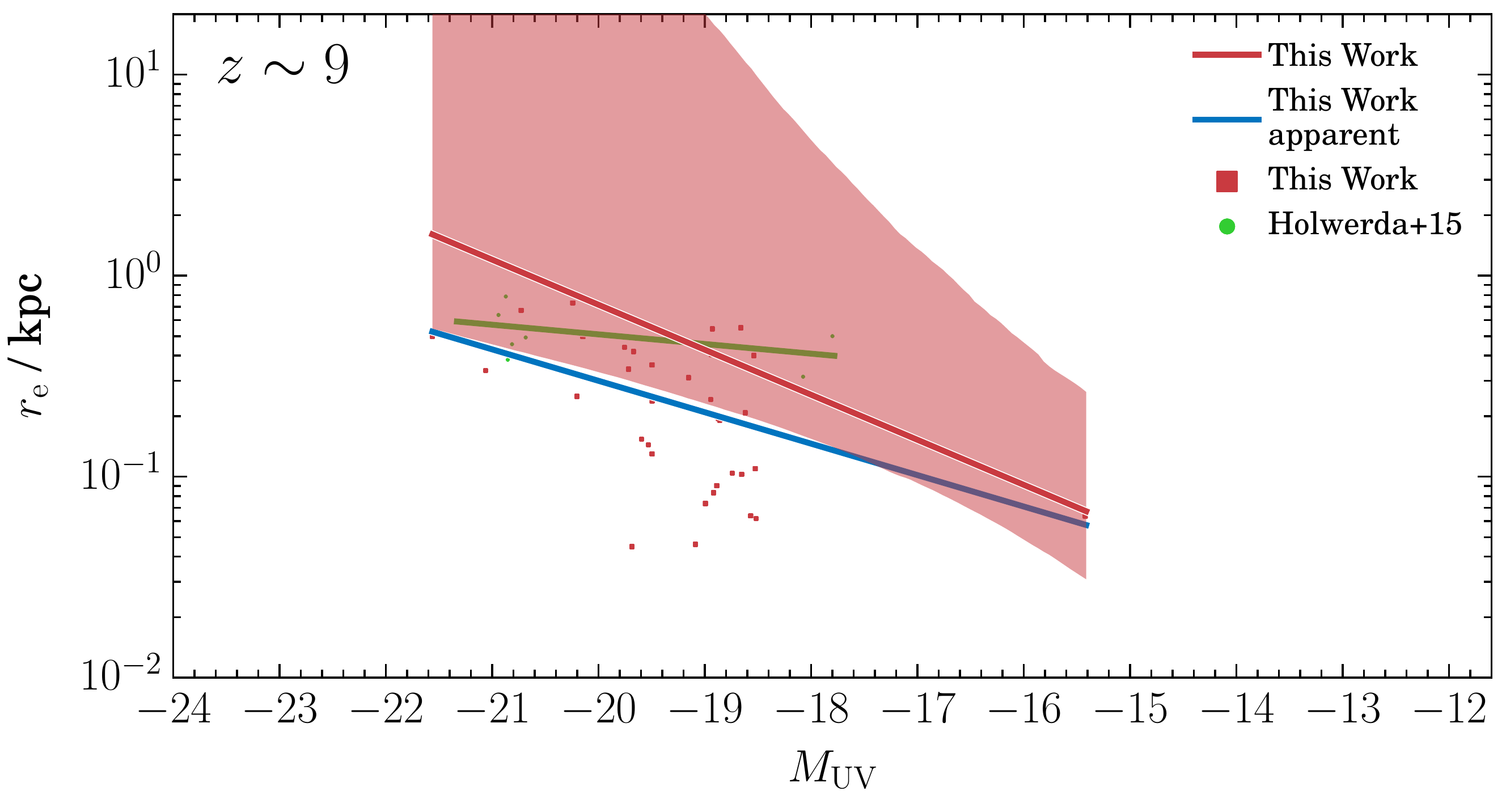}
  \caption{Galaxy distributions on the size--luminosity plane 
  at $z\sim6-7$ (\textit{top}), 8 (\textit{middle}), and 9 (\textit{bottom}), 
  respectively. 
  The red and green points represent, respectively, our galaxies  
  and those from previous studies (\citealt{shibuya15}, for $z\sim6-7$ and $8$;
  \citealt{holwerda15} for $z\sim9$).
  The red and blue solid lines represent the size--luminosity relations by 
  the completeness-corrected 
  and completeness-uncorrected fittings to our samples, respectively.
  The $1 \sigma$ distribution of the completeness-corrected
  size--luminosity relation is shown by the red shaded region.
  While the green solid lines show the best-fit power laws obtained 
  by \citet{shibuya15} and \citet{holwerda15}, the orange solid line is 
  for the result obtained by \citet{bouw17size} based on 
  two-dimensional size measurements.
  }
  \label{fig:rlrelation_compare}
\end{figure*}

\begin{figure}[t]
  \centering
      \includegraphics[width=\linewidth]{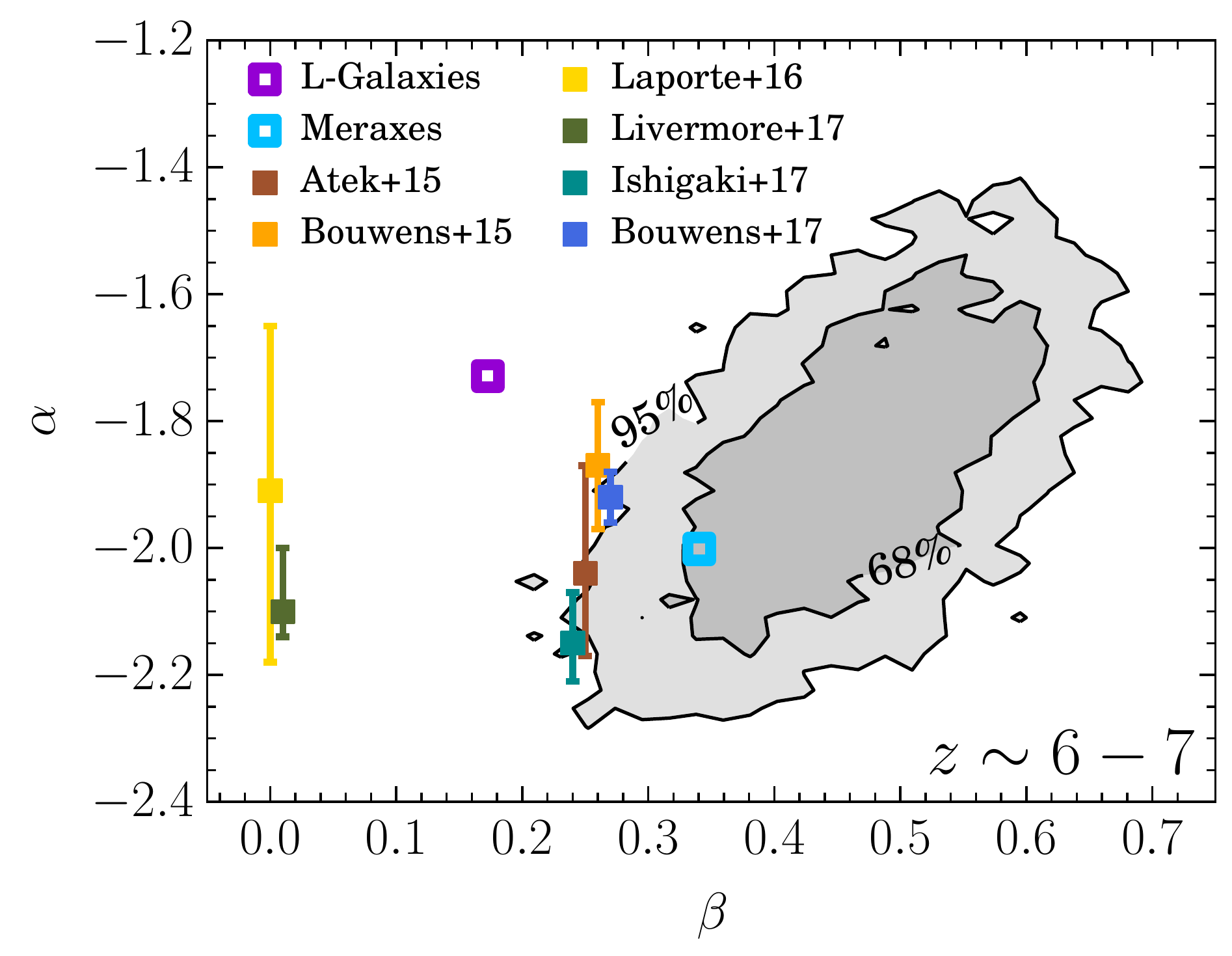}
      \includegraphics[width=\linewidth]{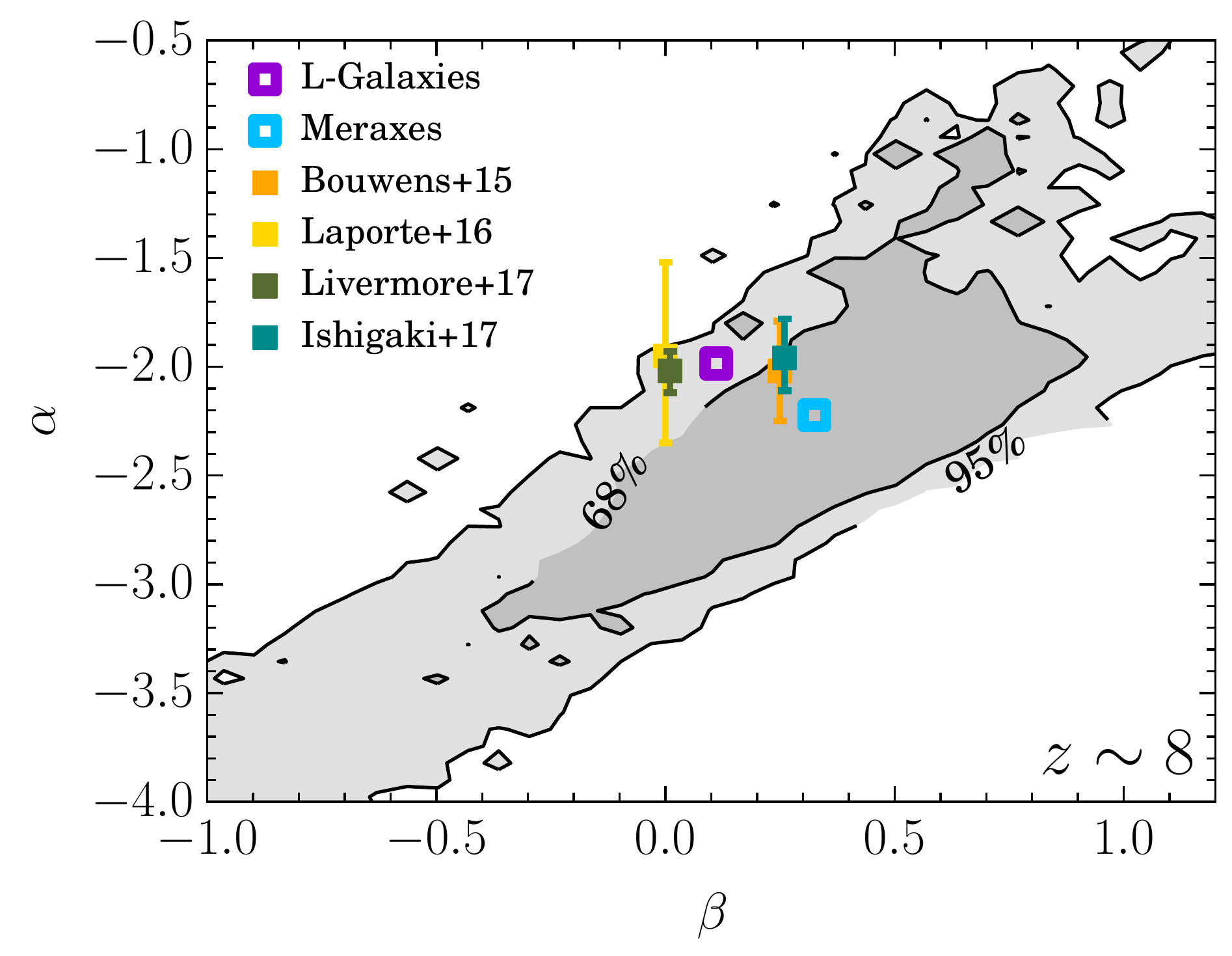}
      \includegraphics[width=\linewidth]{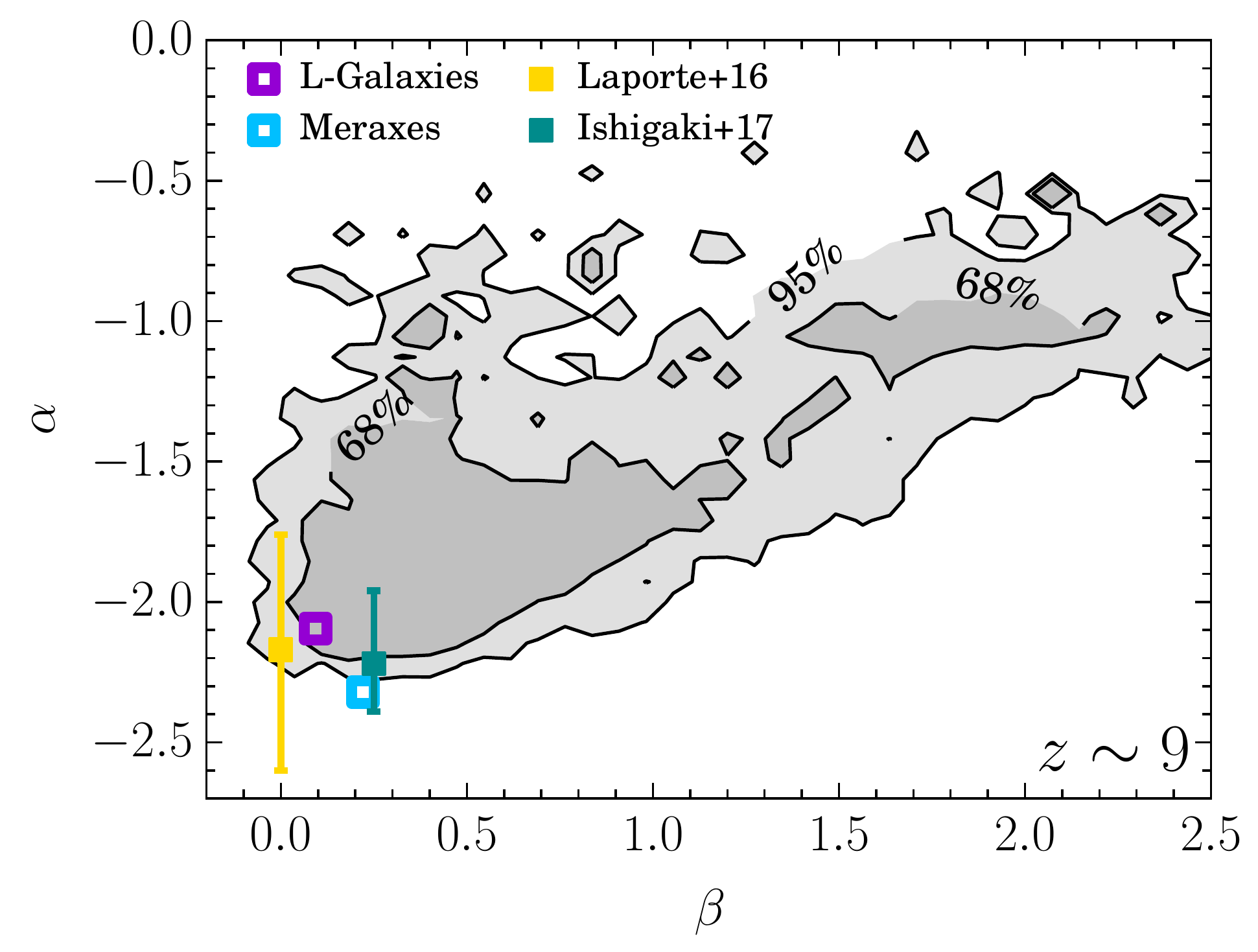}
  \caption{Correlations between the faint-end slope of the luminosity function,
$\alpha$, and the slope of the size--luminosity relation,
$\beta$, overplotted with
 the observational results presented in Table~\ref{tab:bestparams}
 (filled squares) and simulation results (open squares).
 The top, middle, and bottom panels show the results at $z\sim6-7$,
 $8$, and $9$, respectively.
  }
  \label{fig:alpha_beta}
\end{figure}

We discuss here the intrinsic size--luminosity relation
and UV luminosity function
at $z\sim6-7$, which are reliably estimated because of the large sample.
The best-fit size--luminosity relation and its $1 \sigma$ uncertainty
are presented in the top panel of Figure~\ref{fig:rlrelation_compare},
together with the results of previous work.

First, to evaluate the impact of detection incompleteness on the 
estimation of the size--luminosity relation, we fit the apparent
size--luminosity distribution without correcting for completeness.
In this process, 
as an alternative to $\Psi'_{i}$ in Equation~(\ref{eqn:psiobserved}), 
we use a distribution model of
\begin{align}
&\Psi_{\mathrm{apparent}}(r_{\rm e}, M_\mathrm{UV}; r_{0}, \sigma, \beta)
\nonumber\\&= P(r_{\rm e}, M_\mathrm{UV}; r_{0}, 
\sigma, \beta),
\label{eqn:psiapparent}\end{align}
where $P(r_{\rm e}, M_\mathrm{UV})$ is described in Equation~(\ref{eqn:sizedistribution}).
This implies that we assume a flat distribution for the magnitude distribution.
The best-fit parameter sets estimated using MLE are
presented in Table~\ref{tab:bestparams} as ``This work (apparent)''.
We find that the modal sizes are $\sim0.15$ dex underestimated, 
on average, and as large as $\sim 0.2$ dex at $M_{\mathrm UV} = -16$.
The slope of the intrinsic size--luminosity relation is overestimated 
by $\Delta \beta = 0.06$.
This suggests that incompleteness has a slight contribution
to the apparent steepness.
In contrast, we find that the variance of the size--luminosity
relation $\sigma$ is $\sim 25\%$ underestimated if incompleteness
is not corrected for.

Then, we discuss the incompleteness-corrected results.
Concerning the size--luminosity relation, the marginalized 
value of the slope is $\beta=0.46^{+0.08}_{-0.09}$.
This slope is steeper than 
$\beta=0.25^{+0.25}_{-0.14}$ at $z\sim5$ by \citet{huang13}
(with incompleteness correction)
and $\beta=0.25^{+0.05}_{-0.05}$ at $z\sim6$ by \citet{shibuya15}
(without incompleteness correction),
both of them utilizing brighter ($M_{\mathrm UV} \lesssim -18$) samples.
This is the first time to confirm the steepness of the 
intrinsic size--luminosity relation of $z\sim6-7$ galaxies.
Although a steep slope for galaxies at this redshift range was first 
reported by K15 based on reliable size measurements of the first HFF sample 
and then confirmed with larger samples by \citet{bouw17size, bouw17gc},
none of these studies has applied incompleteness correction.
The differences in the slope from \citet{huang13} and 
\citet{shibuya15} can be due to the differences in the
magnitude range and hence in the physics dominating in galaxies.
We further investigate this physical origin of the steepness in Section~\ref{sec:RLmodel}
using the result of the abundance matching by \citet{behroozi13}.
As described in the next paragraph, the difference from 
\citet{shibuya15} can also be explained by the differences in methods to 
measure magnitudes and to fit the size--luminosity relation.
We note that although it has a steep slope, the best-fit 
intrinsic bivariate distribution predicts the existence of
faint galaxies with large sizes,
for instance, $M_\mathrm{UV} = -16$ galaxies with $r_\mathrm{e}\sim 1$ kpc, 
(see the top panel of Figure~\ref{fig:2Ddistributions}).

\citet{shibuya15} have found remarkably shallower slopes 
of $\beta \simeq 0.25\pm0.05$ for brighter galaxies at $z\sim6$ and $7$
even without correcting for incompleteness.
Interestingly, in Figures~\ref{fig:rlrelations}~and~\ref{fig:rlrelation_compare}, 
their galaxies appear to have a similar slope to ours.
In fact, while their samples made public and plotted here use
\texttt{GALFIT} magnitudes, they have used  
\texttt{SExtractor} magnitudes to derive 
the slope (T. Shibuya 2017, private communication). 
Applying the same method (Equation~\ref{eqn:psiapparent}) to their sample, 
we find that using \texttt{SExtractor} magnitudes gives slopes $0.13$ and $0.21$ 
shallower than those based on \texttt{GALFIT} magnitudes at $z\sim6$ and $7$, 
respectively.  
This may suggest that using \texttt{SExtractor} magnitudes
leads them to derive the shallower slopes.
In addition, our fitting method is different from theirs. 
They use a least-squares method that minimizes residuals 
only in size, which can bias the slope toward shallower values.

The modal size at $M_{\mathrm UV} = -21$ is $0.94^{+0.20}_{-0.15}\>\mathrm{kpc}$
at $z\sim6-7$.
This size can be slightly larger than the incompleteness-uncorrected
sizes by the previous studies \citep{bouw04, oesch10b, shibuya15}.
We note that the sizes in \citet{bouw04} and \citet{oesch10b} are averages 
in the range of
$-21 \leq M_{\mathrm UV} \leq -19.7$, which means
the sizes at  $M_{\mathrm UV} = -21$ should be larger 
(see also Figure~\ref{fig:size_evolution}).

The variance of the log-normal size distribution is 
$\sigma=0.87^{+0.10}_{-0.09}$.
This is in good accordance with the values of $\sigma=0.83^{+0.046}_{-0.044}$
and $0.90^{+0.15}_{-0.065}$ at $z\sim4$ and 5, respectively, 
in \citet{huang13}.
According to the analytical model by \citet{mmw98} 
\citep[see also][]{fallefstathiou80}, galaxy sizes 
are basically proportional to their halo sizes and spin parameters.
The distribution of the spin parameter is log-normal at a fixed 
halo mass and thus also approximately log-normal 
at a fixed luminosity.
Its variance was 
estimated to be $\sigma_{\mathrm h} = 0.60$ at $z=0$ and revealed to
scarcely evolve toward higher redshifts by \citet{zjupaspringel17} 
with the dark matter-only Illustris simulation.
Since the observed variance of the galaxy-size distribution is larger than 
that of the spin parameter, there may be some elements 
that broaden the galaxy-size distribution.
For example, a scatter in halo mass at fixed luminosity would 
result in a broader size distribution.
This scatter was recently suggested at low redshifts in \citet{charlton17}.
Another explanation is a disk-to-halo ratio of specific angular 
momentum depending on the spin parameter, which means the 
galaxy size is no longer proportional to the spin parameter.
We note that the derived variance $\sigma$ has been corrected
for errors in size and magnitude measurements as described 
in Equation~(\ref{eqn:gaussian}).

We find a shallow faint-end slope of the luminosity function of 
$\alpha=-1.86^{+0.17}_{-0.18}$, consistent with the slopes in
\citet{bouw15, bouw17magnif} and \citet{laporte16} but
slightly incompatible with recently suggested 
steep slopes of $\alpha \simeq-2.00$ to $-2.15$ \citep[e.g.,][]
{livermore16, ishigaki17}.
The reason for this is that our size--luminosity relation is steeper 
than those utilized in the previous studies.
With a steeper size--luminosity relation, galaxies are easier to 
detect, and a smaller amount of incompleteness correction
is needed in luminosity function derivation, especially at 
faint magnitude ranges.
Thus, the faint-end slope becomes shallower.
The effects of the size--luminosity relation on the
luminosity function are further discussed in Section~\ref{sec:RLrelationsforLFs}.

The characteristic magnitude, $M^{*} = -20.73^{+0.46}_{-0.81}$ 
is consistent with those of previous work.
Since the marginalized distribution has a long tail toward the brighter 
magnitude, the mode of it is slightly larger, $M^{*} \simeq -20.56$.
The uncertainty in $M^{*}$ is relatively large, probably because
we do not use bright-galaxy samples from large-area surveys.

The parameters of the size--luminosity relation strongly correlate
with those of the luminosity function.
The most important may be the correlation between
$\alpha$ and $\beta$, which has been pointed out 
by several works, including \citet{graz11}
and \citet{bouw17size, bouw17magnif}.
The top panel of Figure~\ref{fig:alpha_beta} shows the correlation
between $\alpha$ and $\beta$ obtained in this work together with
the previous measurements of these parameters 
presented in Table~\ref{tab:bestparams}.
We find that the steeper $\alpha$ in \citet{atek15b} and \citet{ishigaki17} 
will become further consistent with ours if steeper size--luminosity relations 
are assumed.
Even with our large and deep sample, at $z\sim6-7$ there still remains 
a moderate uncertainty in $\alpha$ due to the uncertainty in 
the size--luminosity relation.
This uncertainty in $\alpha$ is propagated to the UV luminosity density,
a key quantity to calculating the number density of ionizing photons, 
although no previous studies on cosmic reionization have considered 
this uncertainty.
We note that although the values of $\alpha$ obtained in \citet{laporte16} and 
\citet{livermore16} are consistent with our value, 
their $\alpha$--$\beta$ combinations are outside (with a large margin) 
of the 95\% confidence ellipse obtained in this study.
This demonstrates that these parameters must not be determined independently.

We also compare our $\alpha$ and $\beta$ measurements with the results 
of the semi-analytical model of galaxy formation \textsc{L-Galaxies} \citep{henriques15}.
We run the \textsc{L-Galaxies} code on two $N$-body dark matter simulations
of different resolutions, the Millennium \citep{springel05nat} and 
Millennium-II \citep{boylankolchin09}, and combine the two galaxy 
catalogs to probe a wide halo mass range. 
Applying Equation~(\ref{eqn:intrinsicmodel})
to the combined catalog finds that 
the \textsc{L-Galaxies} predicts an $\alpha$ consistent with our value 
but a significantly flatter $\beta$.
Results of the semi-analytical model of galaxy formation
\textsc{meraxes} \citep{mutch16, liu17} are also compared.
We find a good agreement with our results for $z\sim6-7$ and 8
and an acceptable agreement for $z\sim9$.
Note that the values of $\beta$ obtained here are 
different from those obtained in \citet{liu17} because of 
different fitting methods.

However, we find that the two models tend to predict 
relatively flatter size--luminosity relations, especially at $z\sim6-7$
and 9.
Their sizes are calculated essentially based on the analytical model
by \citet{mmw98}.
The flatter size--luminosity relations than observed may suggest
the importance of careful calculations of the exchange of angular momentum
between the dark matter halo and the stellar disk.
Indeed, \textsc{meraxes} assumes a constant specific angular momentum
of $j_\mathrm{d} / m_\mathrm{d} = 1$, which disagrees with
our result in Section~\ref{sec:RLmodel}.
In \textsc{L-Galaxies}, specific angular momenta are calculated 
and compared with those by other semi-analytical models
and hydrodynamical simulations \citep[e.g.][]{guo16, hou17}. 
However, we do not discuss their results 
because they provide only the specific angular momenta of cooled gas,
which may be systematically different from the specific angular momenta of disks, 
$j_\mathrm{d}/m_\mathrm{d}$.
Further comparison between the observations and simulations is 
beyond the scope of this paper.

Another parameter set that shows a strong correlation is 
$\alpha$ and $M^{*}$, as seen in Figure~\ref{fig:mcmcz7}
and as has been reported in previous studies.
We confirm that the uncertainty in $\alpha$ decreases from $\sim 0.2$ to
$\lesssim 0.1$ if $M^{*}$ is virtually fixed to, for instance,  $M^{*} = -21$.
The slope $\beta$ also correlates with the modal size $r_{0}$ and 
weakly with the width of the size distribution $\sigma$;
both correlations originate from a requirement 
to reproduce small faint galaxies (except for the 
$\beta$--$\sigma$ correlation at $z\sim9$).

Since $\alpha$ strongly correlates with $M^{*}$ and $\beta$,
a more accurate measurement of $\alpha$ requires a larger sample
containing bright objects (to better constrain $M^{*}$) 
accompanied by a completeness estimation on the size--luminosity plane 
(to obtain an unbiased $\beta$ value).

\subsection{The Intrinsic Size--luminosity Relation \\and Luminosity Function at $z\sim8$ and 9}\label{sec:89relations}
The fitting results of the intrinsic size--luminosity distributions at $z\sim8$ and $9$
are presented in the middle and bottom panels of Figure~\ref{fig:rlrelation_compare},
respectively.
Since the samples are smaller than that at $z\sim6-7$, 
the uncertainties in the parameters are typically $\gtrsim2-3$ times larger.

Similar to that at $z\sim6-7$, we find steep slopes of the size--luminosity relations
of $\beta = 0.38^{+0.28}_{-0.78}$ and $0.56^{+1.01}_{-0.27}$ at 
$z\sim 8$ and $9$, respectively.
These are steeper than the slope of $\beta=0.19^{+0.25}_{-0.25}$ at $z\sim8$ by \citet{shibuya15},
although the differences are within the $1\sigma$ errors.
However, the distributions of our galaxies on the size--luminosity plane 
appear to be consistent with theirs, as is the case for $z\sim6-7$.

The modal sizes at $M_{\mathrm UV} = -21$ are $0.81^{+5.28}_{-0.26}\>\mathrm{kpc}$ 
and $1.20^{+367.64}_{-0.74}\>\mathrm{kpc}$ at $z\sim8$ and 9, respectively.
If incompleteness is not corrected for, the sizes become
0.2--0.3 dex smaller at $z\sim8$ and 9,
a slightly larger amount of decrease than that at $z\sim6-7$.
These are consistent with the incompleteness-uncorrected
sizes of $r_{\mathrm e} = 0.419 ^{+1.981}_{-0.262}$ at $z\sim8$ 
by \citet{shibuya15} and $r_{\mathrm e} = 0.6 ^{+0.3}_{-0.3}$ at $z\sim9$ 
by \citet{holwerda15}.

The variance of the size distribution $\sigma$ is
$0.80^{+1.07}_{-0.26}$ and $1.04^{+1.52}_{-0.46}$ at $z\sim8$
and 9, respectively, being almost constant at $z\sim6-9$.
While we do not find any indication of the evolution of
$\sigma$ over this redshift range, the modal value
of the variance distribution may decrease with redshift.
Further discussion needs larger samples.

While the faint-end slope of the luminosity function at $z\sim9$
is relatively shallow ($\alpha=-1.64^{+0.61}_{-0.28}$), that at $z\sim8$
may be steep ($\alpha=-2.26^{+0.49}_{-0.99}$).
However, both values are consistent with the value at $z\sim6-7$ 
due to the large uncertainties.

At $z\sim8-9$, the probability distributions of $M^{*}$
have tails toward the brighter magnitudes, and thus
the median values are remarkably brighter than that at $z\sim6-7$.
This is because our samples do not have enough bright
galaxies due to the small cosmic volume the HFF program is probing.
We note that the $M^{*}$ values at $z\sim8-9$
are close to typical magnitudes at these redshifts
of $M^{*}\sim-21$ within the uncertainties.
Furthermore, the modes are $M^{*} = -19.95$ at $z\sim8$
and $-19.80$ at $z\sim9$.

We also calculate $r_{0}$, $\sigma$, $\beta$, and $\alpha$
by fixing $M^{*}$ to $-20.73$, the best-fit value at $z\sim6-7$, 
and obtain 
$(r_{0} / \mathrm{kpc}, \sigma, \beta, \alpha) = 
(0.75^{+0.53}_{-0.16}, 0.65^{+0.35}_{-0.14}, 0.50^{+0.16}_{-0.21}, -1.80^{+0.22}_{-0.30})$
at $z\sim8$ and 
$(0.59^{+0.61}_{-0.16}, 
0.69^{+0.40}_{-0.20}, 0.42^{+0.17}_{-0.15}, -1.59^{+0.19}_{-0.18})$
at $z\sim9$, as presented in Table~\ref{tab:bestparams}.
These $\alpha$ values are even shallower than those from the full modeling, 
with the uncertainties being reduced to be comparable to those of 
previous studies.

The middle and bottom panels of Figure~\ref{fig:alpha_beta}  are the same as 
the top panel but for $z\sim8$ and $9$, respectively.
In contrast to the case for $z \sim6-7$, all the $\alpha$--$\beta$ combinations 
from previous observations and \textsc{L-Galaxies} are within 
the 95\% confidence contour of our results.
Besides the parameter sets of $(\alpha, \beta)$, $(\alpha, M^{*})$, $(r_{0}, \beta)$,
and $(\sigma, \beta)$ that show correlations at $z\sim6-7$, 
$r_{0}$ and $\sigma$ also correlate strongly at $z\sim8-9$.
This correlation is to reproduce the smaller galaxies and may indicate that
we still do not trace the peak of the size distributions at $z\gtrsim8$.

We find that the parameters of the size--luminosity relations and luminosity functions
at $z\sim8-9$ are still not well constrained.
Thus, there are significant uncertainties in the luminosity function
the faint-end slope of the luminosity function $\alpha$ 
and hence in discussions of reionization based on the UV luminosity density.

\subsection{The Modeling of the Size--luminosity Relation}\label{sec:RLmodel}
\begin{figure}[t]
  \centering
      \includegraphics[width=\linewidth]{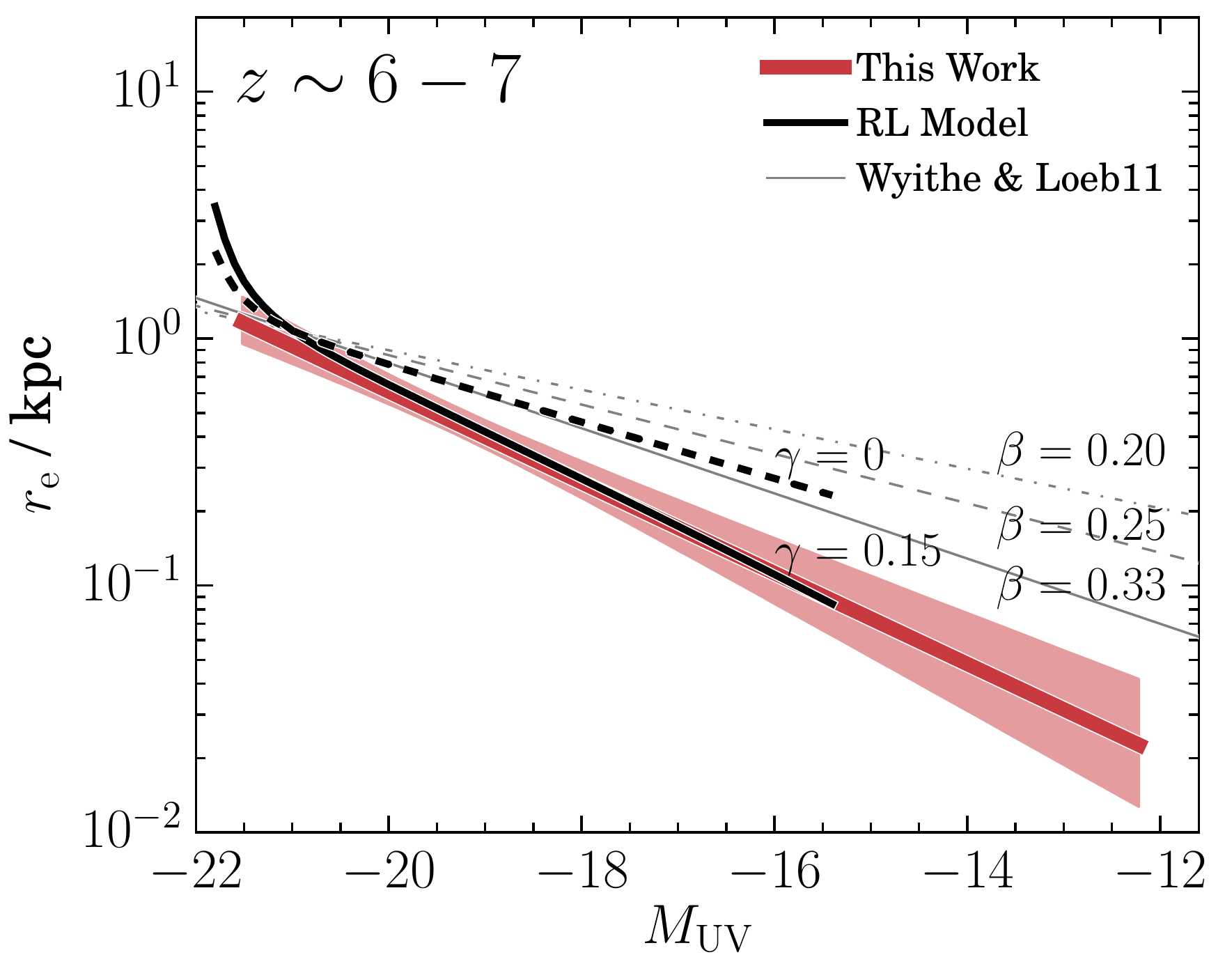}
  \caption{Model predictions of the size--luminosity relation at $z\sim6-7$, 
  overplotted with the fitting result to the observation.
  The black solid and dashed lines show the predictions by the RL model
  with $\gamma=0.15$ and $0$, respectively.
  The gray dash-dotted, dashed, and solid lines represent 
  the predictions by the \citet{wyithe11} model with $\beta=0.20$,
  $0.25$, and $0.33$, respectively, normalized to agree
  with those by the RL model at $M_\mathrm{UV} = -21$.
  }
  \label{fig:RLrelation_model}
\end{figure}

We construct a model to predict the normalization and 
slope of the size--luminosity relation at $z\sim6-7$ in the 
following process, which is referred to as the RL model.
(1) We calculate the average stellar mass of galaxies
as a function of luminosity using
the stellar mass--luminosity relation by \citet{gonz11}.
(2) Combining step (1) with the stellar mass--halo mass relation by \citet{behroozi13}, 
we evaluate the average halo mass of galaxies as a function
of luminosity\footnote{There may be a logical inconsistency 
that we model the steep size--luminosity relation using the results 
in \citet{behroozi13}, where a luminosity function derived assuming
a shallower size--luminosity relation is used.
However, we consider this effect to be of secondary importance.}.
Note that an extrapolated relation covering a wider mass 
range than that presented in their paper is utilized 
(P. Behroozi 2016, private communication).
(3) We calculate the virial radius of halos by 
\begin{align}
r_{\mathrm{vir}} = \left[\frac{2GM_{\mathrm{vir}}}{\Delta_{\mathrm{vir}} {H(z)}^{2}}\right]^{1/3}.
\label{eqn:virialsize}
\end{align}
In the calculation of the virial overdensity $\Delta_{\mathrm{vir}}$, 
we use the fitted form of
$\Delta_{\mathrm{vir}} = 18\pi^{2} + 82x -39x^{2}$
with $x=\Omega_{\mathrm{m}}(z) -1$ by \citet{bryannorman98}.
(4) From the halo radius, we calculate the galaxy size based on the 
equation in \citet{mmw98},
\begin{align}
r_{\mathrm{e}} = \frac{1.678}{\sqrt{2}} f_{j} \lambda f_{\mathrm{c}}^{-1/2} f_{\mathrm{R}} r_{\mathrm{vir}},
\label{eqn:modelratio}
\end{align}
where $\lambda$ is the spin parameter of the halo 
defined in \citet{peebles69}.
The factor $f_{j}(M_{\mathrm{vir}})$ represents 
the ratio of the specific angular momentum in the galaxy
against that in the halo, 
 \begin{align}
f_{j}(M_{\mathrm{vir}}) 
&= \frac{j_{\mathrm{d}}}{m_{\mathrm{d}}}(M_{\mathrm{vir}}) \\
&= \left(\frac{j_{\mathrm{d}}}{m_{\mathrm{d}}}\right)_{M_{\mathrm{UV}}=-21} \left(\frac{M_{\mathrm{vir}}}{M_{\mathrm{vir,0}}}\right)^{\gamma},
\end{align}
where $j_{\mathrm{d}}$ and $m_{\mathrm{d}}$ are the ratio of the
angular momentum and mass, respectively, in the galaxy
against those in the halo.
In contrast to the original equation in \citet{mmw98},
we allow $j_{\mathrm{d}} / m_{\mathrm{d}}$ to vary as a function of the halo mass,
whose dependence was suggested in several observational studies 
\citep[e.g.,][]{somerville17, okamura17} and 
simulations of galaxy formation \citep[e.g., ][]{sales10}.
The factor $(j_{\mathrm{d}} / m_{\mathrm{d}})_{M_{\mathrm{UV}}=-21}$
and $M_{\mathrm{vir, 0}}$ represent the $j_{\mathrm{d}} / m_{\mathrm{d}}$
and the halo mass of galaxies with $M_{\mathrm{UV}} = -21$, respectively.
The index $\gamma$ is the exponent of the mass dependence 
of $j_{\mathrm{d}} / m_{\mathrm{d}}$.
The factor $f_{j}$ equates to the original constant 
 $j_{\mathrm{d}} / m_{\mathrm{d}}$ when $\gamma =0$.
The factor $f_{\mathrm{c}}(c)$, depending only on
the concentration parameter of the halo $c$,
is to correct for the
effect caused by the change in the density profile
from the isothermal sphere to the NFW profile.
The other factor $f_{\mathrm{R}}(j_{\mathrm{d}}/m_{\mathrm{d}}, 
m_{\mathrm{d}}, \lambda, c)$ is to correct for 
effects caused by the change in the density profile
and the gravitational effect by the disk.
We need the factor of $1.678$ to convert 
the scale length of the exponential profile to
the half-light radius $r_{\mathrm{e}}$.
Thus, we obtain the model of the size--luminosity relation.

Except for $\gamma$, there are four parameters that 
are needed to calculate the size; while $\lambda$ and 
$c$ are reliably determined in simulations 
\citep[e.g.,][]{bullock01, vitvitska02, davisnatarajan09, prada12},
the parameters $j_{\mathrm{d}}$
and $m_{\mathrm{d}}$, depending on baryonic physics, are 
difficult to predict. 
In the calculation of the size, we assume $\lambda = 0.04$
that is independent of redshift, which is consistent
with the recent result in \citet{zjupaspringel17}.
For the concentration parameter $c$, we utilize the
fitting function for the $c$--$M_{\mathrm{vir}}$ relation
for Planck cosmology in \citet{correa15}.
We assume the typical values of 
$(j_{\mathrm{d}} / m_{\mathrm{d}})_{M_{\mathrm{UV}}=-21} = 1.0$ 
\citep[e.g.,][]{fallefstathiou80, mmw98, romanowskyfall12, fall13} 
and $m_{\mathrm{d}} = 0.05$ \citep[e.g.,][]{sales10}.
These values of $j_{\mathrm{d}}$ and $m_{\mathrm{d}}$ 
are shown to be consistent with observations in 
Section~\ref{subsec:redshift_evolution}.

The calculated size--luminosity relation at $z\sim6-7$ is 
presented in Figure~\ref{fig:RLrelation_model} as the 
RL model.
We find that the RL model predicts a shallow slope 
of $\beta\simeq0.3$ when $\gamma=0$.
While this shallow slope is consistent with observed slopes
at lower redshifts of $\beta\sim0.25$ 
\citep[e.g.,][see also Figure~\ref{fig:beta_evolution}]
{dejonglacey00, huang13, shibuya15},
it is inconsistent with
our steep slope at $z\sim6-7$.
However, when we change $f_{j}$ as a function of halo mass 
with $\gamma=0.15$, the model 
predicts a steeper slope that is consistent with the observed value 
at $z\sim6-7$.
This may suggests that  $j_{\mathrm{d}} / m_{\mathrm{d}}$, 
that is, the fraction of the specific angular momentum in the galaxy,
is smaller in fainter galaxies at higher redshifts.
In the beginning stage of galaxy formation, 
stars are formed preferentially from 
gas with lower angular momenta.
The halo mass dependence of $f_{j}$ obtained here may suggest that 
the faint galaxies are indeed in such a stage.

Stellar feedback may be another explanation because it redistributes
the angular momentum between the galaxy and the halo, thus changing $f_{j}$.
\citet{genel15} have used the Illustris cosmological simulation
to find that stellar feedback increases the specific angular momentum
of galaxies, although the halo mass dependence is equivalent to 
$\gamma<0$, opposite to what we find here 
(see also \citealt{sales10} for a contradictory result).

Another possibility is that in low-mass halos, only those with relatively 
small spin parameters can form disks, thus making the slope steeper 
even with $\gamma=0$.
If this is the case, the shape and variance of the log-normal size distribution
at faint magnitudes can be different from those at bright magnitudes.

We also compare the obtained intrinsic slope with analytical predictions
by \citet{wyithe11}, which are shown in Figure~\ref{fig:RLrelation_model}
with gray lines.
They construct a simple analytical model that describes 
the relation between the size and luminosity \citep[see also][]{liu17}.
The predicted relation depends on the feedback that dominates 
in galaxies.
They test three kinds of feedback: energy conserving,
momentum conserving, and no feedback.
The predicted slopes are $\beta=0.20$, $0.25$, and $0.33$, respectively,
all of which are shallower than the observed value at $>1\sigma$ levels.
We note that they assume a constant $f_{j}$, which corresponds 
to $\gamma=0$.

Very recently, \citet{ma17} have suggested that UV light 
does not necessarily trace the main part of galaxies using high-resolution
cosmological zoom-in simulations from the FIRE project.
This observational bias might affect our discussion presented here
and might lead to a smaller $\gamma$ 
\citep[see also][]{huang17, somerville17}.

\subsection{The Size--luminosity relations for Derivations of Luminosity Functions}\label{sec:RLrelationsforLFs}
\begin{figure}[t]
  \centering
      \includegraphics[width=0.95\linewidth]{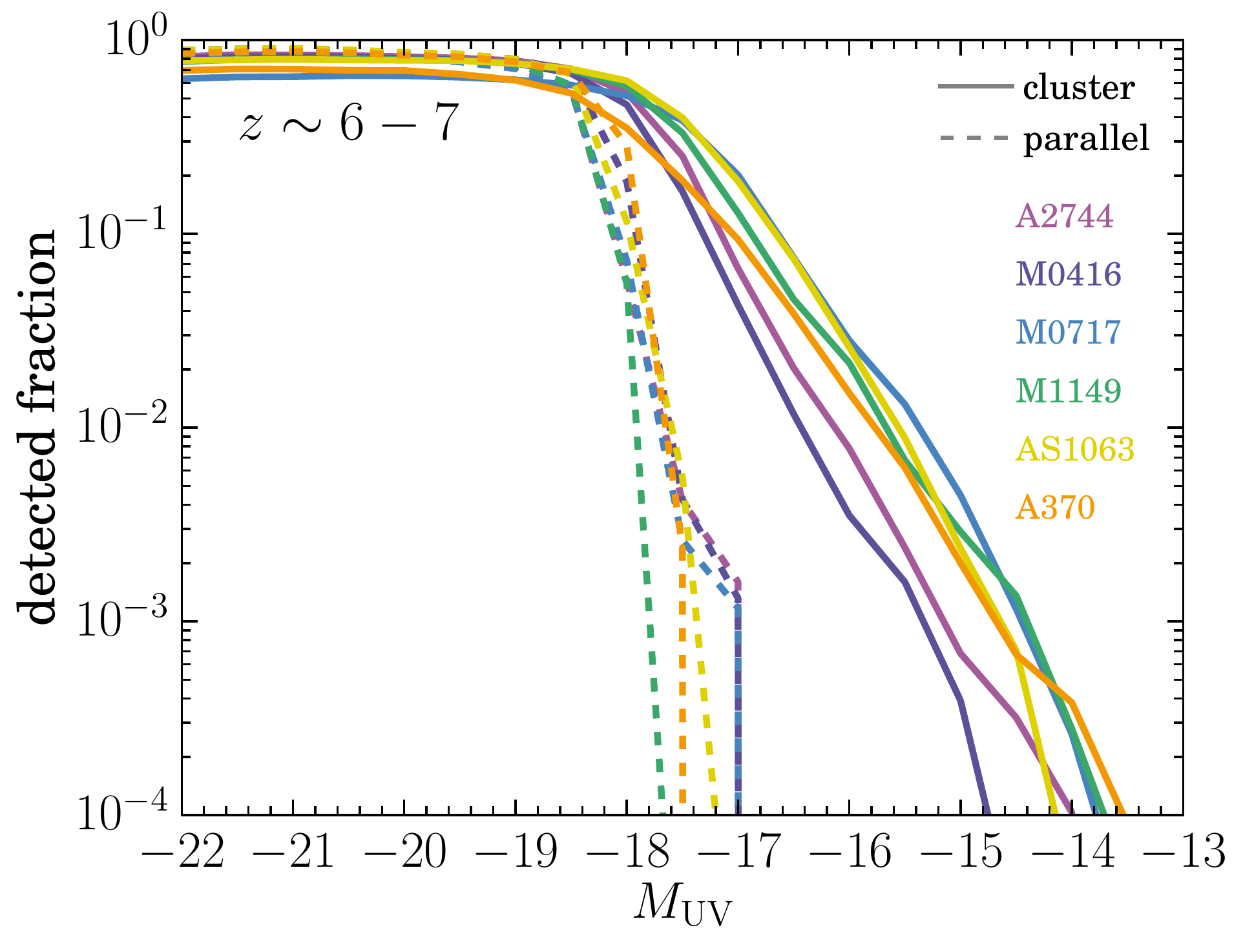}
      \includegraphics[width=0.95\linewidth]{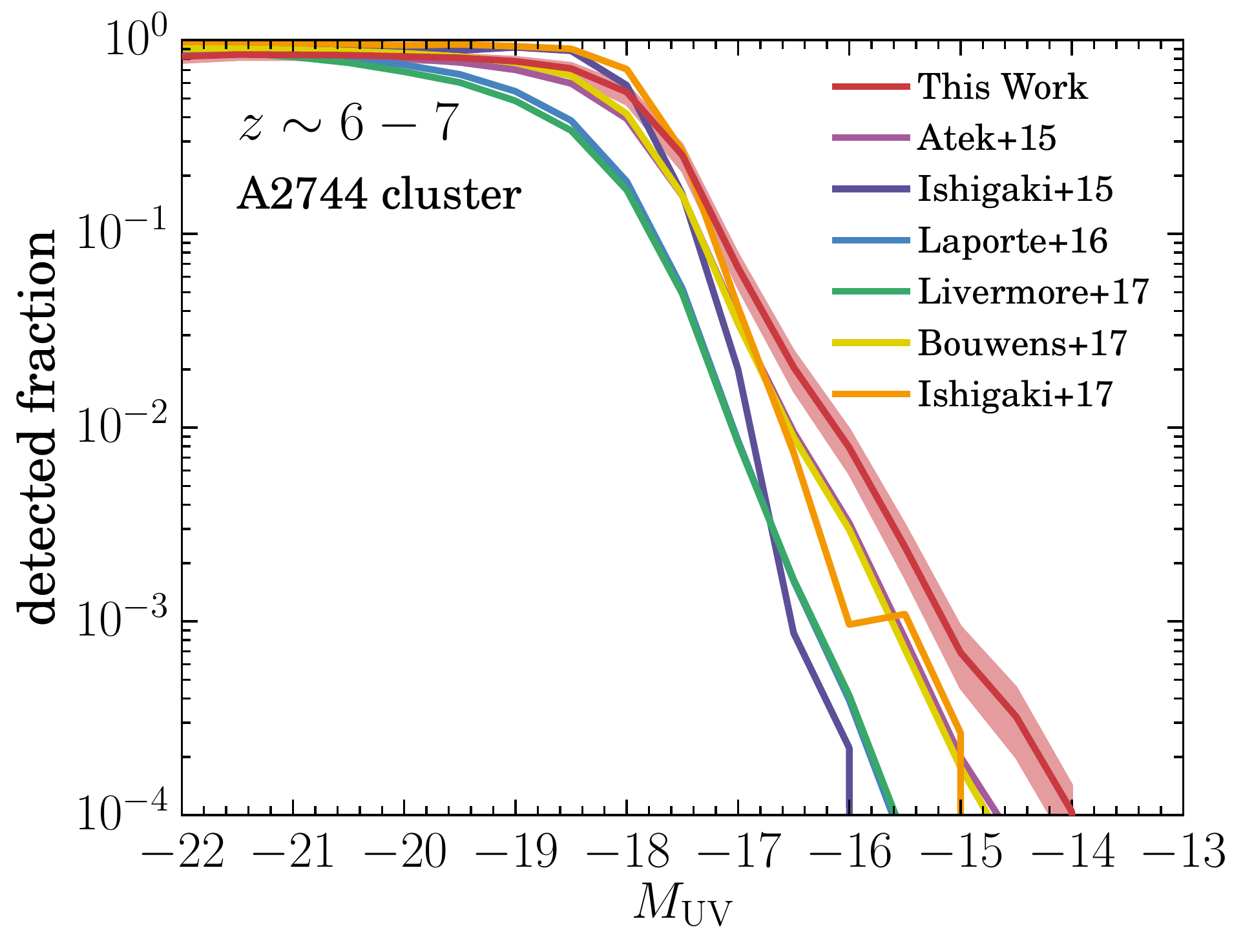}
      \includegraphics[width=0.95\linewidth]{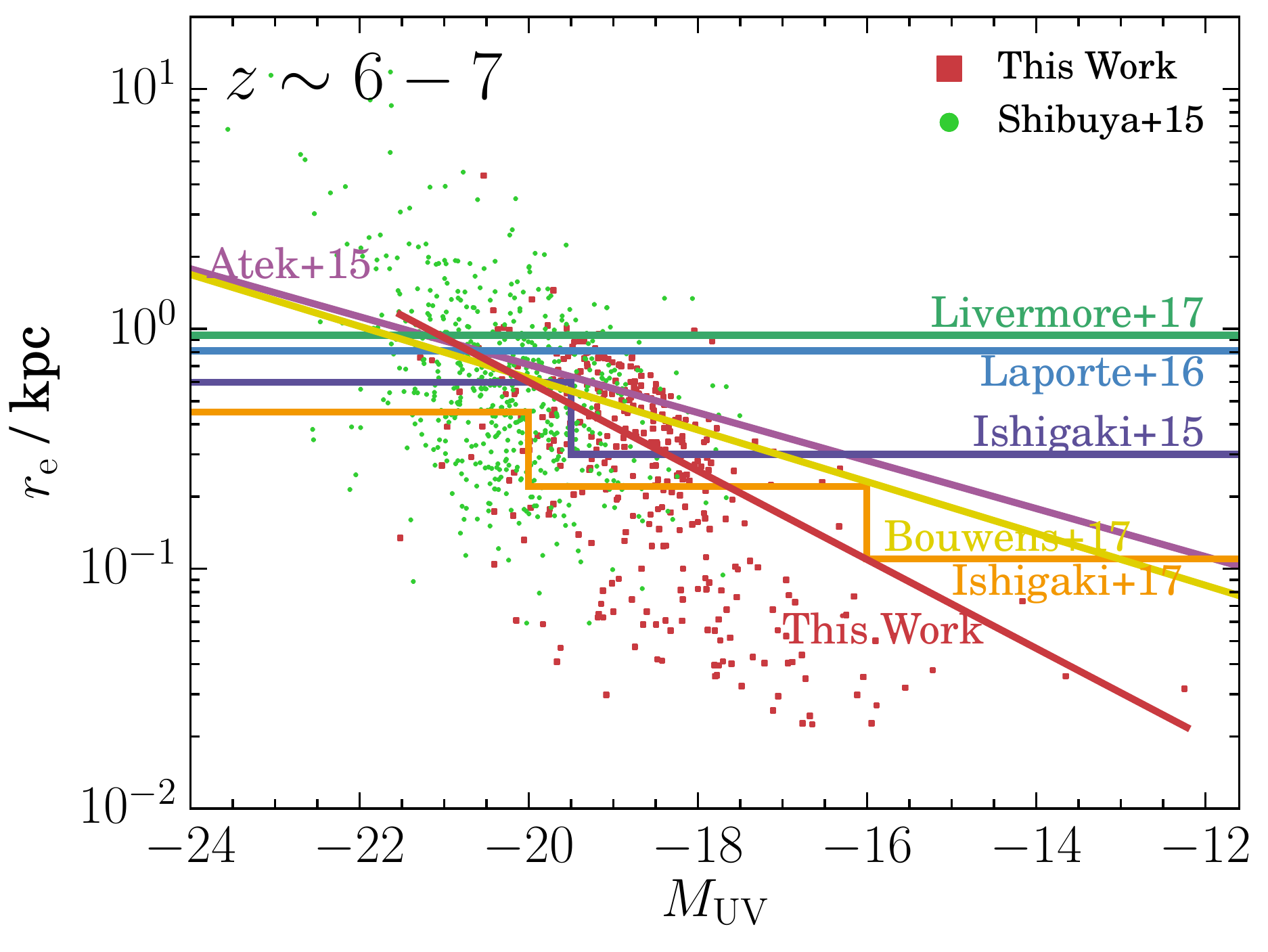}
  \caption{(\textit{Top}) Detected fraction against UV absolute magnitude 
  in each field at $z\sim6-7$ calculated using the completeness map of the field
  and the best-fit size--luminosity relation
  at $z\sim6-7$. 
  The solid and dashed lines correspond to the cluster 
  and parallel fields, respectively.   
  (\textit{Middle}) Variation in the detected fractions at $z\sim6-7$
  in the \clone\ cluster field calculated with size--luminosity relations
  given in previous studies.
  The uncertainty estimated in this work is also plotted by the red shaded 
  region.
  (\textit{Bottom}) Size--luminosity relations
  in the previous studies utilized to calculate 
  the detected fractions in the middle panel, overplotted with
  the galaxy distributions from this work (red points) and \citet{shibuya15}
  (green points).
  }
  \label{fig:detected_fraction}
\end{figure}

In this subsection, we examine the effects of the size--luminosity relation 
on the estimation of the detected fraction of galaxies, and thus of the luminosity function.

The top panel of Figure~\ref{fig:detected_fraction} shows the detected fraction
against UV magnitude for $z\sim6-7$ calculated for all of the HFF cluster and parallel fields
using the best-fit size--luminosity relation.
As shown in this figure, the detected fraction at the faintest magnitudes
$M_{\mathrm UV} \simeq -15$ to $-14$ is extremely low.
This implies that
the luminosity function is calculated from only a small part 
of galaxies in the field of view 
with a large ($\sim 10^3$) incompleteness correction.

We calculate the detected fractions, as an example, in the \clone\ cluster
field assuming 
the six size--luminosity relations utilized in the previous studies
at $z\sim6-7$ \citep{atek15b, ishigaki15, ishigaki17, laporte16, livermore16, bouw17magnif}.
These fractions, together with that calculated assuming our size--luminosity 
relation considering its uncertainty, are shown in the middle panel of 
Figure~\ref{fig:detected_fraction}.
The assumed size--luminosity relations are presented in the bottom panel
of Figure~\ref{fig:detected_fraction}.
Whereas the relations in \citet{ishigaki15,ishigaki17}
have delta-function-like size distributions, those in \citet{atek15b}, 
\citet{laporte16}, \citet{livermore16},
and \citet{bouw17magnif} have variances of 
$\sigma\simeq 0.9$, 0.9, 1.0, and 0.69, respectively.
As shown in the bottom panel, all of the size--luminosity relations 
in the previous studies are considerably 
flatter than ours, which results in underestimation of the detected 
fraction and a steeper faint-end slope of the luminosity function.
Furthermore, there is a considerable difference between the relations,
which introduces a significant uncertainty in the detected fraction
and, consequently, in the luminosity function.
In contrast, the uncertainty in the detected fraction 
calculated by our size--luminosity relation is smaller than 
the scatter of the detected fractions by the relations 
in the previous studies.
This means that we reduce the 
uncertainty in the luminosity 
function that originates from the size--luminosity relation 
(the middle panel of Figure~\ref{fig:detected_fraction}).

Our size--luminosity relations are more accurate
than those in previous studies at $z\sim6-9$
for three reasons:
they are not extrapolations from low-redshift results but are
determined directly from large samples with accurate size measurements,
they are corrected for detection incompleteness, and proper statistics
are utilized.

\subsection{Redshift Evolution of Size}\label{subsec:redshift_evolution}

\begin{figure*}[t]
  \centering
      \includegraphics[width=\linewidth]{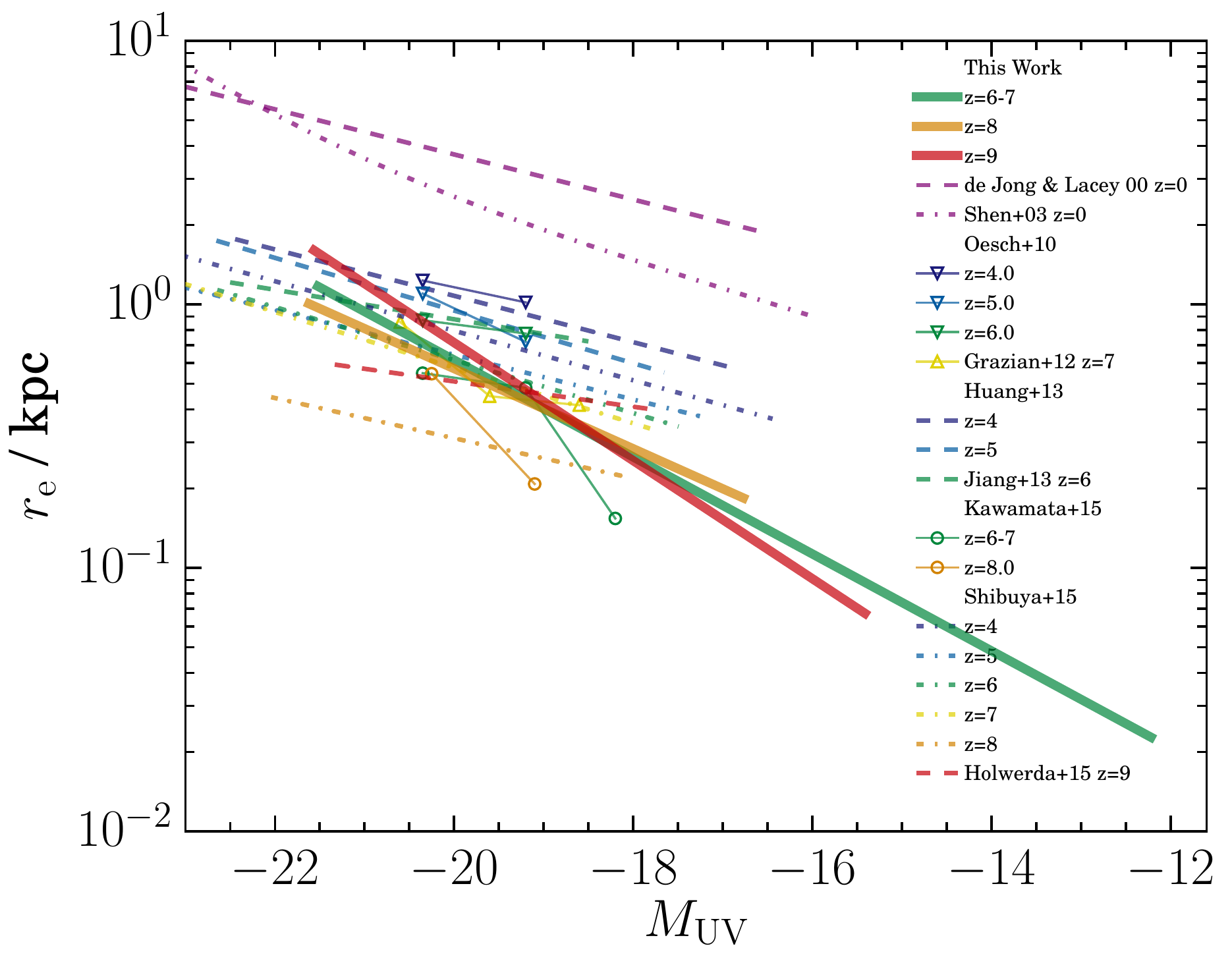}
  \caption{Compilation of size--luminosity relations of LBGs
  from $z\sim4$ to $z\sim9-10$ (\citealt{huang13}, \citealt{jiang13b} [LAEs+LBGs],
  \citealt{oesch10b}, \citealt{graz12}, \citealt{kawamata15}, \citealt{shibuya15},
  and \citealt{holwerda15}) and spiral galaxies at $z=0$ (\citealt{dejonglacey00}, \citealt{shen03}),
  with our results plotted by thick solid lines.
  Redshift is coded by color: purple, $z=0$; violet, $z\sim4$; blue, $z\sim5$; 
  green, $z\sim6$ ($z\sim6-7$ for our result); yellow, $z\sim7$; orange, $z\sim8$; 
  and red, $z\sim9-10$
  ($z\sim9$ for our result).  
  Different symbols represent the average absolute magnitudes and sizes of different samples: 
  inverse triangles, $z\sim4-6$ samples by \citet{oesch10b}; triangles, $z\sim7$ sample by \citet{graz12}; and 
  circles, $z\sim6-8$ samples by \citet{kawamata15}.
  The purple dashed and dot-dashed lines represent the relations
  of $z\sim0$ disk galaxies obtained by the measurements in the $i$ band
  by \citet{dejonglacey00} and $r$ band by \citet{shen03}, respectively.
  The violet, blue, and green dashed lines represent the 
  fitting results to the sample of $z\sim4$ and $5$ LBGs by \citet{huang13}
  and $z\sim5.7-6.5$ Ly$\alpha$ emitters and LBGs by \citet{jiang13b},
  respectively.
  }
  \label{fig:rlrelation_evolution}
\end{figure*}

\begin{figure}[t]
  \centering
      \includegraphics[width=\linewidth]{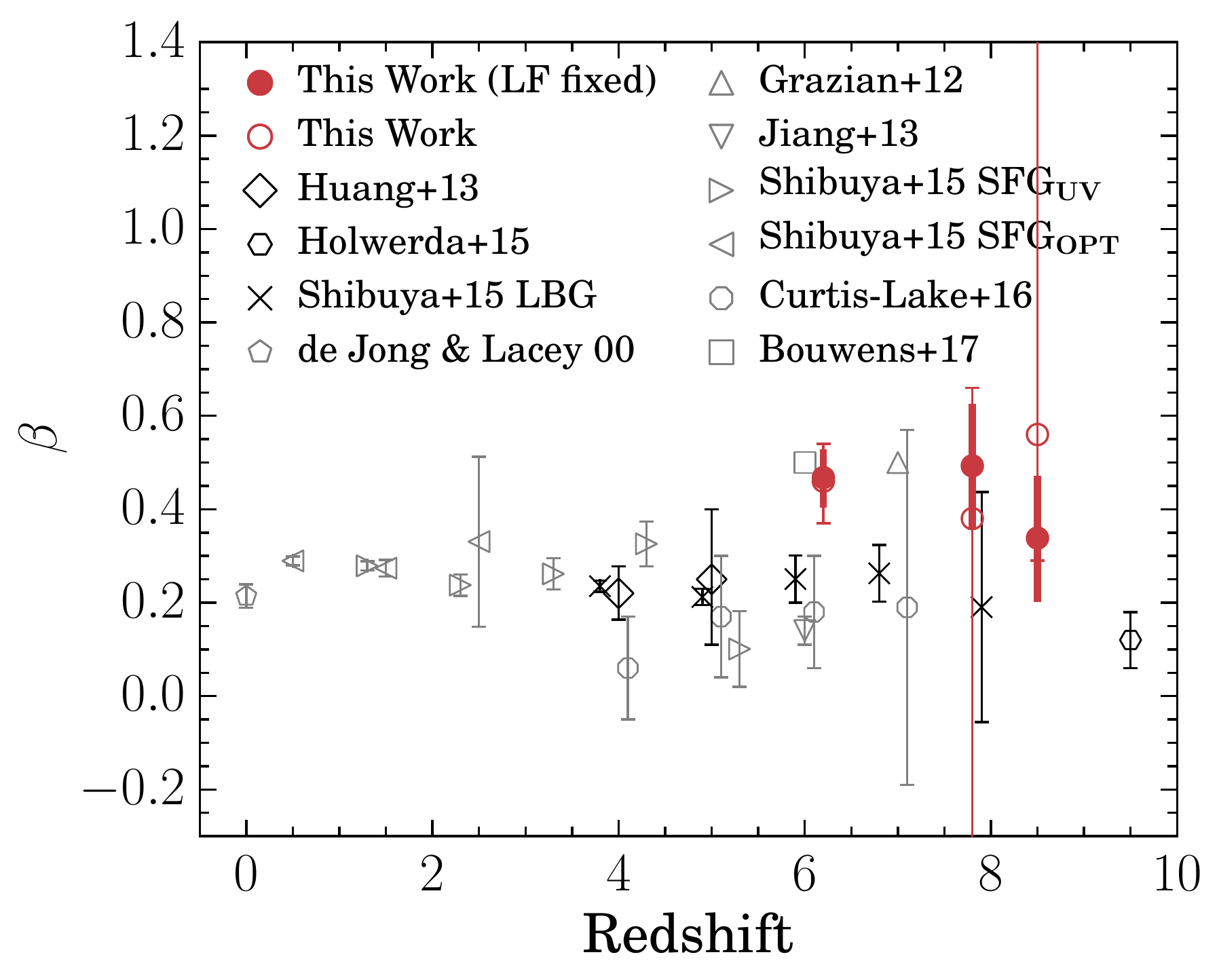}
  \caption{Redshift evolution of the slope of the size--luminosity
  relation.
  The red circles show our measurements, while
  black symbols those of LBGs obtained by previous studies with
  two-dimensional profile fitting.
  The gray symbols represent results for non-LBG samples 
  or those not based on two-dimensional profile fitting.
  The error bars correspond to the $1 \sigma$ standard errors.
  The bold error bars of our samples show the $1 \sigma$ standard errors
  where the parameters of the luminosity functions are fixed to the
  $z\sim6-7$ best-fit values.
  }
  \label{fig:beta_evolution}
\end{figure}

\begin{figure}[t]
  \centering
      \includegraphics[width=\linewidth]{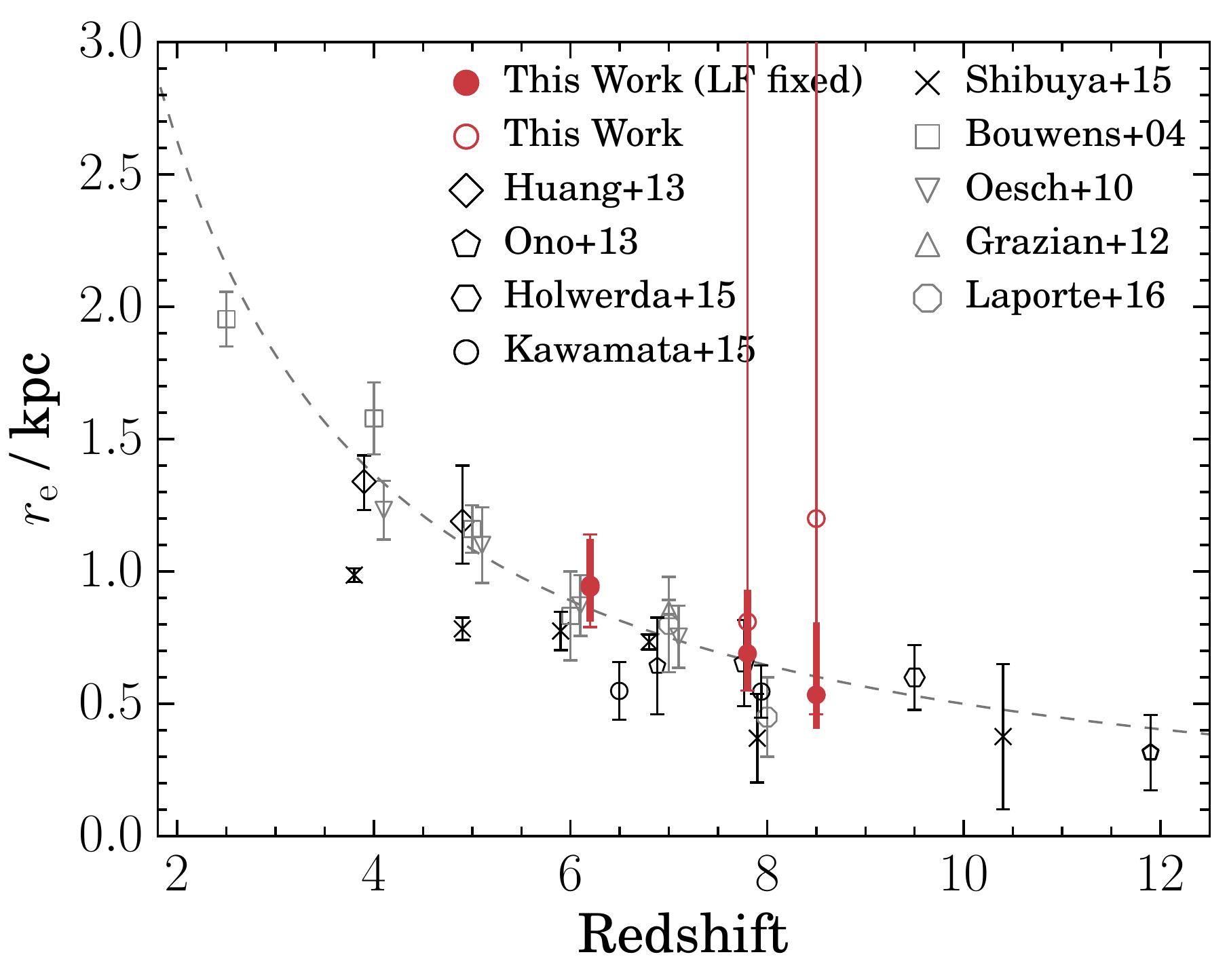}
  \caption{Redshift evolution of the average size of bright galaxies
  at $-21\lesssim M_{\mathrm{UV}} \lesssim -19.7$.
  The red circles show our measurements, while the
  black symbols show those of LBGs obtained by previous studies.
  The gray dashed line represents the best-fit function of
  $r_{\mathrm e} \propto (1+z)^{-m}$ with $m=1.28$.
  The error bars correspond to the $1 \sigma$ standard errors.
  The bold error bars of our samples show the $1 \sigma$ standard errors
  where the parameters of the luminosity functions are fixed to the 
  $z\sim6-7$ best-fit values.
  }
  \label{fig:size_evolution}
\end{figure}

\begin{deluxetable}{ll}
\tablecolumns{2}
\tabletypesize{\scriptsize}
\tablecaption{Luminosities Where the Average Sizes in 
Figure~\ref{fig:size_evolution} Are Calculated
\label{tab:averageluminosities}}
\tablewidth{0pt}
\tablehead{
\colhead{References} &
\colhead{$\overline{M_{\mathrm{UV}}}$} 
}
\startdata
This work & $-21$\\
\citet{bouw04} & $-20.35$\\
\citet{oesch10b} & $-20.35$\\
\citet{graz12} & $-20.50$\\
\citet{huang13} & $-21$\\
\citet{ono13} at $z\sim7$ & $-20.2$\\
\citet{ono13} at $z\sim8$ & $-20.15$\\
\citet{kawamata15} at $z\sim6-7$ & $-20.2$\\
\citet{kawamata15} at $z\sim8$ & $-20.30$\\
\citet{holwerda15} & $-20.87$\\
\citet{shibuya15} & $-21$\\
\citet{laporte16} & $-20.35$
\enddata
\end{deluxetable}

\begin{figure}[t]
  \centering
      \includegraphics[width=\linewidth]{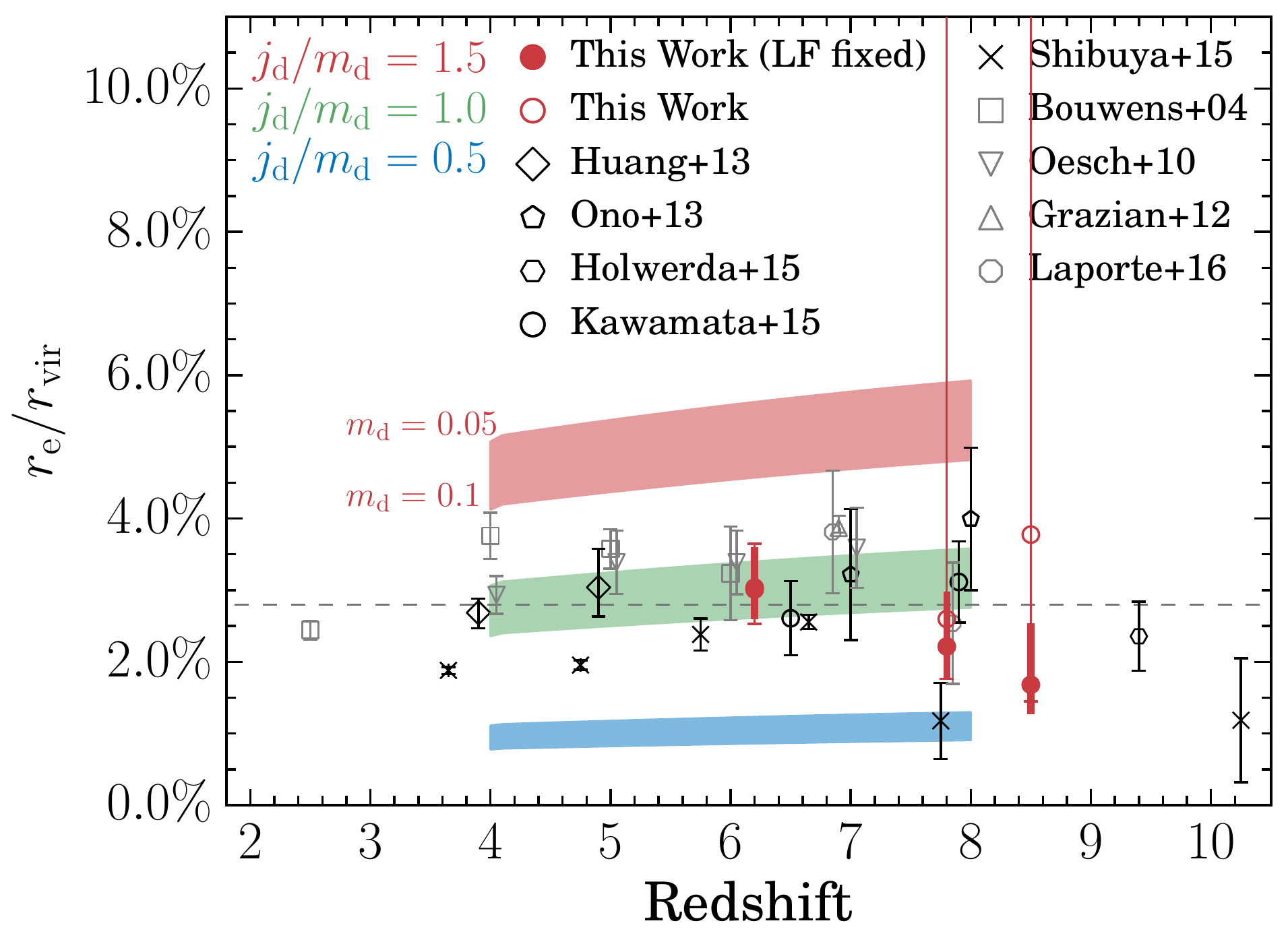}
  \caption{Redshift evolution of the galaxy size--halo size ratio.
  Our samples are shown with red circles, and those from 
  previous studies are in black.
  The errors in disk sizes only (plotted in Figure~\ref{fig:size_evolution}) 
  are considered.
  The gray dashed line shows the average size ratio.
  The red, green, and blue shaded bands represent the ratio
  predicted by the model described in Section~\ref{subsec:redshift_evolution} 
  with $\left({j_{\mathrm{d}}}/
  {m_{\mathrm{d}}}\right)_{M_{\mathrm{UV}}=-21} = 1.5$,
  1.0, and 0.5, respectively, where the width of each band indicates 
  a weak dependence of 
  the ratio on $m_{\mathrm{d}}$ and the upper and lower edges 
  of each band correspond to $m_{\mathrm{d}} = 0.05$ and 0.1,
  respectively.
  }
  \label{fig:ratio_evolution}
\end{figure}

Figure~\ref{fig:rlrelation_evolution} shows the redshift evolution
of the size--luminosity relation. 
While \citet{oesch10b}, \citet{graz12}, \citet{huang13}, \citet{holwerda15}, 
\citet{kawamata15}, and \citet{shibuya15} showed the relations of 
LBGs, \citet{roche96}, \citet{dejonglacey00}, and \citet{jiang13b}
showed those of irregular galaxies, local spiral galaxies, 
and a combined sample of Ly$\alpha$ emitters (LAEs) and LBGs,
respectively.
The slopes at $z\sim6-9$ are slightly steeper than those at $z\lesssim5$
and those derived from bright samples at $z\gtrsim6$.
This may suggest that physical processes that affect the slopes,
such as the formation stage, feedback, and transfers and 
redistributions of angular momentum,
differ at around $z\sim6$, especially for faint galaxies.

Figure~\ref{fig:beta_evolution} shows the redshift evolution
of $\beta$ based on LBG samples by two-dimensional 
profile size measurements.
While our fiducial values, where all uncertainties are considered, 
are plotted with red open circles and thin error bars, values where
the parameters of the luminosity functions are fixed to the $z\sim6-7$ 
best-fit values are plotted with red filled circles and bold error bars
and presented in Table~\ref{tab:bestparams}.
For comparison, we also plot results from samples of non-LBGs and 
samples based on other size measurement methods.
This figure shows that the slopes of our faint LBGs at $z\gtrsim6$
are steeper than those of bright or lower-redshift galaxies, which
are almost constant at $\beta \simeq 0.2$--0.3.

The redshift evolution of sizes at 
$-21 \lesssim M_{\mathrm UV} \lesssim -19.7$ ($(0.3-1)L^{*}_{z=3}$)
is presented in Figure~\ref{fig:size_evolution}, where $L^{*}_{z=3}$ 
is the characteristic UV luminosity of $z\sim3$ LBGs
obtained in \citet{steidel99}.
Similar to Figure~\ref{fig:beta_evolution}, 
we plot our fiducial values and values where the parameters 
of the luminosity functions are fixed.
Our samples give consistent results with previous measurements.
We fit ${r_{\mathrm e}} \propto (1+z)^{-m}$ to data that are 
based on two-dimensional size measurements
at $4<z<9.5$
(except for those by \citealt{shibuya15}, because 
they seem to be considerably smaller than the others).
For our data, we use the ones
where the parameters of the luminosity functions 
are fixed for consistency with the previous studies.
We obtain $m=1.28\pm0.11$, which is 
consistent within the errors with previous work 
\citep{bouw04, oesch10b, ono13, kawamata15, holwerda15, shibuya15}.
The index is predicted by analytical models to be 
$m=1.0$ for halos with a fixed mass
and $m=1.5$ for halos with a fixed circular velocity \citep[e.g.,][]{ferg04}.
We find that we trace halos in the middle of the two states,
as reported in previous work.

We note that the difference in the luminosity range 
makes the comparison between the 
samples difficult.
The average luminosities of individual samples plotted in 
Figure~\ref{fig:rlrelation_evolution} have some variance,
as shown in Table~\ref{tab:averageluminosities}.
For instance, at $z=7$, a difference of 0.5 mag in luminosity
corresponds to a difference in stellar mass of 
$\Delta M^{*}/M^{*} = 54\%$, assuming the mass-luminosity 
relation in \citet{gonz11}.
Based on the stellar mass--halo mass relation by 
\citet{behroozi13},
the difference in stellar mass at $M_{\mathrm{UV}}=-21$ 
is equivalent to those in halo mass and 
halo radius of $\Delta M_{\mathrm{vir}}/M_{\mathrm{vir}} = 52\%$ 
and $\Delta r_{\mathrm{vir}}/r_{\mathrm{vir}} = 21\%$, 
respectively.
Since the galaxy size is fundamentally proportional 
to the halo size, the expected galaxy size would
differ $\Delta r_{\mathrm{e}}/r_{\mathrm{e}} = 21\%$.
This means that the difference between samples in the 
luminosity range introduces a systematic uncertainty into 
the discussion of the evolution of the average size,
which is conventional in previous studies.

In order to resolve the above problem and further investigate 
the size evolution of galaxies, we calculate the evolution of 
the galaxy size--halo size ratio following K15 
\citep[see also][]{shibuya15, okamura17}.
We calculate size ratios with a similar method to 
that for the model construction described in Section~\ref{sec:RLmodel}.
In order to estimate the average halo size of each sample from its 
average luminosity, 
we make use of the stellar mass--luminosity relation in
\citet{reddysteidel09} at $z\sim2.5$ and that in 
\citet{gonz11} at $z\sim4-9.5$,
the stellar mass--halo mass relation \citep{behroozi13}, and 
Equation~(\ref{eqn:virialsize}).
Then, we obtain the size ratio by dividing the galaxy size by the halo size.
The stellar mass--luminosity relation in \citet{gonz11} is originally obtained 
at $z\sim4-7$, but we also apply the relation at $z\sim8-9.5$.
In the above process, the variance in luminosity between the 
samples is corrected for 
because fainter samples are assigned smaller halo sizes.
The result is shown in Figure~\ref{fig:ratio_evolution}.
We confirm that the size ratio is roughly constant over the wide 
redshift range of $2.5 \lesssim z \lesssim 7.0$, 
and the average ratio is $2.80\%\pm0.10\%$ over this redshift range.
This value is in good agreement with those obtained 
in previous studies \citep{kawamata15, shibuya15, huang17, okamura17, somerville17}.

It appears from Figure~\ref{fig:size_evolution} that the average
size continues to decrease with redshift at $z\gtrsim7$.
This trend, if true, predicts that the size ratio starts to decrease at $z\gtrsim7$
because the denominator (halo mass and hence halo size 
of $M_\mathrm{UV} \simeq -21$ galaxies) increases with redshift at $z\gtrsim7$
according to the stellar mass--halo mass relation by \citet{behroozi13}.
This prediction is consistent with our size ratio measurements at $z\sim8$ and 9
within the errors.
This decreasing trend in the size ratio was not observed in our previous work, K15,
because K15 linearly extrapolated the stellar mass--halo mass relation
at $M_{\mathrm h} \simeq 10^{11} M_{\odot}$, while in reality, it has 
a knee at $M_{\mathrm{h}} \simeq 10^{11.3} M_{\odot}$, thus resulting in 
underestimation of the halo masses.

We compare the observed size ratios with those 
predicted by the model constructed in Section~\ref{sec:RLmodel}
with $\gamma=0$, 
\begin{align}
\frac{r_{\mathrm{e}}}{r_{\mathrm{vir}}} = \frac{1.678}{\sqrt{2}}\left(\frac{j_{\mathrm{d}}}{m_{\mathrm{d}}}\right)_{M_{\mathrm{UV}}=-21} \lambda f_{\mathrm{c}}^{-1/2} f_{\mathrm{R}}.
\end{align}
Since $f_{\mathrm{R}}$ strongly depends 
on $j_{\mathrm{d}}/m_{\mathrm{d}}$
and weakly on $m_{\mathrm{d}}$, the only uncertain
parameter to calculate the size ratio is
$\left({j_{\mathrm{d}}}/{m_{\mathrm{d}}}\right)_{M_{\mathrm{UV}}=-21}$.
Following K15, we change $\left({j_{\mathrm{d}}}/{m_{\mathrm{d}}}\right)_{M_{\mathrm{UV}}=-21}$
with the updated size measurements and simulation
results of $\lambda$ and $c$.
Model-predicted size ratios are presented in 
Figure~\ref{fig:ratio_evolution}.
Since they weakly depend on $m_{\mathrm{d}}$,
we show with bands the uncertainty due to 
$m_{\mathrm{d}}$ within the range of $0.05-0.1$. 
If we assume the typical value of 0.05 for $m_{\mathrm{d}}$
\citep[e.g.,][]{sales10},
we confirm that the observed size ratios are in good 
accordance with the model ratios premised on 
$j_{\mathrm{d}} \sim m_{\mathrm{d}}$ at $M_{\mathrm{UV}} = -21$.
This is why we have assumed
$\left({j_{\mathrm{d}}}/{m_{\mathrm{d}}}\right)_{M_{\mathrm{UV}}=-21}=1.0$
and $m_{\mathrm{d}}=0.05$, when modeling the size--luminosity relation
in Section~\ref{sec:RLmodel}.

Using a mass-complete sample at $z\sim1-7$ from the 
FourStar Galaxy Evolution Survey,
\citet{allen17} have found a slower size evolution of 
$r_\mathrm{e} \propto (1+z)^{-0.97\pm0.02}$.
Since the size evolution of LBGs is faster, they have concluded that
LBGs do not represent the entire galaxy population.
Considering their results, it should be noted that 
this study also might not be tracing
the entire galaxy population at $z\sim6-9$.

\section{Conclusion}\label{sec:conclusion}
We have measured the intrinsic sizes and magnitudes of 
334, 61, and 37 faint dropout galaxies at $z\sim6-7$, 8, and 9,
respectively, from the complete HFF data, 
properly correcting for the lensing effects by fitting 
the lensed images with lensing-distorted S\'ersic profiles.
These represent the largest samples, especially at faint magnitudes
of $M_\mathrm{UV}= -18$ to $-12$, where luminosity function
measurements have been made possible only recently.
Systematic and random errors in sizes and magnitudes have been
carefully estimated using Monte Carlo 
simulations.

Although the HFF observations reach the faintest galaxies 
with the help of cluster lensing,
our samples still suffer from the incompleteness
that faint but large galaxies are not detected in observations.
Since the degree of incompleteness strongly depends on 
the intrinsic size--luminosity relation, 
we have conducted simultaneous maximum-likelihood estimation 
of the luminosity function and size--luminosity relation 
from the observed distribution of 
galaxies on the size--luminosity plane and examined correlations 
between the luminosity function and size--luminosity relation.

We have also updated our mass models for \clone\ and \cltwo,
as well as newly constructed models for \clfive\ and \clsix, all of which are 
publicly available through the STScI website.  
The following are the main results of this paper.
\begin{enumerate}[i.]
\item We have found that the slope of the intrinsic size--luminosity
relation of faint galaxies at $z\sim6-7$ is considerably steeper 
($\beta \simeq 0.46$)
than those ($\beta \simeq 0.22$--0.25) at $z\sim4-5$
and those ($\beta \simeq 0.25$) assumed in previous studies 
of the luminosity function at $z\sim6-7$.
As a result of the steep size--luminosity relation, 
a shallow faint-end slope of the luminosity function 
of $\alpha = -1.86^{+0.17}_{-0.18}$ has been derived.
The values of $\beta$ and $\alpha$ at $z\sim8$ and 9 are consistent 
with those at $z\sim6-7$ but have large errors due to small sample sizes.
Thus, at $z\sim8$ and 9, the UV luminosity density is still highly uncertain, 
which has to be taken into account in the discussion of cosmic reionization.
\item We have quantified the correlation between the
parameters of the size--luminosity relation and luminosity function.
Among the parameter pairs, we have found strong correlations 
between the faint-end slope of the luminosity function and 
the slope of the size--luminosity relation, ($\alpha$, $\beta$),
and between the faint-end slope and the characteristic magnitude of 
the luminosity function, ($\alpha$, $M_{\star}$).
Although the values of $\alpha$ in several previous studies are consistent 
with our measurements, 
some of the previous results have been found to be located 
outside our confidence region in the $\alpha$--$\beta$ plane.
\item We have constructed an analytical model to reproduce 
the steep slope of the size--luminosity relation at $z\sim6-7$
utilizing the result of the abundance matching in \citet{behroozi13}.
We have found that the steepness is not reproduced when 
$j_\mathrm{d} / m_\mathrm{d}$ is constant within the magnitude range
studied here.
One possible explanation for the steepness is that 
a smaller fraction of the 
specific angular momentum
is transferred to the disk from its halo at fainter magnitudes.
Another possible explanation is that low-mass halos can 
host galaxies only when they have relatively small halo spin 
parameters.
\item The average size at ($0.3-1)L^{*}_{z=3}$ gradually decreases
with redshift
with $(1+z)^{-m}$, where $m=1.28\pm0.11$ over a redshift range of $4\lesssim z\lesssim9.5$.
However, we have pointed out that this conventional discussion
of the size evolution 
suffers from systematic biases due to 
a variance in average luminosity between the samples.
In order to overcome this issue, we have calculated the disk-to-halo
size ratio to find $j_\mathrm{d} / m_\mathrm{d} \sim1$ 
at $M_\mathrm{UV} = -21$. 
\end{enumerate}

\section*{Acknowledgments}
We would like to thank the anonymous referee for valuable comments that improved our paper.
We would like to thank Michael Fall, Rychard Bouwens, Rebecca Bowler, Henry Ferguson, Kentaro Nagamine, Pascal Oesch, Yoshiaki Ono, Takatoshi Shibuya, Tsutomu Takeuchi, Haruka Kusakabe, Taku Okamura, and Kazushi Irikura for their helpful comments.
We are grateful to Peter Behroozi, Chuanwu Liu, and Takatoshi Shibuya for kindly providing us with their results.
This work was supported in part by a Grant-in-Aid
for JSPS Research Fellow (JP16J01302, JP16J03727) and 
by a KAKENHI (JP16K05286) Grant-in-Aid for Scientific Research (C) 
through the Japan Society for the Promotion of Science (JSPS).
This work was supported in part by the World Premier International
Research Center Initiative (WPI Initiative), MEXT, Japan, and JSPS
KAKENHI Grant Number JP26800093 and JP15H05892.
This work used the 2015 public version of the Munich model of galaxy formation and evolution: L-Galaxies. The source code and a full description of the model are available at \url{http://galformod.mpa-garching.mpg.de/public/LGalaxies/}.
The Millennium and Millennium-II simulation databases used in this paper and the web application providing online access to them were constructed as part of the activities of the German Astrophysical Virtual Observatory (GAVO).


\bibliographystyle{apj}
\bibliography{bibtex.bib}

\begin{thebibliography}{}
\expandafter\ifx\csname natexlab\endcsname\relax\def\natexlab#1{#1}\fi

\bibitem[{{Allen} {et~al.}(2017){Allen}, {Kacprzak}, {Glazebrook}, {Labb{\'e}},
  {Tran}, {Spitler}, {Cowley}, {Nanayakkara}, {Papovich}, {Quadri},
  {Straatman}, {Tilvi}, \& {van Dokkum}}]{allen17}
{Allen}, R.~J., {Kacprzak}, G.~G., {Glazebrook}, K., {et~al.} 2017, \apjl, 834,
  L11

\bibitem[{{Atek} {et~al.}(2014){Atek}, {Richard}, {Kneib}, {Clement}, {Egami},
  {Ebeling}, {Jauzac}, {Jullo}, {Laporte}, {Limousin}, \& {Natarajan}}]{atek14}
{Atek}, H., {Richard}, J., {Kneib}, J.-P., {et~al.} 2014, \apj, 786, 60

\bibitem[{{Atek} {et~al.}(2015{\natexlab{a}}){Atek}, {Richard}, {Jauzac},
  {Kneib}, {Natarajan}, {Limousin}, {Schaerer}, {Jullo}, {Ebeling}, {Egami}, \&
  {Clement}}]{atek15b}
{Atek}, H., {Richard}, J., {Jauzac}, M., {et~al.} 2015{\natexlab{a}}, \apj,
  814, 69

\bibitem[{{Atek} {et~al.}(2015{\natexlab{b}}){Atek}, {Richard}, {Kneib},
  {Jauzac}, {Schaerer}, {Clement}, {Limousin}, {Jullo}, {Natarajan}, {Egami},
  \& {Ebeling}}]{atek15a}
{Atek}, H., {Richard}, J., {Kneib}, J.-P., {et~al.} 2015{\natexlab{b}}, \apj,
  800, 18

\bibitem[{{Balestra} {et~al.}(2013){Balestra}, {Vanzella}, {Rosati}, {Monna},
  {Grillo}, {Nonino}, {Mercurio}, {Biviano}, {Bradley}, {Coe}, {Fritz},
  {Postman}, {Seitz}, {Scodeggio}, {Tozzi}, {Zheng}, {Ziegler}, {Zitrin},
  {Annunziatella}, {Bartelmann}, {Benitez}, {Broadhurst}, {Bouwens}, {Czoske},
  {Donahue}, {Ford}, {Girardi}, {Infante}, {Jouvel}, {Kelson}, {Koekemoer},
  {Kuchner}, {Lemze}, {Lombardi}, {Maier}, {Medezinski}, {Melchior},
  {Meneghetti}, {Merten}, {Molino}, {Moustakas}, {Presotto}, {Smit}, \&
  {Umetsu}}]{balestra13}
{Balestra}, I., {Vanzella}, E., {Rosati}, P., {et~al.} 2013, \aap, 559, L9

\bibitem[{{Behroozi} {et~al.}(2013){Behroozi}, {Wechsler}, \&
  {Conroy}}]{behroozi13}
{Behroozi}, P.~S., {Wechsler}, R.~H., \& {Conroy}, C. 2013, \apj, 770, 57

\bibitem[{{Ben{\'{\i}}tez}(2000)}]{benitez00}
{Ben{\'{\i}}tez}, N. 2000, \apj, 536, 571

\bibitem[{{Bertin} \& {Arnouts}(1996)}]{bertin96}
{Bertin}, E., \& {Arnouts}, S. 1996, \aaps, 117, 393

\bibitem[{{Bertin} {et~al.}(2002){Bertin}, {Mellier}, {Radovich}, {Missonnier},
  {Didelon}, \& {Morin}}]{bertin02}
{Bertin}, E., {Mellier}, Y., {Radovich}, M., {et~al.} 2002, in Astronomical
  Society of the Pacific Conference Series, Vol. 281, Astronomical Data
  Analysis Software and Systems XI, ed. D.~A. {Bohlender}, D.~{Durand}, \&
  T.~H. {Handley}, 228

\bibitem[{{Bouwens} {et~al.}(2004){Bouwens}, {Illingworth}, {Blakeslee},
  {Broadhurst}, \& {Franx}}]{bouw04}
{Bouwens}, R.~J., {Illingworth}, G.~D., {Blakeslee}, J.~P., {Broadhurst},
  T.~J., \& {Franx}, M. 2004, \apjl, 611, L1

\bibitem[{{Bouwens} {et~al.}(2017{\natexlab{a}}){Bouwens}, {Illingworth},
  {Oesch}, {Atek}, {Lam}, \& {Stefanon}}]{bouw17size}
{Bouwens}, R.~J., {Illingworth}, G.~D., {Oesch}, P.~A., {et~al.}
  2017{\natexlab{a}}, \apj, 843, 41

\bibitem[{{Bouwens} {et~al.}(2017{\natexlab{b}}){Bouwens}, {Oesch},
  {Illingworth}, {Ellis}, \& {Stefanon}}]{bouw17magnif}
{Bouwens}, R.~J., {Oesch}, P.~A., {Illingworth}, G.~D., {Ellis}, R.~S., \&
  {Stefanon}, M. 2017{\natexlab{b}}, \apj, 843, 129

\bibitem[{{Bouwens} {et~al.}(2017{\natexlab{c}}){Bouwens}, {van Dokkum},
  {Illingworth}, {Oesch}, {Maseda}, {Ribeiro}, {Stefanon}, \& {Lam}}]{bouw17gc}
{Bouwens}, R.~J., {van Dokkum}, P.~G., {Illingworth}, G.~D., {et~al.}
  2017{\natexlab{c}}, ArXiv e-prints, arXiv:1711.02090

\bibitem[{{Bouwens} {et~al.}(2015){Bouwens}, {Illingworth}, {Oesch}, {Trenti},
  {Labb{\'e}}, {Bradley}, {Carollo}, {van Dokkum}, {Gonzalez}, {Holwerda},
  {Franx}, {Spitler}, {Smit}, \& {Magee}}]{bouw15}
{Bouwens}, R.~J., {Illingworth}, G.~D., {Oesch}, P.~A., {et~al.} 2015, \apj,
  803, 34

\bibitem[{{Bowler} {et~al.}(2017){Bowler}, {Dunlop}, {McLure}, \&
  {McLeod}}]{bowler17}
{Bowler}, R.~A.~A., {Dunlop}, J.~S., {McLure}, R.~J., \& {McLeod}, D.~J. 2017,
  \mnras, 466, 3612

\bibitem[{{Boylan-Kolchin} {et~al.}(2009){Boylan-Kolchin}, {Springel}, {White},
  {Jenkins}, \& {Lemson}}]{boylankolchin09}
{Boylan-Kolchin}, M., {Springel}, V., {White}, S.~D.~M., {Jenkins}, A., \&
  {Lemson}, G. 2009, \mnras, 398, 1150

\bibitem[{{Brook} {et~al.}(2012){Brook}, {Stinson}, {Gibson}, {Ro{\v s}kar},
  {Wadsley}, \& {Quinn}}]{brook12}
{Brook}, C.~B., {Stinson}, G., {Gibson}, B.~K., {et~al.} 2012, \mnras, 419, 771

\bibitem[{{Brooks} {et~al.}(2011){Brooks}, {Solomon}, {Governato}, {McCleary},
  {MacArthur}, {Brook}, {Jonsson}, {Quinn}, \& {Wadsley}}]{brooks11}
{Brooks}, A.~M., {Solomon}, A.~R., {Governato}, F., {et~al.} 2011, \apj, 728,
  51

\bibitem[{{Bryan} \& {Norman}(1998)}]{bryannorman98}
{Bryan}, G.~L., \& {Norman}, M.~L. 1998, \apj, 495, 80

\bibitem[{{Bullock} {et~al.}(2001){Bullock}, {Kolatt}, {Sigad}, {Somerville},
  {Kravtsov}, {Klypin}, {Primack}, \& {Dekel}}]{bullock01}
{Bullock}, J.~S., {Kolatt}, T.~S., {Sigad}, Y., {et~al.} 2001, \mnras, 321, 559

\bibitem[{{Caminha} {et~al.}(2016{\natexlab{a}}){Caminha}, {Grillo}, {Rosati},
  {Balestra}, {Karman}, {Lombardi}, {Mercurio}, {Nonino}, {Tozzi}, {Zitrin},
  {Biviano}, {Girardi}, {Koekemoer}, {Melchior}, {Meneghetti}, {Munari},
  {Suyu}, {Umetsu}, {Annunziatella}, {Borgani}, {Broadhurst}, {Caputi}, {Coe},
  {Delgado-Correal}, {Ettori}, {Fritz}, {Frye}, {Gobat}, {Maier}, {Monna},
  {Postman}, {Sartoris}, {Seitz}, {Vanzella}, \& {Ziegler}}]{caminha16a_as1063}
{Caminha}, G.~B., {Grillo}, C., {Rosati}, P., {et~al.} 2016{\natexlab{a}},
  \aap, 587, A80

\bibitem[{{Caminha} {et~al.}(2016{\natexlab{b}}){Caminha}, {Karman}, {Rosati},
  {Caputi}, {Arrigoni Battaia}, {Balestra}, {Grillo}, {Mercurio}, {Nonino}, \&
  {Vanzella}}]{caminha16b_as1063}
{Caminha}, G.~B., {Karman}, W., {Rosati}, P., {et~al.} 2016{\natexlab{b}},
  \aap, 595, A100

\bibitem[{{Caminha} {et~al.}(2017){Caminha}, {Grillo}, {Rosati}, {Balestra},
  {Mercurio}, {Vanzella}, {Biviano}, {Caputi}, {Delgado-Correal}, {Karman},
  {Lombardi}, {Meneghetti}, {Sartoris}, \& {Tozzi}}]{caminha16}
{Caminha}, G.~B., {Grillo}, C., {Rosati}, P., {et~al.} 2017, \aap, 600, A90

\bibitem[{{Castellano} {et~al.}(2016){Castellano}, {Yue}, {Ferrara}, {Merlin},
  {Fontana}, {Amor{\'{\i}}n}, {Grazian}, {M{\'a}rmol-Queralto},
  {Micha{\l}owski}, {Mortlock}, {Paris}, {Parsa}, {Pilo}, \&
  {Santini}}]{castellano16b}
{Castellano}, M., {Yue}, B., {Ferrara}, A., {et~al.} 2016, \apjl, 823, L40

\bibitem[{{Charlton} {et~al.}(2017){Charlton}, {Hudson}, {Balogh}, \&
  {Khatri}}]{charlton17}
{Charlton}, P.~J.~L., {Hudson}, M.~J., {Balogh}, M.~L., \& {Khatri}, S. 2017,
  \mnras, 472, 2367

\bibitem[{{Christensen} {et~al.}(2012){Christensen}, {Richard}, {Hjorth},
  {Milvang-Jensen}, {Laursen}, {Limousin}, {Dessauges-Zavadsky}, {Grillo}, \&
  {Ebeling}}]{christensen12}
{Christensen}, L., {Richard}, J., {Hjorth}, J., {et~al.} 2012, \mnras, 427,
  1953

\bibitem[{{Correa} {et~al.}(2015){Correa}, {Wyithe}, {Schaye}, \&
  {Duffy}}]{correa15}
{Correa}, C.~A., {Wyithe}, J.~S.~B., {Schaye}, J., \& {Duffy}, A.~R. 2015,
  \mnras, 452, 1217

\bibitem[{{Curtis-Lake} {et~al.}(2016){Curtis-Lake}, {McLure}, {Dunlop},
  {Rogers}, {Targett}, {Dekel}, {Ellis}, {Faber}, {Ferguson}, {Grogin},
  {Kocevski}, {Koekemoer}, {Lai}, {M{\'a}rmol-Queralt{\'o}}, \&
  {Robertson}}]{curtislake16}
{Curtis-Lake}, E., {McLure}, R.~J., {Dunlop}, J.~S., {et~al.} 2016, \mnras,
  457, 440

\bibitem[{{Danovich} {et~al.}(2015){Danovich}, {Dekel}, {Hahn}, {Ceverino}, \&
  {Primack}}]{danovich15}
{Danovich}, M., {Dekel}, A., {Hahn}, O., {Ceverino}, D., \& {Primack}, J. 2015,
  \mnras, 449, 2087

\bibitem[{{Davis} \& {Natarajan}(2009)}]{davisnatarajan09}
{Davis}, A.~J., \& {Natarajan}, P. 2009, \mnras, 393, 1498

\bibitem[{{de Jong} \& {Lacey}(2000)}]{dejonglacey00}
{de Jong}, R.~S., \& {Lacey}, C. 2000, \apj, 545, 781

\bibitem[{{Diego} {et~al.}(2015){Diego}, {Broadhurst}, {Molnar}, {Lam}, \&
  {Lim}}]{diego15a}
{Diego}, J.~M., {Broadhurst}, T., {Molnar}, S.~M., {Lam}, D., \& {Lim}, J.
  2015, \mnras, 447, 3130

\bibitem[{{Diego} {et~al.}(2016{\natexlab{a}}){Diego}, {Broadhurst}, {Wong},
  {Silk}, {Lim}, {Zheng}, {Lam}, \& {Ford}}]{diego15c}
{Diego}, J.~M., {Broadhurst}, T., {Wong}, J., {et~al.} 2016{\natexlab{a}},
  \mnras, 459, 3447

\bibitem[{{Diego} {et~al.}(2016{\natexlab{b}}){Diego}, {Schmidt}, {Broadhurst},
  {Lam}, {Vega-Ferrero}, {Zheng}, {Lee}, {Morishita}, {Bernstein}, {Lim},
  {Silk}, \& {Ford}}]{diego16}
{Diego}, J.~M., {Schmidt}, K.~B., {Broadhurst}, T., {et~al.}
  2016{\natexlab{b}}, ArXiv e-prints, arXiv:1609.04822

\bibitem[{{Ellis} {et~al.}(2013){Ellis}, {McLure}, {Dunlop}, {Robertson},
  {Ono}, {Schenker}, {Koekemoer}, {Bowler}, {Ouchi}, {Rogers}, {Curtis-Lake},
  {Schneider}, {Charlot}, {Stark}, {Furlanetto}, \& {Cirasuolo}}]{ellis13}
{Ellis}, R.~S., {McLure}, R.~J., {Dunlop}, J.~S., {et~al.} 2013, \apjl, 763, L7

\bibitem[{{Fall}(1983)}]{fall83}
{Fall}, S.~M. 1983, in IAU Symposium, Vol. 100, Internal Kinematics and
  Dynamics of Galaxies, ed. E.~{Athanassoula}, 391--398

\bibitem[{{Fall} \& {Efstathiou}(1980)}]{fallefstathiou80}
{Fall}, S.~M., \& {Efstathiou}, G. 1980, \mnras, 193, 189

\bibitem[{{Fall} \& {Romanowsky}(2013)}]{fall13}
{Fall}, S.~M., \& {Romanowsky}, A.~J. 2013, \apjl, 769, L26

\bibitem[{{Ferguson} {et~al.}(2004){Ferguson}, {Dickinson}, {Giavalisco},
  {Kretchmer}, {Ravindranath}, {Idzi}, {Taylor}, {Conselice}, {Fall},
  {Gardner}, {Livio}, {Madau}, {Moustakas}, {Papovich}, {Somerville},
  {Spinrad}, \& {Stern}}]{ferg04}
{Ferguson}, H.~C., {Dickinson}, M., {Giavalisco}, M., {et~al.} 2004, \apjl,
  600, L107

\bibitem[{Foreman-Mackey(2016)}]{foremanmackey16}
Foreman-Mackey, D. 2016, The Journal of Open Source Software, 24, 1

\bibitem[{{Foreman-Mackey} {et~al.}(2013){Foreman-Mackey}, {Hogg}, {Lang}, \&
  {Goodman}}]{foremanmackey13}
{Foreman-Mackey}, D., {Hogg}, D.~W., {Lang}, D., \& {Goodman}, J. 2013, \pasp,
  125, 306

\bibitem[{{Genel} {et~al.}(2015){Genel}, {Fall}, {Hernquist}, {Vogelsberger},
  {Snyder}, {Rodriguez-Gomez}, {Sijacki}, \& {Springel}}]{genel15}
{Genel}, S., {Fall}, S.~M., {Hernquist}, L., {et~al.} 2015, \apjl, 804, L40

\bibitem[{{Gonz{\'a}lez} {et~al.}(2011){Gonz{\'a}lez}, {Labb{\'e}}, {Bouwens},
  {Illingworth}, {Franx}, \& {Kriek}}]{gonz11}
{Gonz{\'a}lez}, V., {Labb{\'e}}, I., {Bouwens}, R.~J., {et~al.} 2011, \apjl,
  735, L34

\bibitem[{{Grazian} {et~al.}(2011){Grazian}, {Castellano}, {Koekemoer},
  {Fontana}, {Pentericci}, {Testa}, {Boutsia}, {Giallongo}, {Giavalisco}, \&
  {Santini}}]{graz11}
{Grazian}, A., {Castellano}, M., {Koekemoer}, A.~M., {et~al.} 2011, \aap, 532,
  A33

\bibitem[{{Grazian} {et~al.}(2012){Grazian}, {Castellano}, {Fontana},
  {Pentericci}, {Dunlop}, {McLure}, {Koekemoer}, {Dickinson}, {Faber},
  {Ferguson}, {Galametz}, {Giavalisco}, {Grogin}, {Hathi}, {Kocevski}, {Lai},
  {Newman}, \& {Vanzella}}]{graz12}
{Grazian}, A., {Castellano}, M., {Fontana}, A., {et~al.} 2012, \aap, 547, A51

\bibitem[{{Grillo} {et~al.}(2015){Grillo}, {Suyu}, {Rosati}, {Mercurio},
  {Balestra}, {Munari}, {Nonino}, {Caminha}, {Lombardi}, {De Lucia}, {Borgani},
  {Gobat}, {Biviano}, {Girardi}, {Umetsu}, {Coe}, {Koekemoer}, {Postman},
  {Zitrin}, {Halkola}, {Broadhurst}, {Sartoris}, {Presotto}, {Annunziatella},
  {Maier}, {Fritz}, {Vanzella}, \& {Frye}}]{grillo15}
{Grillo}, C., {Suyu}, S.~H., {Rosati}, P., {et~al.} 2015, \apj, 800, 38

\bibitem[{{Grogin} {et~al.}(2011){Grogin}, {Kocevski}, {Faber}, {Ferguson},
  {Koekemoer}, {Riess}, {Acquaviva}, {Alexander}, {Almaini}, {Ashby}, {Barden},
  {Bell}, {Bournaud}, {Brown}, {Caputi}, {Casertano}, {Cassata}, {Castellano},
  {Challis}, {Chary}, {Cheung}, {Cirasuolo}, {Conselice}, {Roshan Cooray},
  {Croton}, {Daddi}, {Dahlen}, {Dav{\'e}}, {de Mello}, {Dekel}, {Dickinson},
  {Dolch}, {Donley}, {Dunlop}, {Dutton}, {Elbaz}, {Fazio}, {Filippenko},
  {Finkelstein}, {Fontana}, {Gardner}, {Garnavich}, {Gawiser}, {Giavalisco},
  {Grazian}, {Guo}, {Hathi}, {H{\"a}ussler}, {Hopkins}, {Huang}, {Huang},
  {Jha}, {Kartaltepe}, {Kirshner}, {Koo}, {Lai}, {Lee}, {Li}, {Lotz}, {Lucas},
  {Madau}, {McCarthy}, {McGrath}, {McIntosh}, {McLure}, {Mobasher},
  {Moustakas}, {Mozena}, {Nandra}, {Newman}, {Niemi}, {Noeske}, {Papovich},
  {Pentericci}, {Pope}, {Primack}, {Rajan}, {Ravindranath}, {Reddy}, {Renzini},
  {Rix}, {Robaina}, {Rodney}, {Rosario}, {Rosati}, {Salimbeni}, {Scarlata},
  {Siana}, {Simard}, {Smidt}, {Somerville}, {Spinrad}, {Straughn}, {Strolger},
  {Telford}, {Teplitz}, {Trump}, {van der Wel}, {Villforth}, {Wechsler},
  {Weiner}, {Wiklind}, {Wild}, {Wilson}, {Wuyts}, {Yan}, \& {Yun}}]{grogin11}
{Grogin}, N.~A., {Kocevski}, D.~D., {Faber}, S.~M., {et~al.} 2011, \apjs, 197,
  35

\bibitem[{{Guo} {et~al.}(2016){Guo}, {Gonzalez-Perez}, {Guo}, {Schaller},
  {Furlong}, {Bower}, {Cole}, {Crain}, {Frenk}, {Helly}, {Lacey}, {Lagos},
  {Mitchell}, {Schaye}, \& {Theuns}}]{guo16}
{Guo}, Q., {Gonzalez-Perez}, V., {Guo}, Q., {et~al.} 2016, \mnras, 461, 3457

\bibitem[{{Henriques} {et~al.}(2015){Henriques}, {White}, {Thomas}, {Angulo},
  {Guo}, {Lemson}, {Springel}, \& {Overzier}}]{henriques15}
{Henriques}, B.~M.~B., {White}, S.~D.~M., {Thomas}, P.~A., {et~al.} 2015,
  \mnras, 451, 2663

\bibitem[{{Hoag} {et~al.}(2016){Hoag}, {Huang}, {Treu}, {Brada{\v c}},
  {Schmidt}, {Wang}, {Brammer}, {Broussard}, {Amorin}, {Castellano}, {Fontana},
  {Merlin}, {Schrabback}, {Trenti}, \& {Vulcani}}]{hoag16}
{Hoag}, A., {Huang}, K.-H., {Treu}, T., {et~al.} 2016, \apj, 831, 182

\bibitem[{{Holwerda} {et~al.}(2015){Holwerda}, {Bouwens}, {Oesch}, {Smit},
  {Illingworth}, \& {Labbe}}]{holwerda15}
{Holwerda}, B.~W., {Bouwens}, R., {Oesch}, P., {et~al.} 2015, \apj, 808, 6

\bibitem[{{Hou} {et~al.}(2017){Hou}, {Lacey}, \& {Frenk}}]{hou17}
{Hou}, J., {Lacey}, C.~G., \& {Frenk}, C.~S. 2017, ArXiv e-prints,
  arXiv:1708.02950

\bibitem[{{Huang} {et~al.}(2013){Huang}, {Ferguson}, {Ravindranath}, \&
  {Su}}]{huang13}
{Huang}, K.-H., {Ferguson}, H.~C., {Ravindranath}, S., \& {Su}, J. 2013, \apj,
  765, 68

\bibitem[{{Huang} {et~al.}(2017){Huang}, {Fall}, {Ferguson}, {van der Wel},
  {Grogin}, {Koekemoer}, {Lee}, {P{\'e}rez-Gonz{\'a}lez}, \& {Wuyts}}]{huang17}
{Huang}, K.-H., {Fall}, S.~M., {Ferguson}, H.~C., {et~al.} 2017, \apj, 838, 6

\bibitem[{{Illingworth} {et~al.}(2013){Illingworth}, {Magee}, {Oesch},
  {Bouwens}, {Labb{\'e}}, {Stiavelli}, {van Dokkum}, {Franx}, {Trenti},
  {Carollo}, \& {Gonzalez}}]{illingworth13}
{Illingworth}, G.~D., {Magee}, D., {Oesch}, P.~A., {et~al.} 2013, \apjs, 209, 6

\bibitem[{{Ishigaki} {et~al.}(2015){Ishigaki}, {Kawamata}, {Ouchi}, {Oguri},
  {Shimasaku}, \& {Ono}}]{ishigaki15}
{Ishigaki}, M., {Kawamata}, R., {Ouchi}, M., {et~al.} 2015, \apj, 799, 12

\bibitem[{{Ishigaki} {et~al.}(2018){Ishigaki}, {Kawamata}, {Ouchi}, {Oguri},
  {Shimasaku}, \& {Ono}}]{ishigaki17}
---. 2018, \apj, 854, 73

\bibitem[{{Jauzac} {et~al.}(2014){Jauzac}, {Cl{\'e}ment}, {Limousin},
  {Richard}, {Jullo}, {Ebeling}, {Atek}, {Kneib}, {Knowles}, {Natarajan},
  {Eckert}, {Egami}, {Massey}, \& {Rexroth}}]{jauzac14}
{Jauzac}, M., {Cl{\'e}ment}, B., {Limousin}, M., {et~al.} 2014, \mnras, 443,
  1549

\bibitem[{{Jauzac} {et~al.}(2015){Jauzac}, {Richard}, {Jullo}, {Cl{\'e}ment},
  {Limousin}, {Kneib}, {Ebeling}, {Natarajan}, {Rodney}, {Atek}, {Massey},
  {Eckert}, {Egami}, \& {Rexroth}}]{jauzac15}
{Jauzac}, M., {Richard}, J., {Jullo}, E., {et~al.} 2015, \mnras, 452, 1437

\bibitem[{{Jiang} {et~al.}(2013){Jiang}, {Egami}, {Fan}, {Windhorst}, {Cohen},
  {Dav{\'e}}, {Finlator}, {Kashikawa}, {Mechtley}, {Ouchi}, \&
  {Shimasaku}}]{jiang13b}
{Jiang}, L., {Egami}, E., {Fan}, X., {et~al.} 2013, \apj, 773, 153

\bibitem[{{Johnson} {et~al.}(2014){Johnson}, {Sharon}, {Bayliss}, {Gladders},
  {Coe}, \& {Ebeling}}]{johnson14}
{Johnson}, T.~L., {Sharon}, K., {Bayliss}, M.~B., {et~al.} 2014, \apj, 797, 48

\bibitem[{{Karman} {et~al.}(2015){Karman}, {Caputi}, {Grillo}, {Balestra},
  {Rosati}, {Vanzella}, {Coe}, {Christensen}, {Koekemoer}, {Kr{\"u}hler},
  {Lombardi}, {Mercurio}, {Nonino}, \& {van der Wel}}]{karman15}
{Karman}, W., {Caputi}, K.~I., {Grillo}, C., {et~al.} 2015, \aap, 574, A11

\bibitem[{{Karman} {et~al.}(2017){Karman}, {Caputi}, {Caminha}, {Gronke},
  {Grillo}, {Balestra}, {Rosati}, {Vanzella}, {Coe}, {Dijkstra}, {Koekemoer},
  {McLeod}, {Mercurio}, \& {Nonino}}]{karman16}
{Karman}, W., {Caputi}, K.~I., {Caminha}, G.~B., {et~al.} 2017, \aap, 599, A28

\bibitem[{{Kawamata} {et~al.}(2015){Kawamata}, {Ishigaki}, {Shimasaku},
  {Oguri}, \& {Ouchi}}]{kawamata15}
{Kawamata}, R., {Ishigaki}, M., {Shimasaku}, K., {Oguri}, M., \& {Ouchi}, M.
  2015, \apj, 804, 103

\bibitem[{{Kawamata} {et~al.}(2016){Kawamata}, {Oguri}, {Ishigaki},
  {Shimasaku}, \& {Ouchi}}]{kawamata16}
{Kawamata}, R., {Oguri}, M., {Ishigaki}, M., {Shimasaku}, K., \& {Ouchi}, M.
  2016, \apj, 819, 114

\bibitem[{{Koekemoer} {et~al.}(2011){Koekemoer}, {Faber}, {Ferguson}, {Grogin},
  {Kocevski}, {Koo}, {Lai}, {Lotz}, {Lucas}, {McGrath}, {Ogaz}, {Rajan},
  {Riess}, {Rodney}, {Strolger}, {Casertano}, {Castellano}, {Dahlen},
  {Dickinson}, {Dolch}, {Fontana}, {Giavalisco}, {Grazian}, {Guo}, {Hathi},
  {Huang}, {van der Wel}, {Yan}, {Acquaviva}, {Alexander}, {Almaini}, {Ashby},
  {Barden}, {Bell}, {Bournaud}, {Brown}, {Caputi}, {Cassata}, {Challis},
  {Chary}, {Cheung}, {Cirasuolo}, {Conselice}, {Roshan Cooray}, {Croton},
  {Daddi}, {Dav{\'e}}, {de Mello}, {de Ravel}, {Dekel}, {Donley}, {Dunlop},
  {Dutton}, {Elbaz}, {Fazio}, {Filippenko}, {Finkelstein}, {Frazer}, {Gardner},
  {Garnavich}, {Gawiser}, {Gruetzbauch}, {Hartley}, {H{\"a}ussler},
  {Herrington}, {Hopkins}, {Huang}, {Jha}, {Johnson}, {Kartaltepe},
  {Khostovan}, {Kirshner}, {Lani}, {Lee}, {Li}, {Madau}, {McCarthy},
  {McIntosh}, {McLure}, {McPartland}, {Mobasher}, {Moreira}, {Mortlock},
  {Moustakas}, {Mozena}, {Nandra}, {Newman}, {Nielsen}, {Niemi}, {Noeske},
  {Papovich}, {Pentericci}, {Pope}, {Primack}, {Ravindranath}, {Reddy},
  {Renzini}, {Rix}, {Robaina}, {Rosario}, {Rosati}, {Salimbeni}, {Scarlata},
  {Siana}, {Simard}, {Smidt}, {Snyder}, {Somerville}, {Spinrad}, {Straughn},
  {Telford}, {Teplitz}, {Trump}, {Vargas}, {Villforth}, {Wagner}, {Wandro},
  {Wechsler}, {Weiner}, {Wiklind}, {Wild}, {Wilson}, {Wuyts}, \&
  {Yun}}]{koekemoer11}
{Koekemoer}, A.~M., {Faber}, S.~M., {Ferguson}, H.~C., {et~al.} 2011, \apjs,
  197, 36

\bibitem[{{Koekemoer} {et~al.}(2013){Koekemoer}, {Ellis}, {McLure}, {Dunlop},
  {Robertson}, {Ono}, {Schenker}, {Ouchi}, {Bowler}, {Rogers}, {Curtis-Lake},
  {Schneider}, {Charlot}, {Stark}, {Furlanetto}, {Cirasuolo}, {Wild}, \&
  {Targett}}]{koekemoer13}
{Koekemoer}, A.~M., {Ellis}, R.~S., {McLure}, R.~J., {et~al.} 2013, \apjs, 209,
  3

\bibitem[{{Lagattuta} {et~al.}(2017){Lagattuta}, {Richard}, {Cl{\'e}ment},
  {Mahler}, {Patr{\'{\i}}cio}, {Pell{\'o}}, {Soucail}, {Schmidt}, {Wisotzki},
  {Martinez}, \& {Bina}}]{lagattuta17}
{Lagattuta}, D.~J., {Richard}, J., {Cl{\'e}ment}, B., {et~al.} 2017, \mnras,
  469, 3946

\bibitem[{{Lam} {et~al.}(2014){Lam}, {Broadhurst}, {Diego}, {Lim}, {Coe},
  {Ford}, \& {Zheng}}]{lam14}
{Lam}, D., {Broadhurst}, T., {Diego}, J.~M., {et~al.} 2014, \apj, 797, 98

\bibitem[{{Laporte} {et~al.}(2016){Laporte}, {Infante}, {Troncoso Iribarren},
  {Zheng}, {Molino}, {Bauer}, {Bina}, {Broadhurst}, {Chilingarian}, {Huang},
  {Garcia}, {Kim}, {Marques-Chaves}, {Moustakas}, {Pell{\'o}},
  {P{\'e}rez-Fournon}, {Shu}, {Streblyanska}, \& {Zitrin}}]{laporte16}
{Laporte}, N., {Infante}, L., {Troncoso Iribarren}, P., {et~al.} 2016, \apj,
  820, 98

\bibitem[{{Liu} {et~al.}(2017){Liu}, {Mutch}, {Poole}, {Angel}, {Duffy},
  {Geil}, {Mesinger}, \& {Wyithe}}]{liu17}
{Liu}, C., {Mutch}, S.~J., {Poole}, G.~B., {et~al.} 2017, \mnras, 465, 3134

\bibitem[{{Livermore} {et~al.}(2017){Livermore}, {Finkelstein}, \&
  {Lotz}}]{livermore16}
{Livermore}, R.~C., {Finkelstein}, S.~L., \& {Lotz}, J.~M. 2017, \apj, 835, 113

\bibitem[{{Lotz} {et~al.}(2017){Lotz}, {Koekemoer}, {Coe}, {Grogin}, {Capak},
  {Mack}, {Anderson}, {Avila}, {Barker}, {Borncamp}, {Brammer}, {Durbin},
  {Gunning}, {Hilbert}, {Jenkner}, {Khandrika}, {Levay}, {Lucas}, {MacKenty},
  {Ogaz}, {Porterfield}, {Reid}, {Robberto}, {Royle}, {Smith},
  {Storrie-Lombardi}, {Sunnquist}, {Surace}, {Taylor}, {Williams}, {Bullock},
  {Dickinson}, {Finkelstein}, {Natarajan}, {Richard}, {Robertson}, {Tumlinson},
  {Zitrin}, {Flanagan}, {Sembach}, {Soifer}, \& {Mountain}}]{lotz17}
{Lotz}, J.~M., {Koekemoer}, A., {Coe}, D., {et~al.} 2017, \apj, 837, 97

\bibitem[{{Ma} {et~al.}(2017){Ma}, {Hopkins}, {Boylan-Kolchin},
  {Faucher-Gigu{\`e}re}, {Quataert}, {Feldmann}, {Garrison-Kimmel}, {Hayward},
  {Kere{\v s}}, \& {Wetzel}}]{ma17}
{Ma}, X., {Hopkins}, P.~F., {Boylan-Kolchin}, M., {et~al.} 2017, ArXiv
  e-prints, arXiv:1710.00008

\bibitem[{{Mahler} {et~al.}(2018){Mahler}, {Richard}, {Cl{\'e}ment},
  {Lagattuta}, {Schmidt}, {Patr{\'{\i}}cio}, {Soucail}, {Bacon}, {Pello},
  {Bouwens}, {Maseda}, {Martinez}, {Carollo}, {Inami}, {Leclercq}, \&
  {Wisotzki}}]{mahler17}
{Mahler}, G., {Richard}, J., {Cl{\'e}ment}, B., {et~al.} 2018, \mnras, 473, 663

\bibitem[{{McLeod} {et~al.}(2015){McLeod}, {McLure}, {Dunlop}, {Robertson},
  {Ellis}, \& {Targett}}]{mcleod15}
{McLeod}, D.~J., {McLure}, R.~J., {Dunlop}, J.~S., {et~al.} 2015, \mnras, 450,
  3032

\bibitem[{{Meneghetti} {et~al.}(2017){Meneghetti}, {Natarajan}, {Coe},
  {Contini}, {De Lucia}, {Giocoli}, {Acebron}, {Borgani}, {Bradac}, {Diego},
  {Hoag}, {Ishigaki}, {Johnson}, {Jullo}, {Kawamata}, {Lam}, {Limousin},
  {Liesenborgs}, {Oguri}, {Sebesta}, {Sharon}, {Williams}, \&
  {Zitrin}}]{meneghetti17}
{Meneghetti}, M., {Natarajan}, P., {Coe}, D., {et~al.} 2017, \mnras, 472, 3177

\bibitem[{{Merten} {et~al.}(2011){Merten}, {Coe}, {Dupke}, {Massey}, {Zitrin},
  {Cypriano}, {Okabe}, {Frye}, {Braglia}, {Jim{\'e}nez-Teja}, {Ben{\'{\i}}tez},
  {Broadhurst}, {Rhodes}, {Meneghetti}, {Moustakas}, {Sodr{\'e}}, {Krick}, \&
  {Bregman}}]{merten11}
{Merten}, J., {Coe}, D., {Dupke}, R., {et~al.} 2011, \mnras, 417, 333

\bibitem[{{Mo} {et~al.}(1998){Mo}, {Mao}, \& {White}}]{mmw98}
{Mo}, H.~J., {Mao}, S., \& {White}, S.~D.~M. 1998, \mnras, 295, 319

\bibitem[{{Monna} {et~al.}(2014){Monna}, {Seitz}, {Greisel}, {Eichner},
  {Drory}, {Postman}, {Zitrin}, {Coe}, {Halkola}, {Suyu}, {Grillo}, {Rosati},
  {Lemze}, {Balestra}, {Snigula}, {Bradley}, {Umetsu}, {Koekemoer}, {Kuchner},
  {Moustakas}, {Bartelmann}, {Ben{\'{\i}}tez}, {Bouwens}, {Broadhurst},
  {Donahue}, {Ford}, {Host}, {Infante}, {Jimenez-Teja}, {Jouvel}, {Kelson},
  {Lahav}, {Medezinski}, {Melchior}, {Meneghetti}, {Merten}, {Molino},
  {Moustakas}, {Nonino}, \& {Zheng}}]{monna14}
{Monna}, A., {Seitz}, S., {Greisel}, N., {et~al.} 2014, \mnras, 438, 1417

\bibitem[{{Mutch} {et~al.}(2016){Mutch}, {Geil}, {Poole}, {Angel}, {Duffy},
  {Mesinger}, \& {Wyithe}}]{mutch16}
{Mutch}, S.~J., {Geil}, P.~M., {Poole}, G.~B., {et~al.} 2016, \mnras, 462, 250

\bibitem[{{Oesch} {et~al.}(2010{\natexlab{a}}){Oesch}, {Bouwens}, {Carollo},
  {Illingworth}, {Trenti}, {Stiavelli}, {Magee}, {Labb{\'e}}, \&
  {Franx}}]{oesch10b}
{Oesch}, P.~A., {Bouwens}, R.~J., {Carollo}, C.~M., {et~al.}
  2010{\natexlab{a}}, \apjl, 709, L21

\bibitem[{{Oesch} {et~al.}(2010{\natexlab{b}}){Oesch}, {Bouwens},
  {Illingworth}, {Carollo}, {Franx}, {Labb{\'e}}, {Magee}, {Stiavelli},
  {Trenti}, \& {van Dokkum}}]{oesch10a}
{Oesch}, P.~A., {Bouwens}, R.~J., {Illingworth}, G.~D., {et~al.}
  2010{\natexlab{b}}, \apjl, 709, L16

\bibitem[{{Oesch} {et~al.}(2013){Oesch}, {Bouwens}, {Illingworth}, {Labb{\'e}},
  {Franx}, {van Dokkum}, {Trenti}, {Stiavelli}, {Gonzalez}, \&
  {Magee}}]{oesch13}
---. 2013, \apj, 773, 75

\bibitem[{{Oguri}(2010)}]{oguri10}
{Oguri}, M. 2010, \pasj, 62, 1017

\bibitem[{{Okamura} {et~al.}(2018){Okamura}, {Shimasaku}, \&
  {Kawamata}}]{okamura17}
{Okamura}, T., {Shimasaku}, K., \& {Kawamata}, R. 2018, \apj, 854, 22

\bibitem[{{Oke} \& {Gunn}(1983)}]{okeg83}
{Oke}, J.~B., \& {Gunn}, J.~E. 1983, \apj, 266, 713

\bibitem[{{Ono} {et~al.}(2013){Ono}, {Ouchi}, {Curtis-Lake}, {Schenker},
  {Ellis}, {McLure}, {Dunlop}, {Robertson}, {Koekemoer}, {Bowler}, {Rogers},
  {Schneider}, {Charlot}, {Stark}, {Shimasaku}, {Furlanetto}, \&
  {Cirasuolo}}]{ono13}
{Ono}, Y., {Ouchi}, M., {Curtis-Lake}, E., {et~al.} 2013, \apj, 777, 155

\bibitem[{{Ono} {et~al.}(2017){Ono}, {Ouchi}, {Harikane}, {Toshikawa}, {Rauch},
  {Yuma}, {Sawicki}, {Shibuya}, {Shimasaku}, {Oguri}, {Willott}, {Akhlaghi},
  {Akiyama}, {Coupon}, {Kashikawa}, {Komiyama}, {Konno}, {Lin}, {Matsuoka},
  {Miyazaki}, {Nagao}, {Nakajima}, {Silverman}, {Tanaka}, \& {Wang}}]{ono17}
{Ono}, Y., {Ouchi}, M., {Harikane}, Y., {et~al.} 2017, ArXiv e-prints,
  arXiv:1704.06004

\bibitem[{{Peebles}(1969)}]{peebles69}
{Peebles}, P.~J.~E. 1969, \apj, 155, 393

\bibitem[{{Peng} {et~al.}(2002){Peng}, {Ho}, {Impey}, \& {Rix}}]{peng02}
{Peng}, C.~Y., {Ho}, L.~C., {Impey}, C.~D., \& {Rix}, H.-W. 2002, \aj, 124, 266

\bibitem[{{Peng} {et~al.}(2010){Peng}, {Ho}, {Impey}, \& {Rix}}]{peng10}
---. 2010, \aj, 139, 2097

\bibitem[{{Postman} {et~al.}(2012){Postman}, {Coe}, {Ben{\'{\i}}tez},
  {Bradley}, {Broadhurst}, {Donahue}, {Ford}, {Graur}, {Graves}, {Jouvel},
  {Koekemoer}, {Lemze}, {Medezinski}, {Molino}, {Moustakas}, {Ogaz}, {Riess},
  {Rodney}, {Rosati}, {Umetsu}, {Zheng}, {Zitrin}, {Bartelmann}, {Bouwens},
  {Czakon}, {Golwala}, {Host}, {Infante}, {Jha}, {Jimenez-Teja}, {Kelson},
  {Lahav}, {Lazkoz}, {Maoz}, {McCully}, {Melchior}, {Meneghetti}, {Merten},
  {Moustakas}, {Nonino}, {Patel}, {Reg{\"o}s}, {Sayers}, {Seitz}, \& {Van der
  Wel}}]{postman12}
{Postman}, M., {Coe}, D., {Ben{\'{\i}}tez}, N., {et~al.} 2012, \apjs, 199, 25

\bibitem[{{Prada} {et~al.}(2012){Prada}, {Klypin}, {Cuesta}, {Betancort-Rijo},
  \& {Primack}}]{prada12}
{Prada}, F., {Klypin}, A.~A., {Cuesta}, A.~J., {Betancort-Rijo}, J.~E., \&
  {Primack}, J. 2012, \mnras, 423, 3018

\bibitem[{{Priewe} {et~al.}(2017){Priewe}, {Williams}, {Liesenborgs}, {Coe}, \&
  {Rodney}}]{priewe17}
{Priewe}, J., {Williams}, L.~L.~R., {Liesenborgs}, J., {Coe}, D., \& {Rodney},
  S.~A. 2017, \mnras, 465, 1030

\bibitem[{{Reddy} \& {Steidel}(2009)}]{reddysteidel09}
{Reddy}, N.~A., \& {Steidel}, C.~C. 2009, \apj, 692, 778

\bibitem[{{Richard} {et~al.}(2010){Richard}, {Kneib}, {Limousin}, {Edge}, \&
  {Jullo}}]{richard10}
{Richard}, J., {Kneib}, J.-P., {Limousin}, M., {Edge}, A., \& {Jullo}, E. 2010,
  \mnras, 402, L44

\bibitem[{{Richard} {et~al.}(2014){Richard}, {Jauzac}, {Limousin}, {Jullo},
  {Cl{\'e}ment}, {Ebeling}, {Kneib}, {Atek}, {Natarajan}, {Egami}, {Livermore},
  \& {Bower}}]{richard14}
{Richard}, J., {Jauzac}, M., {Limousin}, M., {et~al.} 2014, \mnras, 444, 268

\bibitem[{{Roche} {et~al.}(1996){Roche}, {Ratnatunga}, {Griffiths}, {Im}, \&
  {Neuschaefer}}]{roche96}
{Roche}, N., {Ratnatunga}, K., {Griffiths}, R.~E., {Im}, M., \& {Neuschaefer},
  L. 1996, \mnras, 282, 1247

\bibitem[{{Rodney} {et~al.}(2017){Rodney}, {Balestra}, {Bradac}, {Brammer},
  {Broadhurst}, {Caminha}, {Chirivi}, {Diego}, {Filippenko}, {Foley}, {Graur},
  {Grillo}, {Hemmati}, {Hjorth}, {Hoag}, {Jauzac}, {Jha}, {Kawamata}, {Kelly},
  {McCully}, {Mobasher}, {Molino}, {Oguri}, {Richard}, {Riess}, {Rosati},
  {Schmidt}, {Selsing}, {Sharon}, {Strolger}, {Suyu}, {Treu}, {Weiner},
  {Williams}, \& {Zitrin}}]{rodney17}
{Rodney}, S.~A., {Balestra}, I., {Bradac}, M., {et~al.} 2017, ArXiv e-prints,
  arXiv:1707.02434

\bibitem[{{Romanowsky} \& {Fall}(2012)}]{romanowskyfall12}
{Romanowsky}, A.~J., \& {Fall}, S.~M. 2012, \apjs, 203, 17

\bibitem[{{Sales} {et~al.}(2010){Sales}, {Navarro}, {Schaye}, {Dalla Vecchia},
  {Springel}, \& {Booth}}]{sales10}
{Sales}, L.~V., {Navarro}, J.~F., {Schaye}, J., {et~al.} 2010, \mnras, 409,
  1541

\bibitem[{{Schmidt} {et~al.}(2014{\natexlab{a}}){Schmidt}, {Treu}, {Trenti},
  {Bradley}, {Kelly}, {Oesch}, {Holwerda}, {Shull}, \&
  {Stiavelli}}]{schmidt14lf}
{Schmidt}, K.~B., {Treu}, T., {Trenti}, M., {et~al.} 2014{\natexlab{a}}, \apj,
  786, 57

\bibitem[{{Schmidt} {et~al.}(2014{\natexlab{b}}){Schmidt}, {Treu}, {Brammer},
  {Brada{\v c}}, {Wang}, {Dijkstra}, {Dressler}, {Fontana}, {Gavazzi}, {Henry},
  {Hoag}, {Jones}, {Kelly}, {Malkan}, {Mason}, {Pentericci}, {Poggianti},
  {Stiavelli}, {Trenti}, {von der Linden}, \& {Vulcani}}]{schmidt14}
{Schmidt}, K.~B., {Treu}, T., {Brammer}, G.~B., {et~al.} 2014{\natexlab{b}},
  \apjl, 782, L36

\bibitem[{{Shen} {et~al.}(2003){Shen}, {Mo}, {White}, {Blanton}, {Kauffmann},
  {Voges}, {Brinkmann}, \& {Csabai}}]{shen03}
{Shen}, S., {Mo}, H.~J., {White}, S.~D.~M., {et~al.} 2003, \mnras, 343, 978

\bibitem[{{Shibuya} {et~al.}(2015){Shibuya}, {Ouchi}, \&
  {Harikane}}]{shibuya15}
{Shibuya}, T., {Ouchi}, M., \& {Harikane}, Y. 2015, \apjs, 219, 15

\bibitem[{{Somerville} {et~al.}(2017){Somerville}, {Behroozi}, {Pandya},
  {Dekel}, {Faber}, {Fontana}, {Koekemoer}, {Koo}, {P{\'e}rez-Gonz{\'a}lez},
  {Primack}, {Santini}, {Taylor}, \& {van der Wel}}]{somerville17}
{Somerville}, R.~S., {Behroozi}, P., {Pandya}, V., {et~al.} 2017, ArXiv
  e-prints, arXiv:1701.03526

\bibitem[{{Springel} {et~al.}(2005){Springel}, {White}, {Jenkins}, {Frenk},
  {Yoshida}, {Gao}, {Navarro}, {Thacker}, {Croton}, {Helly}, {Peacock}, {Cole},
  {Thomas}, {Couchman}, {Evrard}, {Colberg}, \& {Pearce}}]{springel05nat}
{Springel}, V., {White}, S.~D.~M., {Jenkins}, A., {et~al.} 2005, \nat, 435, 629

\bibitem[{{Steidel} {et~al.}(1999){Steidel}, {Adelberger}, {Giavalisco},
  {Dickinson}, \& {Pettini}}]{steidel99}
{Steidel}, C.~C., {Adelberger}, K.~L., {Giavalisco}, M., {Dickinson}, M., \&
  {Pettini}, M. 1999, \apj, 519, 1

\bibitem[{{Treu} {et~al.}(2015){Treu}, {Schmidt}, {Brammer}, {Vulcani}, {Wang},
  {Brada{\v c}}, {Dijkstra}, {Dressler}, {Fontana}, {Gavazzi}, {Henry}, {Hoag},
  {Huang}, {Jones}, {Kelly}, {Malkan}, {Mason}, {Pentericci}, {Poggianti},
  {Stiavelli}, {Trenti}, \& {von der Linden}}]{treu15a}
{Treu}, T., {Schmidt}, K.~B., {Brammer}, G.~B., {et~al.} 2015, \apj, 812, 114

\bibitem[{{Vitvitska} {et~al.}(2002){Vitvitska}, {Klypin}, {Kravtsov},
  {Wechsler}, {Primack}, \& {Bullock}}]{vitvitska02}
{Vitvitska}, M., {Klypin}, A.~A., {Kravtsov}, A.~V., {et~al.} 2002, \apj, 581,
  799

\bibitem[{{Wang} {et~al.}(2015){Wang}, {Hoag}, {Huang}, {Treu}, {Brada{\v c}},
  {Schmidt}, {Brammer}, {Vulcani}, {Jones}, {Ryan}, {Amor{\'{\i}}n},
  {Castellano}, {Fontana}, {Merlin}, \& {Trenti}}]{wang15}
{Wang}, X., {Hoag}, A., {Huang}, K.-H., {et~al.} 2015, \apj, 811, 29

\bibitem[{{Wyithe} \& {Loeb}(2011)}]{wyithe11}
{Wyithe}, J.~S.~B., \& {Loeb}, A. 2011, \mnras, 413, L38

\bibitem[{{Yue} {et~al.}(2017){Yue}, {Castellano}, {Ferrara}, {Fontana},
  {Merlin}, {Amor{\'{\i}}n}, {Grazian}, {M{\'a}rmol-Queralto},
  {Micha{\l}owski}, {Mortlock}, {Paris}, {Parsa}, {Pilo}, {Santini}, \& {Di
  Criscienzo}}]{yue17}
{Yue}, B., {Castellano}, M., {Ferrara}, A., {et~al.} 2017, ArXiv e-prints,
  arXiv:1711.05130

\bibitem[{{Zitrin} {et~al.}(2013){Zitrin}, {Meneghetti}, {Umetsu},
  {Broadhurst}, {Bartelmann}, {Bouwens}, {Bradley}, {Carrasco}, {Coe}, {Ford},
  {Kelson}, {Koekemoer}, {Medezinski}, {Moustakas}, {Moustakas}, {Nonino},
  {Postman}, {Rosati}, {Seidel}, {Seitz}, {Sendra}, {Shu}, {Vega}, \&
  {Zheng}}]{zitrin13}
{Zitrin}, A., {Meneghetti}, M., {Umetsu}, K., {et~al.} 2013, \apjl, 762, L30

\bibitem[{{Zitrin} {et~al.}(2014){Zitrin}, {Zheng}, {Broadhurst}, {Moustakas},
  {Lam}, {Shu}, {Huang}, {Diego}, {Ford}, {Lim}, {Bauer}, {Infante}, {Kelson},
  \& {Molino}}]{zitrin14}
{Zitrin}, A., {Zheng}, W., {Broadhurst}, T., {et~al.} 2014, \apjl, 793, L12

\bibitem[{{Zjupa} \& {Springel}(2017)}]{zjupaspringel17}
{Zjupa}, J., \& {Springel}, V. 2017, \mnras, 466, 1625

\end{thebibliography}

\clearpage

\begin{appendix}

\section{Mass Models}
\label{sec:appendix_models}

\begin{figure*}[h]
  \centering
      \includegraphics[width=0.4975\linewidth, trim=0 0 0 0, clip]{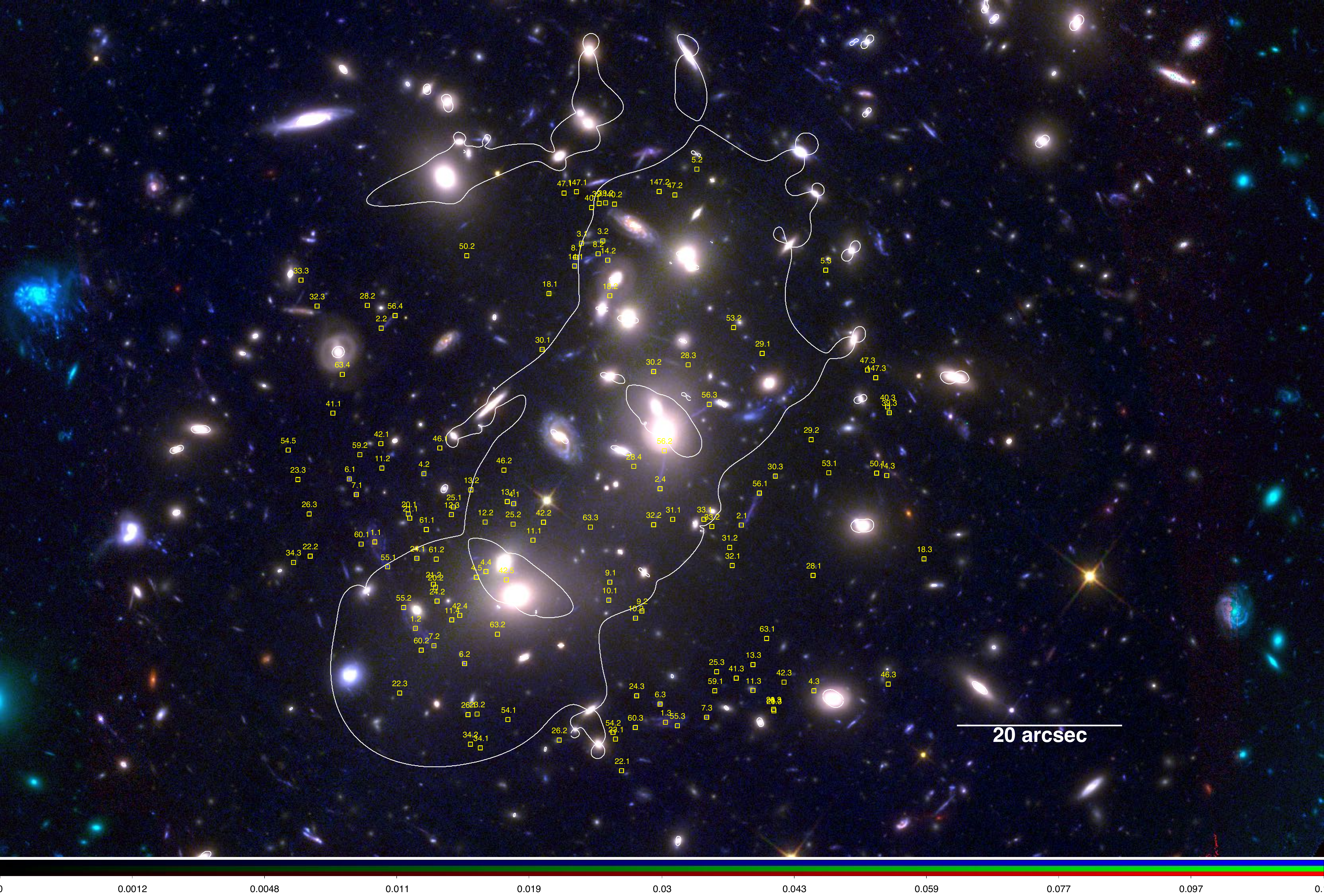}
      \includegraphics[width=0.4975\linewidth, trim=0 0 0 0, clip]{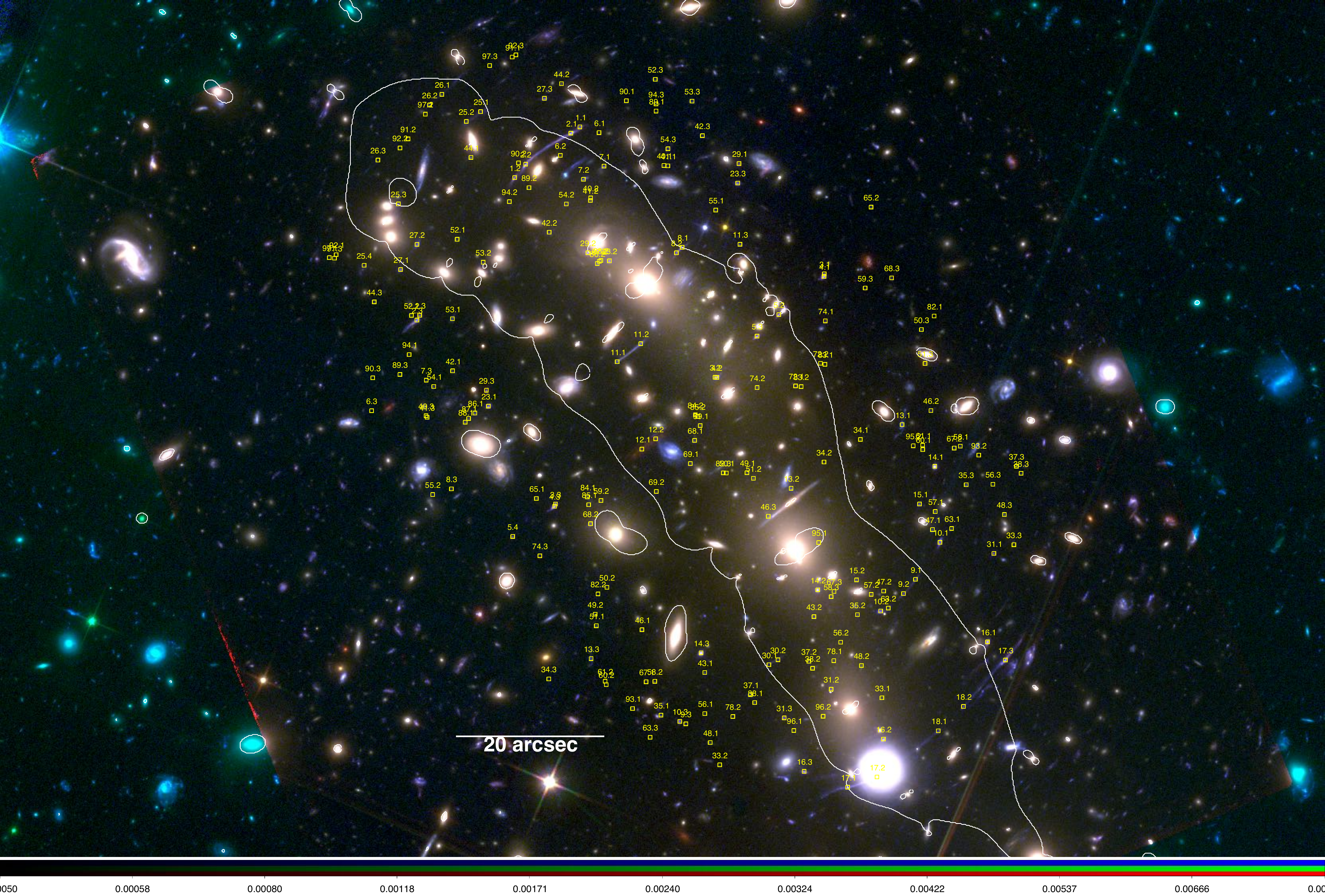}\\
      \vspace{2pt}
      \includegraphics[width=0.4975\linewidth, trim=0 0 0 0, clip]{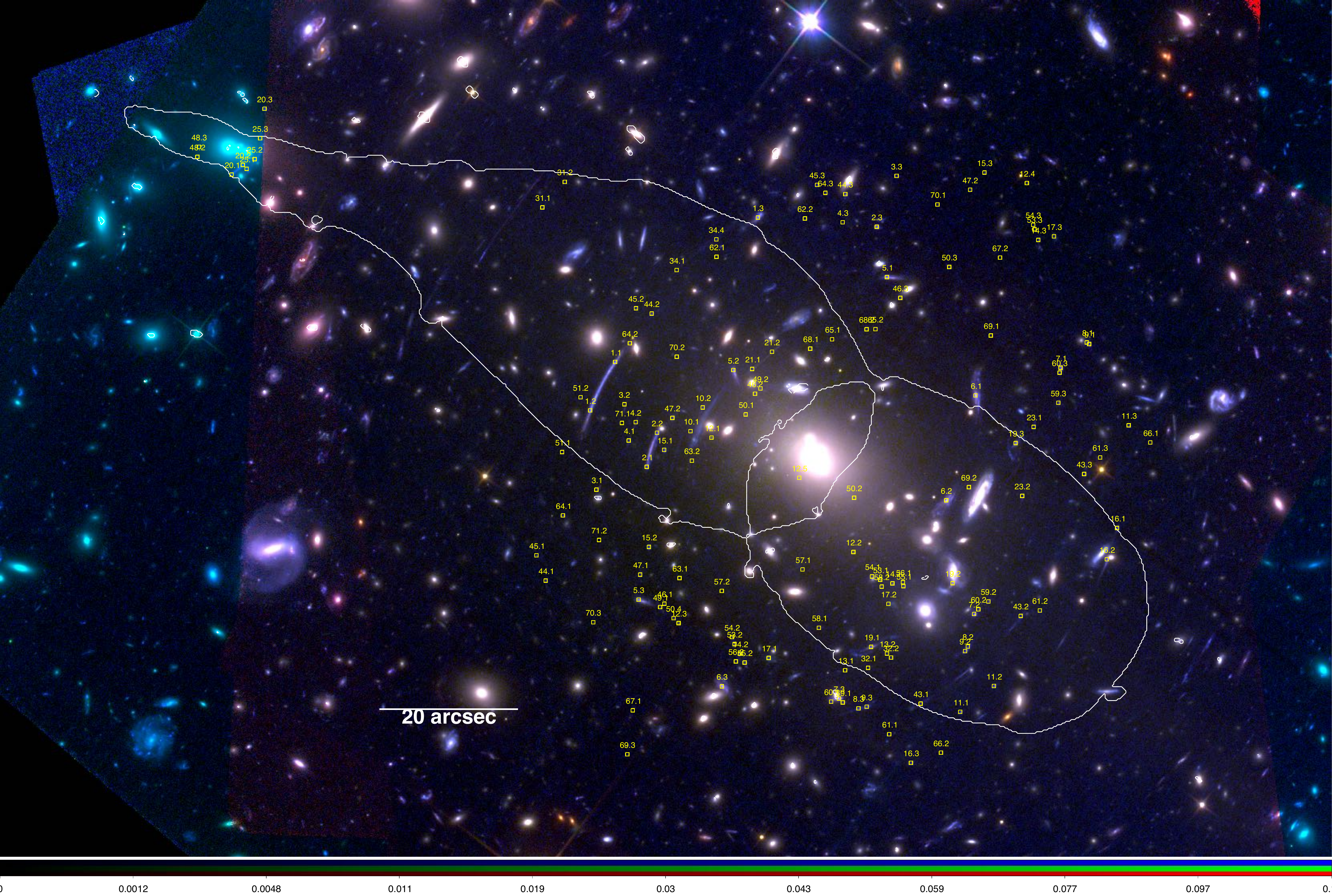}
      \includegraphics[width=0.4975\linewidth, trim=0 0 0 0, clip]{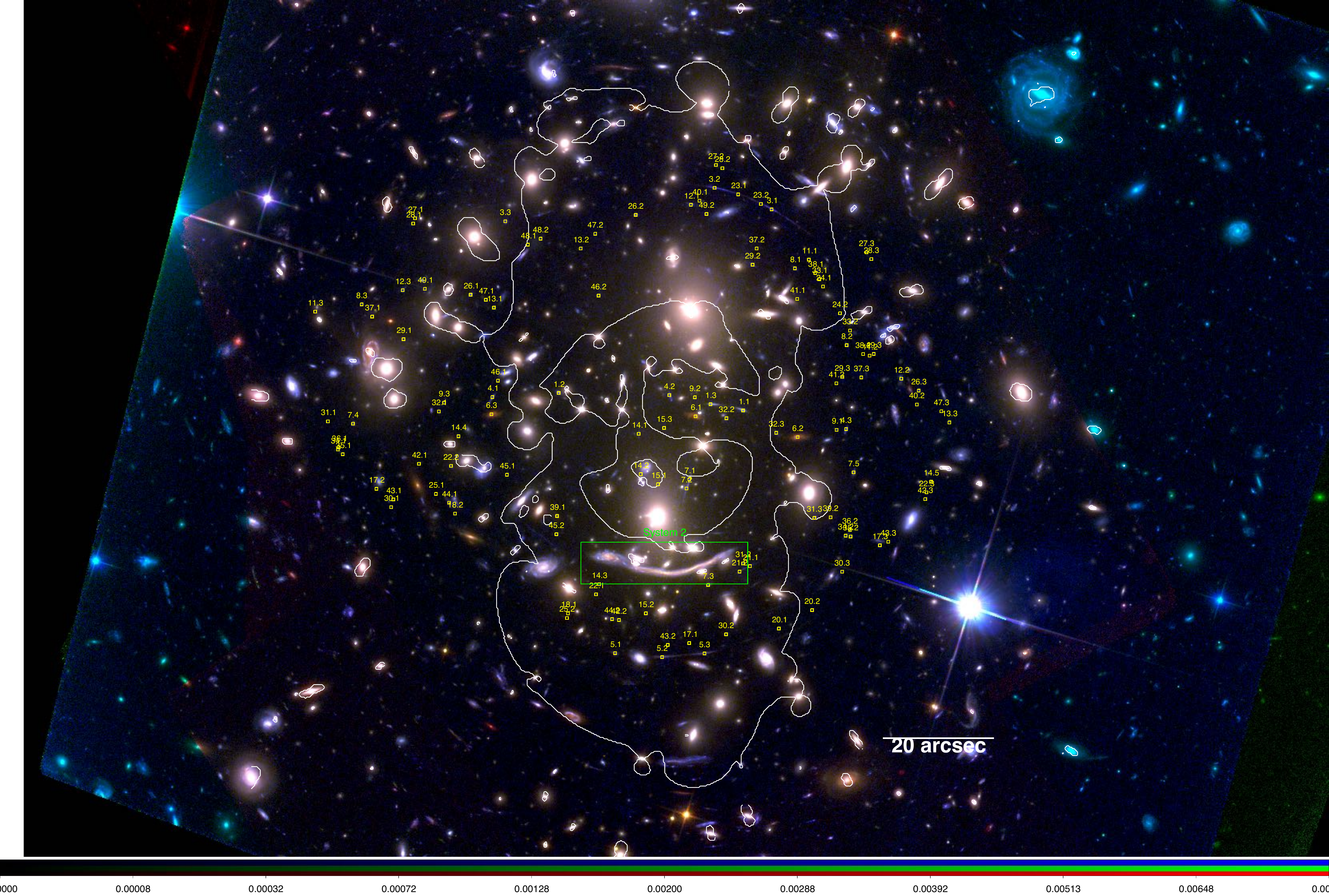}\\
      \raggedleft
      \vspace{2pt}
      \includegraphics[width=0.4975\linewidth, trim=0 0 0 0, clip]{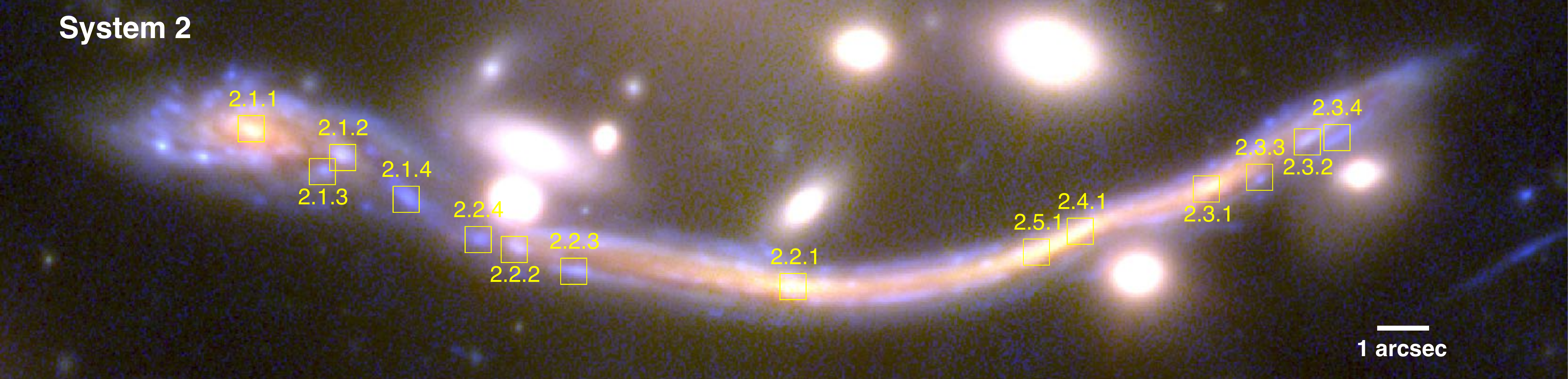}
  \caption{Multiple image systems used for mass modeling for \clone\ ({\it upper left}), 
  \cltwo\ ({\it upper right}), \clfive\ ({\it middle left}), and \clsix\ ({\it middle right}). 
  The bottom panel shows zoomed-in images of System 2 in the \clsix\ field. 
  Underlying color-composite images are created from the {\it HST}
  \bFilter+\vFilter-, \iFilter+\yFilter-, and \jFilter+\jhFilter+\hFilter-band
  images. Small yellow squares show the positions of multiple images (see
  Tables~\ref{tab:a2744multiple}--\ref{tab:a370multiple} for details). 
  Critical curves for a source redshift of $z=8$
   are shown with white solid lines. 
   }
  \label{fig:multipleimages}
\end{figure*}

Among the six HFF clusters, we use our version 3 mass models constructed 
in \citet{kawamata16} for \clthree\ and \clfour;
update the version 3 models for \clone\ and \cltwo,
which are now referred to as version 4 models;
and newly construct mass models for the last two clusters, 
\clfive\ and \clsix, also referred to as version 4.
In this section, we describe the two updated and the two newly 
constructed models, whose modeling strategy is the same
as that established in \citet{kawamata16}.
The lensing calculation is conducted using \texttt{glafic} \citep{oguri10},
which adopts a parametric modeling method.
A summary of the mass modeling is presented in 
Table~\ref{tab:modelsummaries}, including the numbers of multiple images
and the reduced $\chi^{2}$.
The best-fit parameters of the mass components are 
shown in Tables~\ref{tab:modelparams_a2744}--\ref{tab:modelparams_a370}.
Tables~\ref{tab:a2744multiple}--\ref{tab:a370multiple} show 
the position and redshift constraints of the multiple images
for each cluster,
which are utilized in the model constructions.
The positions of the multiple images and the critical curves 
of the best-fit models are shown in Figure~\ref{fig:multipleimages}.
The root mean squares of the distances between positions of observed 
and model-predicted multiple images on the image plane
are presented in Table~\ref{tab:modelsummaries}.
The relatively small root mean squares for our models
indicate that they well reproduce the observations despite 
strict constraints due to large numbers of multiple images and 
spectroscopic redshifts.

\subsection{Updated model for \clone}

After the publication of \citet{kawamata16}, a new 
mass model was published by \citet{mahler17}
that exploits new multiple images and spectroscopic 
redshifts of multiple images revealed by MUSE observation.
Considering the results of the 
MUSE observation, we update our model by incorporating and removing
25 and six positions of multiple images, respectively.
Specifically, we incorporate multiple image systems 
39, 40, 42, 47, 50, 61, 63, 
and 147 and remove IDs 2.3, 5.1, 5.4, and 8.3 and system 62.
The positions of IDs 33.3 and 34.3 are corrected.
In addition, we incorporate 20 MUSE spectroscopic redshifts 
that are considered to be reliable based on their qualities and 
consistency with our mass model.
The spectroscopic redshift of system 1 is updated, and 19 redshifts 
are newly determined for systems 2, 5, 8, 10,
22, 24, 26, 30, 31, 33, 34, 39, 40, 41, 42, 47, 61, 63, and 147.
We add a mass component representing a multipole perturbation
in order to better fit the observations.
While the image plane rms increases by $0\farcs05$, the reduced 
$\chi^{2}$ slightly decreases by 0.01 compared to our version 3 model.

\subsection{Updated model for \cltwo}
After the publication of \citet{kawamata16}, a new 
mass model was published by \citet{caminha16}
that exploits new multiple images and spectroscopic 
redshifts of multiple images revealed by the MUSE observation.
Considering the results of the 
MUSE observation, we update our model by incorporating
22 positions of multiple images.
Specifically, we incorporate systems 26, 58, 67, 92, 94, 95, 96, and 97, 
among which system 97 is found for the first time in this work. 
The position of ID 91.3 is corrected.
In addition, we incorporate 18 MUSE spectroscopic redshifts 
that are considered to be reliable based on their qualities and 
consistency with our mass model.
The spectroscopic redshifts incorporated are 
for systems 26, 33, 25, 38, 44, 47, 48, 49, 51, 55, 58, 67, 86, 
91, 92, 94, 95, and 96.
The image plane rms increases by $0\farcs06$, and the reduced 
$\chi^{2}$ increases by 0.30 compared to our version 3 model.

\subsection{Newly constructed model for \clfive}
In order to construct a mass model for \clfive, 
we use multiple images identified in
\citet{balestra13}, \citet{monna14}, \citet{richard14}, \citet{johnson14}, 
\citet{karman15}, \citet{caminha16a_as1063, caminha16b_as1063}, \citet{diego15c}, and \citet{karman16}.
Spectroscopic redshifts of multiple images were obtained 
in \citet{balestra13}, \citet{richard14}, \citet{johnson14}, 
\citet{karman15}, \citet{caminha16a_as1063}, \citet{karman16},
and the GLASS program \citep{schmidt14, treu15a}.
In addition, we find three new counterimages and 13 new systems,
which sums up to 35 new multiple images.
As a result, we use 40 systems from the literature and 
13 new systems; the total number of multiple images
is 141.
The positional uncertainty in the multiple images is 
assumed to be $\sigma_{x} = 0\farcs4$ for all of them.

\subsection{Newly constructed model for \clsix}
In order to construct a mass model for \clsix, 
we use multiple images identified in
\citet{richard10}, \citet{richard14}, \citet{johnson14}, \citet{diego16}, 
and \citet{lagattuta17}.
Spectroscopic redshifts of multiple images were obtained 
in \citet{richard10,richard14}, \citet{lagattuta17},
and the GLASS program \citep{schmidt14, treu15a}.
We correct the positions of four multiple images of IDs 3.3, 8.3, 13.3, and 26.3.
In addition, we find two new counterimages and 16 new systems,
which sums up to 40 new multiple images.
As a result, we use 33 systems from the literature and 
16 new systems; the total number of multiple images
is 135.
The positional uncertainty in the multiple images is 
assumed to be $\sigma_{x} = 0\farcs4$ for all of them.

\clearpage

\begin{deluxetable*}{lccrc}
\tabletypesize{\scriptsize}
\tablecaption{Summary of Version 4 Mass Modeling for Four Clusters\label{tab:modelsummaries}} 
\tablewidth{0pt}
\tablehead{
  \colhead{Cluster} & \colhead{No. of Multiple Image} & \colhead{No.
    of Multiple Images} & \colhead{$\chi^2$/dof} &  \colhead{Image
        Plane rms}\\
  \colhead{} &  \colhead{Systems (with spec-$z$)} & \colhead{}
  & \colhead{} & \colhead{(arcsec)} 
}
\startdata
\clone   & $ 45\ (24)$  & $132$ & 130.2/134  & 0.42 \\
\cltwo   & $ 75\ (34)$ & $202$ & 240.0/196 & 0.50 \\
\clfive & $ 53\ (19)$  & $141$ & 136.2/138 & 0.38 \\
\clsix  & $ 49\ (19)$ & $135$ & 99.5/140 & 0.50 
\enddata
\end{deluxetable*}

\begin{deluxetable*}{cccccccc} 
\tabletypesize{\scriptsize}
\tablecaption{Mass Model Parameters for \clone\label{tab:modelparams_a2744}}
\tablewidth{0pt}
\tablehead{
      \colhead{Component} & \colhead{Model} & \colhead{Mass} & \colhead{$e$} & \colhead{$\theta_{e}$} & \colhead{$c$} & \colhead{$\Delta x$\tablenotemark{a}} & \colhead{$\Delta y$\tablenotemark{a}}\\
      \colhead{} & \colhead{} & \colhead{$(10^{14}\ h^{-1}M_{\odot})$} & \colhead{} & \colhead{(deg)} & \colhead{} & \colhead{(arcsec)} & \colhead{(arcsec)}
}
\startdata
Cluster halo 1 & NFW & $4.53^{+0.88}_{-0.80}$ & $0.37^{+0.03}_{-0.03}$ & $165.05^{+4.04}_{-2.88}$ & $4.04^{+0.36}_{-0.33}$ & $-0.21^{+0.41}_{-0.37}$ & $3.30^{+0.53}_{-0.52}$ \\ 
Cluster halo 2 & NFW & $1.41^{+0.21}_{-0.22}$ & $0.38^{+0.02}_{-0.02}$ & $129.42^{+1.79}_{-2.02}$ & $9.49^{+0.88}_{-0.69}$ & $-18.69^{+0.18}_{-0.18}$ & $-17.91^{+0.16}_{-0.17}$ \\ 
Cluster halo 3 & NFW & $0.21^{+0.07}_{-0.08}$ & $0.76^{+0.03}_{-0.08}$ & $107.61^{+9.78}_{-3.94}$ & [$10.00$] & [$-26.97$] & [$30.91$] \\ 
\tableline
 &  & $\sigma_{*}\tablenotemark{b}$ & $r_{\rm trun, *}$ & $\eta$ & & \\
 &  & $({\rm km\ s^{-1}})$ & (arcsec) & & & \\ \tableline \\
Member galaxies  & GALS & $190.76^{+9.69}_{-9.94}$ & $71.98^{+31.40}_{-19.06}$ & $1.42^{+0.06}_{-0.11}$ & \\ 
\tableline 
 &  & $z_{s, {\rm fid}}$ & $\gamma \times 10^{2}$ & $\theta_{\gamma}$ & & \\
\enddata
\tablenotetext{a}{Coordinates are relative to the brightest cluster galaxy position in the \clone\ field (R.A. = 3.58611, decl. = $-30.40024$).}
\tablenotetext{b}{The normalization luminosity $L^*$ corresponds to $i_{814} = 18.33$.}
\tablecomments{Numbers in square brackets are fixed during the model optimization.}
\end{deluxetable*}

\begin{deluxetable*}{cccccccc} 
\tabletypesize{\scriptsize}
\tablecaption{Mass Model Parameters for \cltwo\label{tab:modelparams_m0416}}
\tablewidth{0pt}
\tablehead{
      \colhead{Component} & \colhead{Model} & \colhead{Mass} & \colhead{$e$} & \colhead{$\theta_{e}$} & \colhead{$c$} & \colhead{$\Delta x$\tablenotemark{a}} & \colhead{$\Delta y$\tablenotemark{a}}\\
      \colhead{} & \colhead{} & \colhead{$(10^{14}\ h^{-1}M_{\odot})$} & \colhead{} & \colhead{(deg)} & \colhead{} & \colhead{(arcsec)} & \colhead{(arcsec)}
}
\startdata
Cluster halo 1 & NFW & $2.78^{+0.37}_{-0.29}$ & $0.66^{+0.01}_{-0.01}$ & $60.58^{+0.65}_{-1.07}$ & $4.18^{+0.33}_{-0.26}$ & $-1.98^{+0.65}_{-0.49}$ & $1.69^{+0.29}_{-0.37}$ \\ 
Cluster halo 2 & NFW & $2.23^{+0.71}_{-0.58}$ & $0.69^{+0.02}_{-0.02}$ & $42.93^{+0.97}_{-1.10}$ & $4.45^{+0.77}_{-0.42}$ & $21.95^{+0.58}_{-0.48}$ & $-34.32^{+0.39}_{-0.41}$ \\ 
Cluster halo 3 & NFW & $1.18^{+0.52}_{-0.31}$ & $0.61^{+0.04}_{-0.04}$ & $37.66^{+2.24}_{-1.61}$ & $4.19^{+0.63}_{-0.55}$ & $24.40^{+0.57}_{-0.56}$ & $-53.79^{+0.85}_{-0.95}$ \\ 
\tableline
 &  & $\sigma_{*}\tablenotemark{b}$ & $r_{\rm trun, *}$ & $\eta$ & & \\
 &  & $({\rm km\ s^{-1}})$ & (arcsec) & & & \\ \tableline \\
Member galaxies  & GALS & $291.59^{+19.70}_{-16.41}$ & $16.39^{+4.89}_{-3.13}$ & $1.40^{+0.06}_{-0.10}$ & \\ 
\tableline 
 &  & $\sigma$ & $e$ & $\theta_{e}$ & $r_{\rm trun}$ & \colhead{$\Delta x$\tablenotemark{a}} & \colhead{$\Delta y$\tablenotemark{a}} \\
\enddata
\tablenotetext{a}{Coordinates are relative to the brightest cluster galaxy position in the \cltwo\ field (R.A. = 64.0380981, decl. = $-24.0674834$).}
\tablenotetext{b}{The normalization luminosity $L^*$ corresponds to $i_{814} = 18.73$.}
\tablecomments{Numbers in square brackets are fixed during the model optimization.}
\end{deluxetable*}

\begin{deluxetable*}{cccccccc} 
\tabletypesize{\scriptsize}
\tablecaption{Mass Model Parameters for \clfive\label{tab:modelparams_as1063}}
\tablewidth{0pt}
\tablehead{
      \colhead{Component} & \colhead{Model} & \colhead{Mass} & \colhead{$e$} & \colhead{$\theta_{e}$} & \colhead{$c$} & \colhead{$\Delta x$\tablenotemark{a}} & \colhead{$\Delta y$\tablenotemark{a}}\\
      \colhead{} & \colhead{} & \colhead{$(10^{14}\ h^{-1}M_{\odot})$} & \colhead{} & \colhead{(deg)} & \colhead{} & \colhead{(arcsec)} & \colhead{(arcsec)}
}
\startdata
Cluster halo 1 & NFW & $11.47^{+0.99}_{-1.05}$ & $0.46^{+0.01}_{-0.01}$ & $53.38^{+0.33}_{-0.35}$ & $5.15^{+0.25}_{-0.19}$ & $-0.53^{+0.16}_{-0.15}$ & $0.27^{+0.20}_{-0.17}$ \\ 
Cluster halo 2 & NFW & $0.47^{+0.14}_{-0.16}$ & $0.77^{+0.02}_{-0.06}$ & $70.84^{+2.30}_{-2.03}$ & $2.92^{+1.57}_{-0.61}$ & [$-84.67$] & [$45.07$] \\ 
Cluster halo 3 & NFW & $0.04^{+0.02}_{-0.01}$ & $0.62^{+0.06}_{-0.11}$ & $102.80^{+5.63}_{-5.38}$ & $7.94^{+2.48}_{-2.11}$ & $16.02^{+1.23}_{-0.89}$ & $-17.23^{+0.66}_{-0.54}$ \\ 
\tableline
 &  & $\sigma_{*}\tablenotemark{b}$ & $r_{\rm trun, *}$ & $\eta$ & & \\
 &  & $({\rm km\ s^{-1}})$ & (arcsec) & & & \\ \tableline \\
Member galaxies  & GALS & $130.39^{+8.45}_{-7.39}$  & $173.91^{+186.37}_{-72.16}$ & $0.57^{+0.51}_{-0.22}$ & \\ 
\tableline 
 &  & $z_{s, {\rm fid}}$ & $\gamma \times 10^{2}$ & $\theta_{\gamma}$ & & \\
 &  & & & (deg) & & & \\\tableline \\
External perturbation & PERT & [$2.00$] & $3.93^{+0.72}_{-0.69}$ & $37.58^{+3.16}_{-4.20}$ & 
\enddata
\tablenotetext{a}{Coordinates are relative to the brightest cluster galaxy position in the \clfive\ field (R.A. = 342.1832095, decl. = $-44.5308829$).}
\tablenotetext{b}{The normalization luminosity $L^*$ corresponds to $i_{814} = 18.19$.}
\tablecomments{Numbers in square brackets are fixed during the model optimization.}
\end{deluxetable*}

\begin{deluxetable*}{cccccccc} 
\tabletypesize{\scriptsize}
\tablecaption{Mass Model Parameters for \clsix\label{tab:modelparams_a370}}
\tablewidth{0pt}
\tablehead{
      \colhead{Component} & \colhead{Model} & \colhead{Mass} & \colhead{$e$} & \colhead{$\theta_{e}$} & \colhead{$c$} & \colhead{$\Delta x$\tablenotemark{a}} & \colhead{$\Delta y$\tablenotemark{a}}\\
      \colhead{} & \colhead{} & \colhead{$(10^{14}\ h^{-1}M_{\odot})$} & \colhead{} & \colhead{(deg)} & \colhead{} & \colhead{(arcsec)} & \colhead{(arcsec)}
}
\startdata
Cluster halo 1 & NFW & $0.70^{+0.22}_{-0.18}$ & $0.50^{+0.06}_{-0.06}$ & $5.80^{+4.00}_{-3.40}$ & $10.05^{+3.42}_{-2.00}$ & $3.69^{+0.69}_{-0.66}$ & $-5.24^{+1.42}_{-1.93}$ \\ 
Cluster halo 2 & NFW & $16.57^{+6.82}_{-6.27}$ & $0.28^{+0.06}_{-0.07}$ & $154.22^{+11.20}_{-21.52}$ & $1.62^{+0.31}_{-0.28}$ & $-17.10^{+2.97}_{-3.63}$ & $-4.55^{+2.37}_{-2.13}$ \\ 
Cluster halo 3 & NFW & $0.40^{+0.82}_{-0.27}$ & $0.43^{+0.19}_{-0.32}$ & $15.44^{+9.03}_{-6.18}$ & $6.85^{+3.50}_{-1.84}$ & $-6.55^{+0.82}_{-0.82}$ & $-24.14^{+1.14}_{-1.11}$ \\ 
Cluster halo 4 & NFW & $1.52^{+0.67}_{-0.43}$ & $0.15^{+0.05}_{-0.05}$ & $164.13^{+15.39}_{-38.20}$ & $8.80^{+2.11}_{-1.60}$ & [$-5.88$] & [$-37.21$] \\ 
\tableline
 &  & $\sigma_{*}\tablenotemark{b}$ & $r_{\rm trun, *}$ & $\eta$ & & \\
 &  & $({\rm km\ s^{-1}})$ & (arcsec) & & & \\ \tableline \\
Member galaxies  & GALS & $235.92^{+35.58}_{-26.20}$ & $23.60^{+17.36}_{-9.86}$ & $1.41^{+0.07}_{-0.15}$ & \\
\enddata
\tablenotetext{a}{Coordinates are relative to the brightest cluster galaxy position in the \clsix\ field (R.A. = 39.969704, decl. = $-1.571899$).}
\tablenotetext{b}{The normalization luminosity $L^*$ corresponds to $i_{814} = 18.42$.}
\tablecomments{Numbers in square brackets are fixed during the model optimization.}
\end{deluxetable*}

\clearpage
\LongTables
\begin{deluxetable*}{rcccccl}
\tabletypesize{\scriptsize}
\tablecaption{\clone\ Multiple Image Systems\label{tab:a2744multiple}}
\tablewidth{0pt}
\tablehead{
      \colhead{ID} & \colhead{R.A.} & \colhead{decl.} & \colhead{$z_{\rm spec}$} & \colhead{$z_{\rm model}$} & \colhead{Photo-$z$ Prior} & \colhead{References\tablenotemark{a}} 
}
\startdata
 1.1 & $3.597542$ & $-30.403917$ & 1.688 & \nodata & \nodata & JM, GM\\
 1.2 & $3.595958$ & $-30.406822$ &  & \nodata & \nodata & JM\\
 1.3 & $3.586208$ & $-30.409986$ &  & \nodata & \nodata & JM\\
 2.1 & $3.583250$ & $-30.403350$ & 1.888 & \nodata & \nodata & JM, GM\\
 2.2 & $3.597292$ & $-30.396725$ &  & \nodata & \nodata & JM\\
 2.4 & $3.586417$ & $-30.402128$ &  & \nodata & \nodata & JM\\
 3.1 & $3.589480$ & $-30.393867$ & 3.980 & \nodata & \nodata & JM, TJ\\
 3.2 & $3.588647$ & $-30.393788$ &  & \nodata & \nodata & JM\\
\enddata
\tablenotetext{a}{JM---\citet{merten11}, HA14---\citet{atek14}, JR---\citet{richard14}, AZ---\citet{zitrin14}, DL---\citet{lam14}, TJ---\citet{johnson14}, MI---\citet{ishigaki15}, MJ---\citet{jauzac15}, XW---\citet{wang15}, GM---\citet{mahler17}.}
\end{deluxetable*}

\LongTables
\begin{deluxetable*}{rcccccl}
\tabletypesize{\scriptsize}
\tablecaption{\cltwo\ Multiple Image Systems\label{tab:m0416multiple}}
\tablewidth{0pt}
\tablehead{
      \colhead{ID} & \colhead{R.A.} & \colhead{decl.} & \colhead{$z_{\rm spec}$} & \colhead{$z_{\rm model}$} & \colhead{Photo-$z$ Prior} & \colhead{References\tablenotemark{a}} 
}
\startdata
 1.1 & $64.040797$ & $-24.061616$ & 1.896 & \nodata & \nodata & AZ, LC\\
 1.2 & $64.043472$ & $-24.063514$ &  & \nodata & \nodata & AZ\\
 1.3 & $64.047367$ & $-24.068673$ &  & \nodata & \nodata & AZ\\
 2.1 & $64.041168$ & $-24.061851$ & 1.893 & \nodata & \nodata & AZ, CG\\
 2.2 & $64.043021$ & $-24.063010$ &  & \nodata & \nodata & AZ\\
 2.3 & $64.047484$ & $-24.068851$ &  & \nodata & \nodata & AZ\\
 3.1 & $64.030772$ & $-24.067118$ & 1.989 & \nodata & \nodata & AZ, CG\\
 3.2 & $64.035254$ & $-24.070993$ &  & \nodata & \nodata & AZ\\
\enddata
\tablenotetext{a}{LC---\citet{christensen12}, AZ---\citet{zitrin13}, MJ---\citet{jauzac14}, CG---\citet{grillo15}, JD---\citet{diego15a}, RK16---\citet{kawamata16}, AH---\citet{hoag16}, GC---\citet{caminha16}, SR---\citet{rodney17}.}
\end{deluxetable*}

\LongTables
\begin{deluxetable*}{rcccccl}
\tabletypesize{\scriptsize}
\tablecaption{\clfive\ Multiple Image Systems\label{tab:as1063multiple}}
\tablewidth{0pt}
\tablehead{
      \colhead{ID} & \colhead{R.A.} & \colhead{decl.} & \colhead{$z_{\rm spec}$} & \colhead{$z_{\rm model}$} & \colhead{Photo-$z$ Prior} & \colhead{References\tablenotemark{a}} 
}
\startdata
 1.1 & $342.194460$ & $-44.526987$ & 1.228 & \nodata & \nodata & JR, WK16\\
 1.2 & $342.195863$ & $-44.528929$ &  & \nodata & \nodata & JR\\
 1.3 & $342.186429$ & $-44.521196$ &  & \nodata & \nodata & JR\\
 2.1 & $342.192683$ & $-44.531195$ & 1.259 & \nodata & \nodata & JR, WK16\\
 2.2 & $342.192103$ & $-44.529830$ &  & \nodata & \nodata & JR\\
 2.3 & $342.179742$ & $-44.521566$ &  & \nodata & \nodata & JR\\
 3.1 & $342.195513$ & $-44.532112$ & \nodata & $2.15^{+0.05}_{-0.05}$ & \nodata & TJ\\
 3.2 & $342.193934$ & $-44.528687$ & \nodata &  & \nodata & TJ\\
\enddata
\tablenotetext{a}{IB---\citet{balestra13}, JR---\citet{richard14}, WK15---\citet{karman15}, TJ---\citet{johnson14}, GC---\citet{caminha16a_as1063}, JD---\citet{diego15c}, WK16---\citet{karman16}, GLASS---\citet{schmidt14}, \citet{treu15a}}
\end{deluxetable*}

\LongTables
\begin{deluxetable*}{rcccccl}
\tabletypesize{\scriptsize}
\tablecaption{\clsix\ Multiple Image Systems\label{tab:a370multiple}}
\tablewidth{0pt}
\tablehead{
      \colhead{ID} & \colhead{R.A.} & \colhead{decl.} & \colhead{$z_{\rm spec}$} & \colhead{$z_{\rm model}$} & \colhead{Photo-$z$ Prior} & \colhead{References\tablenotemark{a}} 
}
\startdata
 1.1 & $39.967071$ & $-1.576914$ & 0.806 & \nodata & \nodata & JR10, JR10\\
 1.2 & $39.976292$ & $-1.576042$ &  & \nodata & \nodata & JR10\\
 1.3 & $39.968704$ & $-1.576606$ &  & \nodata & \nodata & JR10\\
 3.1 & $39.965647$ & $-1.566871$ & 1.950 & \nodata & \nodata & JR10, DL\\
 3.2 & $39.968504$ & $-1.565792$ &  & \nodata & \nodata & JR10\\
 3.3 & $39.978958$ & $-1.567461$ &  & \nodata & \nodata & \nodata\\
 4.1 & $39.979604$ & $-1.576245$ & 1.273 & \nodata & \nodata & JR10, DL\\
 4.2 & $39.970758$ & $-1.576147$ &  & \nodata & \nodata & JR10\\
\enddata
\tablenotetext{a}{JR10---\citet{richard10}, JR14---\citet{richard14}, JD---\citet{diego16}, DL---\citet{lagattuta17}.}
\end{deluxetable*}


\section{Fitting Results}
\label{sec:fitting_results}

The obtained morphological properties and magnitudes are presented in 
Tables~\ref{tab:i-results}--\ref{tab:yj-results}. The fitting results for galaxies fainter
 than $-18$ mag are graphically shown in Figures~\ref{fig:faceimages_z7_1} and 
 \ref{fig:faceimages_z89}.

\LongTables
\begin{deluxetable*}{lcccccccrc}
\tablecolumns{10}
\tabletypesize{\scriptsize}
\tablecaption{Fitting results for dropouts at $z\sim 6-7$
\label{tab:i-results}}
\tablewidth{0pt}
\tablehead{
\colhead{ID\tablenotemark{a}} &
\colhead{R.A.} & 
\colhead{decl.} &
\colhead{$m_\mathrm{UV}$\tablenotemark{b,c}} &
\colhead{$M_\mathrm{UV}$\tablenotemark{c}} &
\colhead{$r_\mathrm{e}$ / kpc\tablenotemark{c,d}} &
\colhead{$e$} &
\colhead{$\mu_{\rm best}$ \tablenotemark{e}}& 
\colhead{$\mu$\tablenotemark{f}} & 
\colhead{$z_{\rm photo}$}
}
\startdata
\sidehead{HFF1C}
2251-4556 &  $ 3.593804 $ & $ -30.415447 $ & $27.68^{+ 0.10}_{-0.14}$ & $-19.01^{+ 0.10}_{-0.14}$ & $ 0.09^{+ 0.03}_{-0.04}$ & $ 0.38$ & $ 3.44$ & $ 3.85^{+ 0.38}_{- 0.32}$ & $ 6.6^{+ 0.8}_{- 0.7}$ \\  
1695-4527 &  $ 3.570654 $ & $ -30.414659 $ & $26.72^{+ 0.11}_{-0.14}$ & $-19.98^{+ 0.11}_{-0.14}$ & $ 0.18^{+ 0.06}_{-0.08}$ & $ 0.81$ & $ 1.41$ & $ 1.53^{+ 0.09}_{- 0.06}$ & $ 6.0^{+ 0.7}_{- 0.7}$ \\  
2549-3119 &  $ 3.606222 $ & $ -30.386644 $ & $26.71^{+ 0.18}_{-0.23}$ & $-19.99^{+ 0.18}_{-0.23}$ & $ 0.34^{+ 0.11}_{-0.12}$ & $ 0.24$ & $ 1.48$ & $ 1.64^{+ 0.08}_{- 0.06}$ & $ 5.8^{+ 0.7}_{- 0.7}$ \\  
1930-4181\tablenotemark{s1} &  $ 3.580453 $ & $ -30.405043 $ & $28.52^{+ 0.10}_{-0.13}$ & $-18.17^{+ 0.10}_{-0.13}$ & $ 0.09^{+ 0.03}_{-0.04}$ & $ 0.45$ & $ 4.21$ & $ 4.81^{+ 0.41}_{- 0.35}$ & $ 6.8^{+ 0.8}_{- 0.8}$ \\  
2516-4570 &  $ 3.604865 $ & $ -30.415839 $ & $27.15^{+ 0.36}_{-0.41}$ & $-19.54^{+ 0.36}_{-0.41}$ & $ 0.55^{+ 0.19}_{-0.18}$ & $ 0.65$ & $ 2.31$ & $ 2.97^{+ 0.24}_{- 0.22}$ & $ 5.6^{+ 0.7}_{- 0.6}$ \\  
2348-3454\tablenotemark{s1} &  $ 3.597834 $ & $ -30.395961 $ & $28.54^{+ 0.11}_{-0.13}$ & $-18.15^{+ 0.11}_{-0.13}$ & $ 0.03^{+ 0.02}_{-0.01}$ & $ 0.90$ & $ 3.08$ & $ 3.42^{+ 0.26}_{- 0.20}$ & $ 7.0^{+ 0.8}_{- 0.8}$ \\  
2178-2458 &  $ 3.590761 $ & $ -30.379409 $ & $27.61^{+ 0.17}_{-0.22}$ & $-19.08^{+ 0.17}_{-0.22}$ & $ 0.21^{+ 0.07}_{-0.07}$ & $ 0.90$ & $ 1.94$ & $ 1.77^{+ 0.12}_{- 0.09}$ & $ 5.8^{+ 0.7}_{- 0.6}$ \\  
2047-3526\tablenotemark{s1} &  $ 3.585322 $ & $ -30.397958 $ & $28.49^{+ 0.30}_{-0.15}$ & $-18.20^{+ 0.30}_{-0.15}$ & $ 0.18^{+ 0.08}_{-0.05}$ & $ 0.41$ & $ 3.63$ & $ 3.79^{+ 0.36}_{- 0.34}$ & $ 6.8^{+ 0.7}_{- 0.8}$ \\
\enddata
\tablenotetext{a}{Asterisks indicate galaxies with multiple cores.}
\tablenotetext{b}{Total apparent magnitude from light profile fitting with \texttt{glafic}.}
\tablenotetext{c}{Errors are random errors in the fitting procedure.}
\tablenotetext{d}{Circularized effective radius, 
              $r_\mathrm{e}^\mathrm{maj} \sqrt{1-e}$, where 
              $r_\mathrm{e}^\mathrm{maj}$ 
              is the radius along the major axis and 
              $e$ the ellipticity.}
\tablenotetext{e}{Best-fit value of magnification.}
\tablenotetext{f}{Median value and $1\sigma$ error of the magnification factor from the MCMC posterior distribution.}
\tablenotetext{s1--3}{Dropout galaxy that composes a multiple image system.}
\end{deluxetable*}

\LongTables
\begin{deluxetable*}{lcccccccrc}
\tablecolumns{10}
\tabletypesize{\scriptsize}
\tablecaption{Fitting results for dropouts at $z\sim 8$
\label{tab:y-results}}
\tablewidth{0pt}
\tablehead{
\colhead{ID\tablenotemark{a}} &
\colhead{R.A.} & 
\colhead{decl.} &
\colhead{$m_\mathrm{UV}$\tablenotemark{b,c}} &
\colhead{$M_\mathrm{UV}$\tablenotemark{c}} &
\colhead{$r_\mathrm{e}$ / kpc\tablenotemark{c,d}} &
\colhead{$e$} &
\colhead{$\mu_{\rm best}$ \tablenotemark{e}}& 
\colhead{$\mu$\tablenotemark{f}} & 
\colhead{$z_{\rm photo}$}
}
\startdata
\sidehead{HFF1C}
2508-2496 &  $ 3.604518 $ & $ -30.380467 $ & $26.40^{+ 0.07}_{-0.08}$ & $-20.74^{+ 0.07}_{-0.08}$ & $ 0.22^{+ 0.04}_{-0.07}$ & $ 0.74$ & $ 1.34$ & $ 1.38^{+ 0.07}_{- 0.05}$ & $ 8.0^{+ 0.9}_{- 0.9}$ \\  
2481-2561 &  $ 3.603378 $ & $ -30.382255 $ & $26.66^{+ 0.16}_{-0.16}$ & $-20.48^{+ 0.16}_{-0.16}$ & $ 0.70^{+ 0.11}_{-0.10}$ & $ 0.60$ & $ 1.42$ & $ 1.48^{+ 0.08}_{- 0.06}$ & $ 8.2^{+ 0.9}_{- 0.9}$ \\  
2306-3089 &  $ 3.596091 $ & $ -30.385833 $ & $27.40^{+ 0.21}_{-0.45}$ & $-19.74^{+ 0.21}_{-0.45}$ & $ 0.31^{+ 0.09}_{-0.15}$ & $ 0.43$ & $ 2.13$ & $ 2.20^{+ 0.15}_{- 0.12}$ & $ 8.2^{+ 0.9}_{- 0.9}$ \\  
2555-2515 &  $ 3.606461 $ & $ -30.380996 $ & $27.41^{+ 0.18}_{-0.22}$ & $-19.73^{+ 0.18}_{-0.22}$ & $ 0.16^{+ 0.05}_{-0.07}$ & $ 0.89$ & $ 1.33$ & $ 1.40^{+ 0.07}_{- 0.05}$ & $ 8.0^{+ 0.8}_{- 0.9}$ \\  
2492-2561 &  $ 3.603859 $ & $ -30.382264 $ & $27.23^{+ 0.36}_{-0.45}$ & $-19.91^{+ 0.36}_{-0.45}$ & $ 0.49^{+ 0.19}_{-0.19}$ & $ 0.49$ & $ 1.41$ & $ 1.48^{+ 0.08}_{- 0.06}$ & $ 8.4^{+ 0.9}_{- 0.9}$ \\  
2557-2513 &  $ 3.606576 $ & $ -30.380924 $ & $26.79^{+ 0.30}_{-0.77}$ & $-20.35^{+ 0.30}_{-0.77}$ & $ 0.77^{+ 0.13}_{-0.43}$ & $ 0.32$ & $ 1.33$ & $ 1.39^{+ 0.07}_{- 0.05}$ & $ 7.9^{+ 0.9}_{- 6.1}$ \\  
2135-2432 &  $ 3.588980 $ & $ -30.378668 $ & $29.21^{+ 0.20}_{-0.16}$ & $-17.93^{+ 0.20}_{-0.16}$ & $ 0.06^{+ 0.05}_{-0.03}$ & $ 0.76$ & $ 1.98$ & $ 1.79^{+ 0.12}_{- 0.09}$ & $ 8.0^{+ 0.8}_{- 0.9}$ \\  
2495-2562 &  $ 3.603997 $ & $ -30.382304 $ & $27.20^{+ 0.32}_{-0.39}$ & $-19.94^{+ 0.32}_{-0.39}$ & $ 1.04^{+ 0.22}_{-0.33}$ & $ 0.57$ & $ 1.41$ & $ 1.48^{+ 0.08}_{- 0.06}$ & $ 7.9^{+ 0.9}_{- 6.1}$ \\
\enddata
\tablenotetext{a}{Asterisks indicate galaxies with multiple cores.}
\tablenotetext{b}{Total apparent magnitude from light profile fitting with \texttt{glafic}.}
\tablenotetext{c}{Errors are random errors in the fitting procedure.}
\tablenotetext{d}{Circularized effective radius, 
              $r_\mathrm{e}^\mathrm{maj} \sqrt{1-e}$, where 
              $r_\mathrm{e}^\mathrm{maj}$ 
              is the radius along the major axis and 
              $e$ the ellipticity.}
\tablenotetext{e}{Best-fit value of magnification.}
\tablenotetext{f}{Median value and $1\sigma$ error of the magnification factor from the MCMC posterior distribution.}
\tablenotetext{s1--2}{Dropout galaxy that composes a multiple image system.}
\end{deluxetable*}

\LongTables
\begin{deluxetable*}{lcccccccrc}
\tablecolumns{10}
\tabletypesize{\scriptsize}
\tablecaption{Fitting results for dropouts at $z\sim 9$
\label{tab:yj-results}}
\tablewidth{0pt}
\tablehead{
\colhead{ID\tablenotemark{a}} &
\colhead{R.A.} & 
\colhead{decl.} &
\colhead{$m_\mathrm{UV}$\tablenotemark{b,c}} &
\colhead{$M_\mathrm{UV}$\tablenotemark{c}} &
\colhead{$r_\mathrm{e}$ / kpc\tablenotemark{c,d}} &
\colhead{$e$} &
\colhead{$\mu_{\rm best}$ \tablenotemark{e}}& 
\colhead{$\mu$\tablenotemark{f}} & 
\colhead{$z_{\rm photo}$}
}
\startdata
\sidehead{HFF1C}
2481-2561 &  $ 3.603379 $ & $ -30.382255 $ & $26.59^{+ 0.17}_{-0.34}$ & $-20.73^{+ 0.17}_{-0.34}$ & $ 0.67^{+ 0.12}_{-0.16}$ & $ 0.56$ & $ 1.42$ & $ 1.49^{+ 0.08}_{- 0.06}$ & $ 8.2^{+ 0.9}_{- 0.9}$ \\  
2492-2561 &  $ 3.603859 $ & $ -30.382262 $ & $27.17^{+ 0.21}_{-0.21}$ & $-20.15^{+ 0.21}_{-0.21}$ & $ 0.50^{+ 0.12}_{-0.12}$ & $ 0.51$ & $ 1.41$ & $ 1.48^{+ 0.08}_{- 0.06}$ & $ 8.4^{+ 0.9}_{- 0.9}$ \\  
2220-4053 &  $ 3.592512 $ & $ -30.401486 $ & $28.70^{+ 0.42}_{-0.31}$ & $-18.62^{+ 0.42}_{-0.31}$ & $ 0.21^{+ 0.09}_{-0.06}$ & $ 0.50$ & $12.00$ & $14.37^{+ 1.22}_{- 0.90}$ & $ 9.5^{+ 1.1}_{- 7.3}$ \\  

\tableline

\sidehead{HFF1P}
5363-4016 &  $ 3.473469 $ & $ -30.400459 $ & $28.67^{+ 0.18}_{-0.17}$ & $-18.65^{+ 0.18}_{-0.17}$ & $ 0.10^{+ 0.07}_{-0.05}$ & $ 0.75$ & $ 1.08$ & $ 1.02^{+ 0.01}_{- 0.01}$ & $ 8.3^{+ 0.9}_{- 6.6}$ \\  
5364-3024 &  $ 3.473522 $ & $ -30.384024 $ & $28.46^{+ 0.32}_{-0.21}$ & $-18.86^{+ 0.32}_{-0.21}$ & $ 0.19^{+ 0.09}_{-0.07}$ & $ 0.71$ & $ 1.09$ & $ 1.03^{+ 0.01}_{- 0.01}$ & $ 8.7^{+ 1.0}_{- 6.3}$ \\  
5733-3462 &  $ 3.488893 $ & $ -30.396182 $ & $27.82^{+ 0.24}_{-0.23}$ & $-19.50^{+ 0.24}_{-0.23}$ & $ 0.36^{+ 0.16}_{-0.12}$ & $ 0.60$ & $ 1.10$ & $ 1.03^{+ 0.02}_{- 0.01}$ & $ 8.7^{+ 1.0}_{- 0.9}$ \\  
5386-2074 &  $ 3.474446 $ & $ -30.368728 $ & $28.17^{+ 0.32}_{-0.27}$ & $-19.15^{+ 0.32}_{-0.27}$ & $ 0.31^{+ 0.17}_{-0.15}$ & $ 0.90$ & $ 1.08$ & $ 1.05^{+ 0.01}_{- 0.01}$ & $ 8.9^{+ 1.0}_{- 6.9}$ \\  

\tableline

\sidehead{HFF2C}
1151-4540 &  $ 64.047984 $ & $ -24.081671 $ & $27.12^{+ 0.12}_{-0.17}$ & $-20.20^{+ 0.12}_{-0.17}$ & $ 0.25^{+ 0.05}_{-0.06}$ & $ 0.57$ & $ 1.44$ & $ 1.50^{+ 0.04}_{- 0.03}$ & $ 8.4^{+ 0.9}_{- 0.9}$ \\
\enddata
\tablenotetext{a}{Asterisks indicate galaxies with multiple cores.}
\tablenotetext{b}{Total apparent magnitude from light profile fitting with \texttt{glafic}.}
\tablenotetext{c}{Errors are random errors in the fitting procedure.}
\tablenotetext{d}{Circularized effective radius, 
              $r_\mathrm{e}^\mathrm{maj} \sqrt{1-e}$, where 
              $r_\mathrm{e}^\mathrm{maj}$ 
              is the radius along the major axis and 
              $e$ the ellipticity.}
\tablenotetext{e}{Best-fit value of magnification.}
\tablenotetext{f}{Median value and $1\sigma$ error of the magnification factor from the MCMC posterior distribution.}
\tablenotetext{s1}{Dropout galaxy that composes a multiple image system.}
\end{deluxetable*}

\clearpage

\begin{figure}[h]
  \centering
      \includegraphics[width=0.45\linewidth, trim=0 0 0 0, clip]{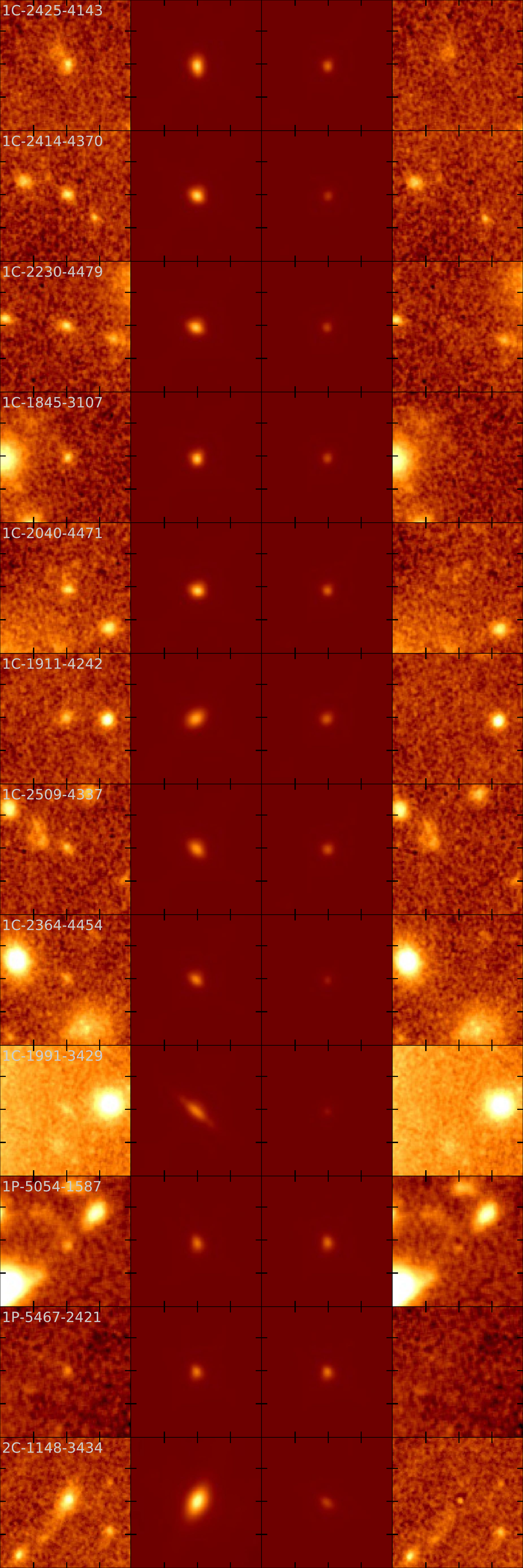}
      \includegraphics[width=0.45\linewidth, trim=0 0 0 0, clip]{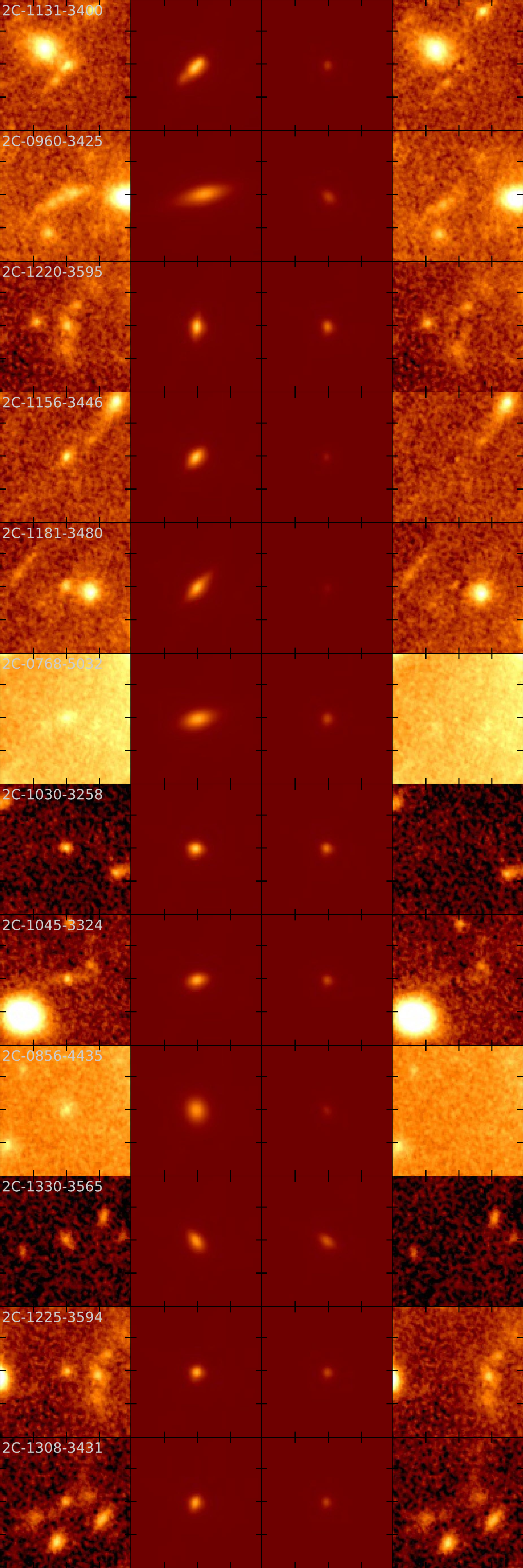}
  \caption{Images for $z\sim6-7$ faint galaxies at $M_{\mathrm{UV}} \gtrsim -18$.
  From left to right, $3'' \times 3''$ cutout images, best-fit S\'ersic profiles on the 
  image plane, best-fit S\'ersic profiles on the source plane, and residual images 
  on the image plane.
   }
  \label{fig:faceimages_z7_1}
\end{figure}

\addtocounter{figure}{-1}
\begin{figure}[h]
  \centering
      \includegraphics[width=0.45\linewidth, trim=0 0 0 0, clip]{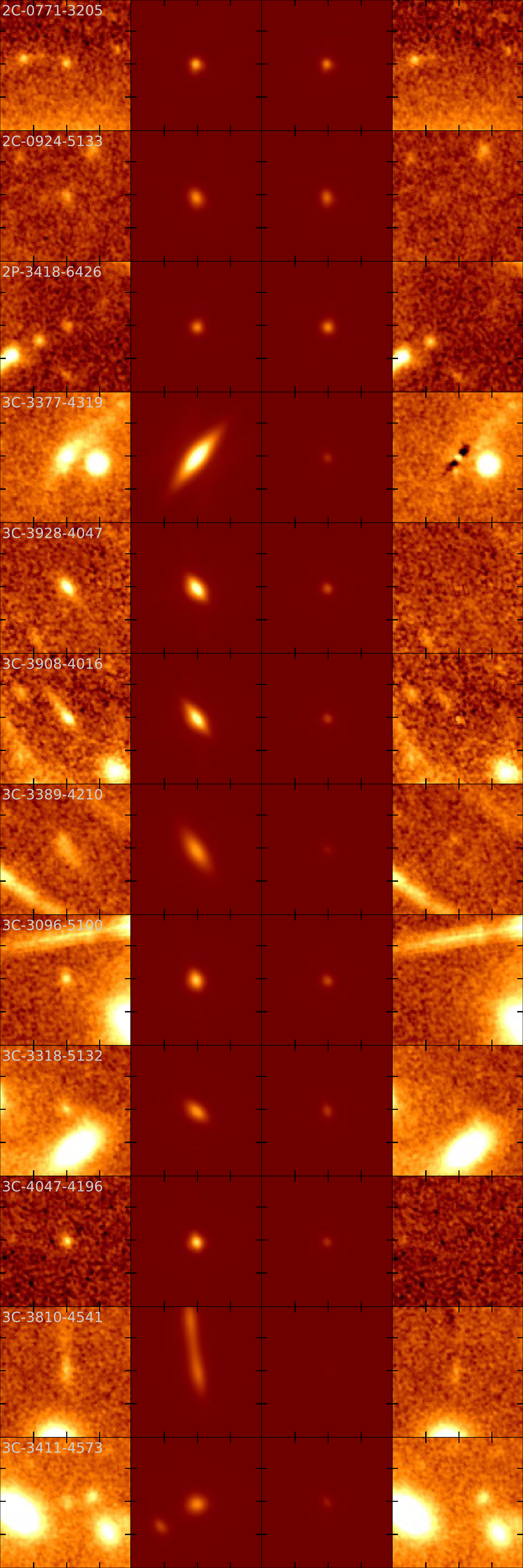}
      \includegraphics[width=0.45\linewidth, trim=0 0 0 0, clip]{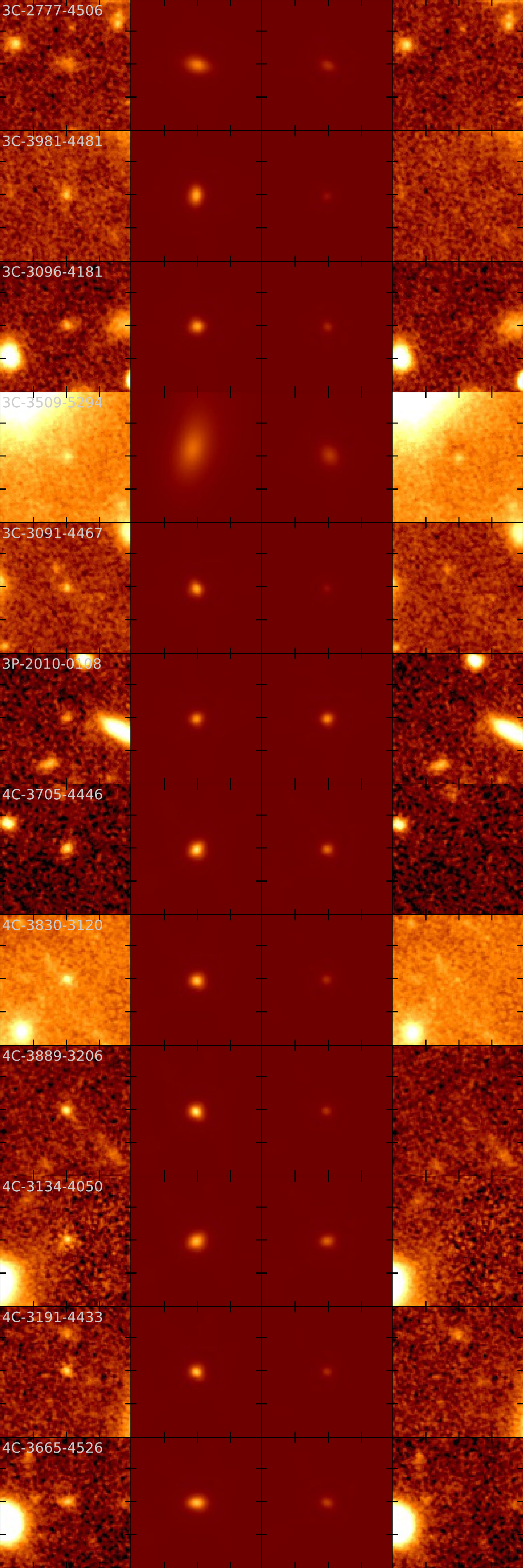}
  \caption{Continued.
   }
  \label{fig:faceimages_z7_2}
\end{figure}

\addtocounter{figure}{-1}
\begin{figure}[h]
  \centering
      \includegraphics[width=0.45\linewidth, trim=0 0 0 0, clip]{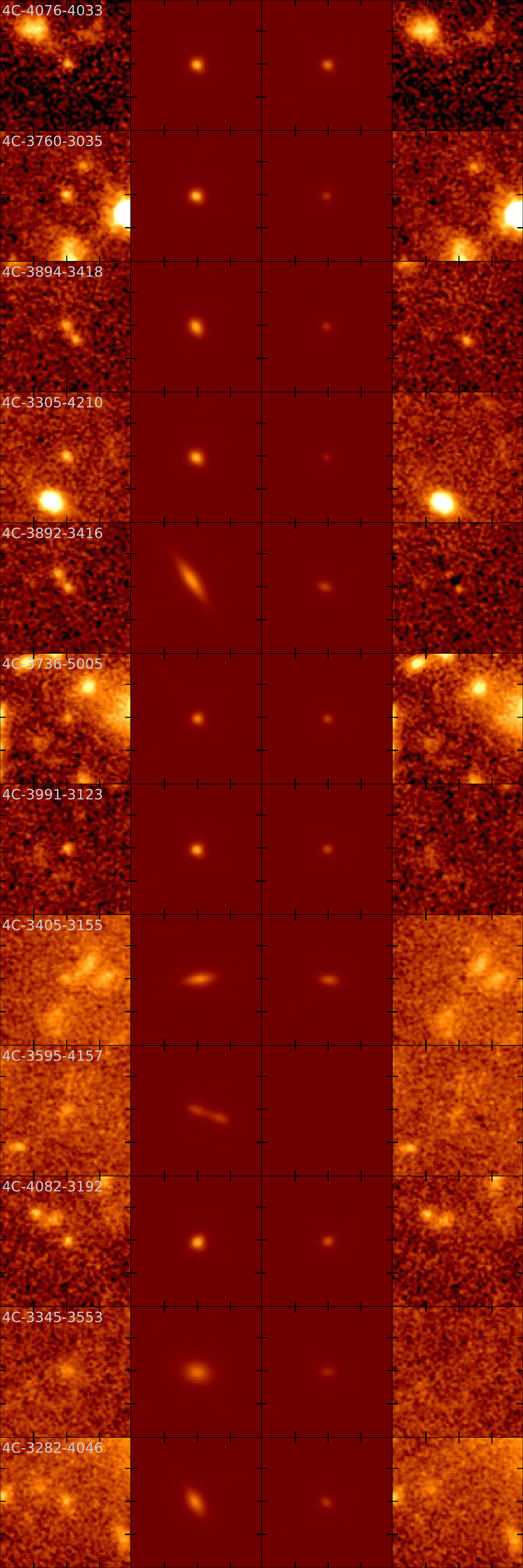}
      \includegraphics[width=0.45\linewidth, trim=0 0 0 0, clip]{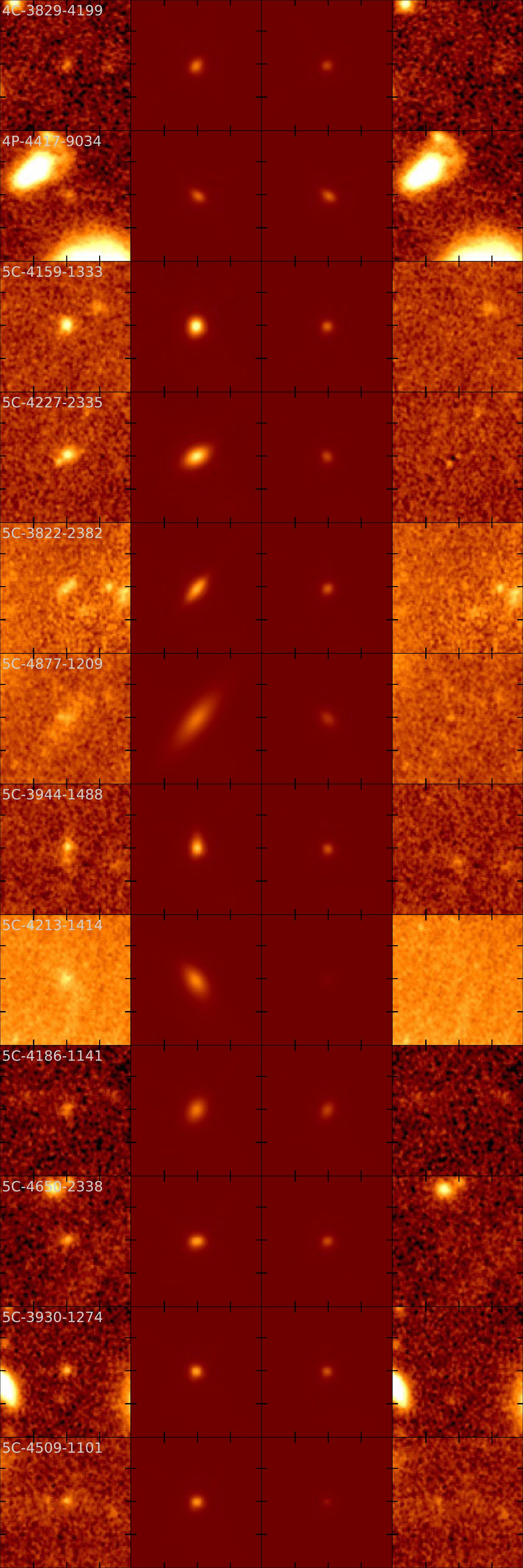}
  \caption{Continued.
   }
  \label{fig:faceimages_z7_3}
\end{figure}

\addtocounter{figure}{-1}
\begin{figure}[t]
  \centering
      \includegraphics[width=0.45\linewidth, trim=0 0 0 0, clip]{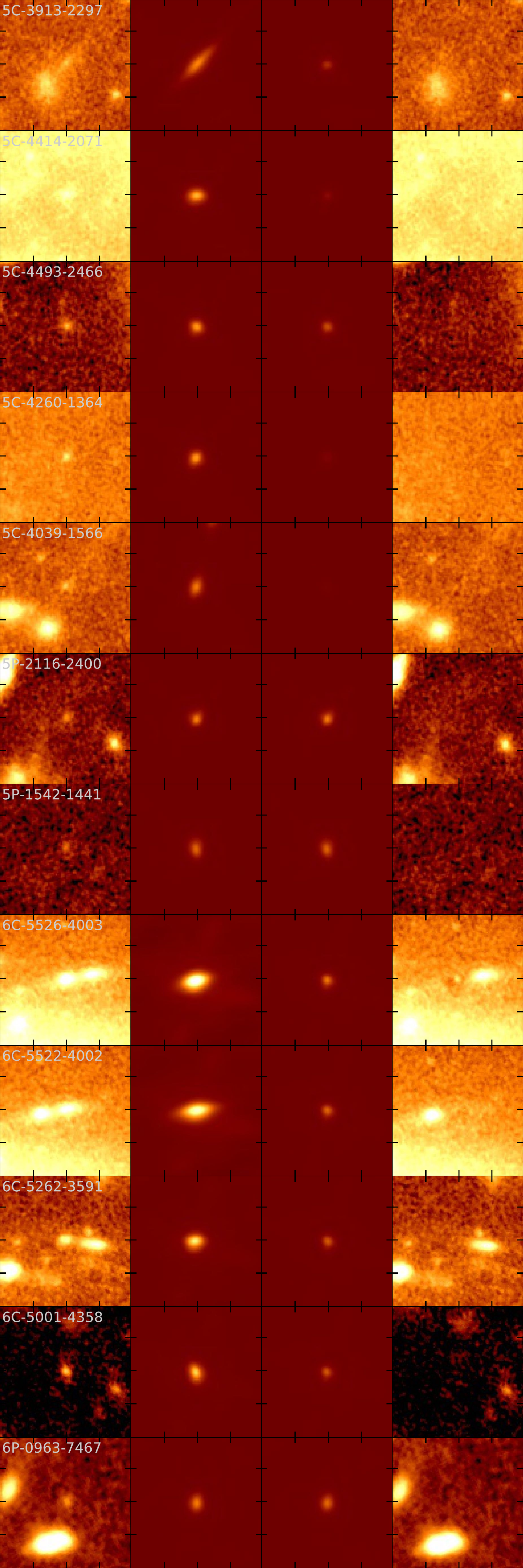}
      \includegraphics[width=0.45\linewidth, trim=0 0 0 0, clip]{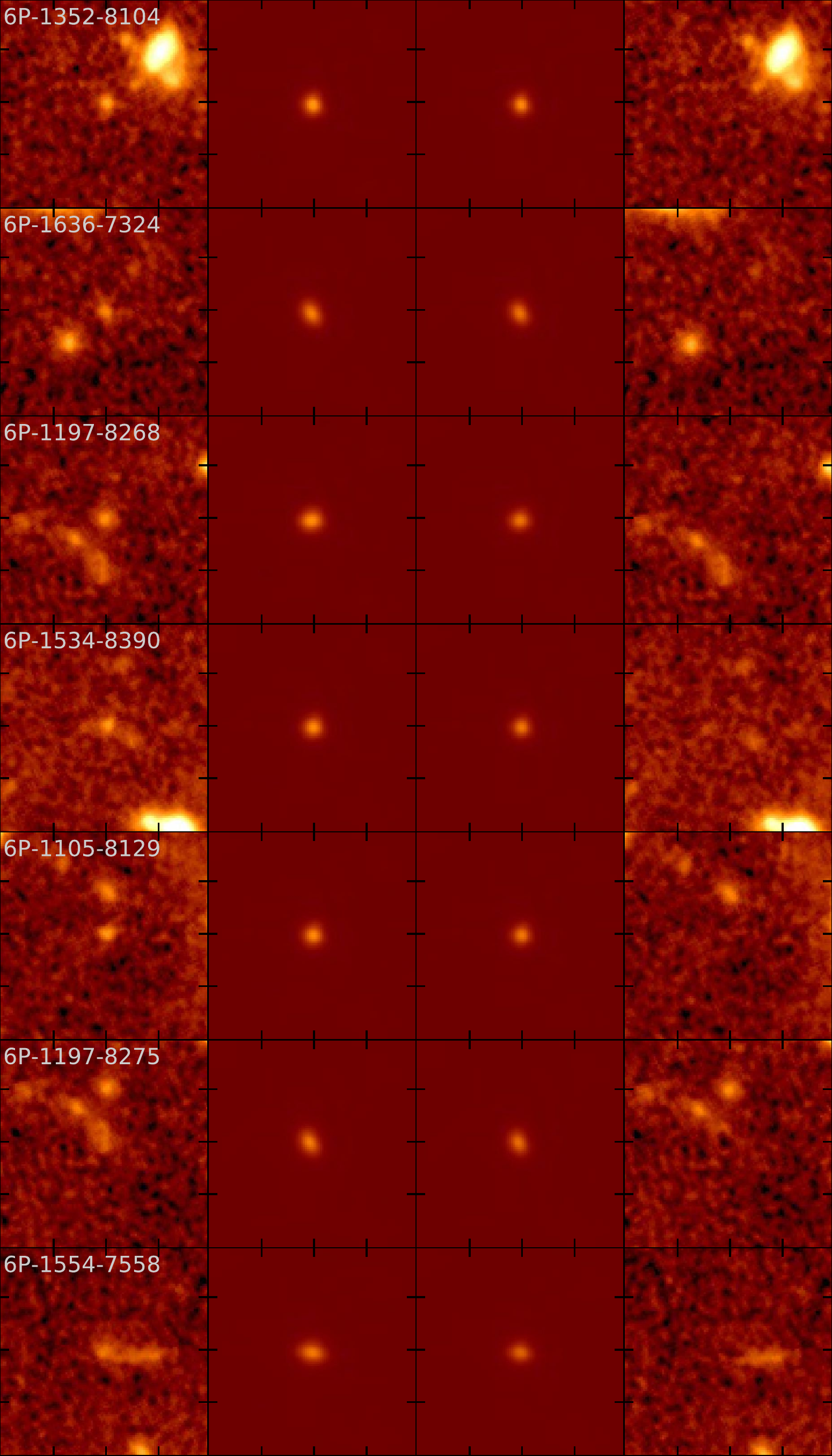}
  \caption{Continued.
   }
  \label{fig:faceimages_z7_4}
\end{figure}

\begin{figure}[h]
  \centering
      \includegraphics[width=0.45\linewidth, trim=0 0 0 0, clip]{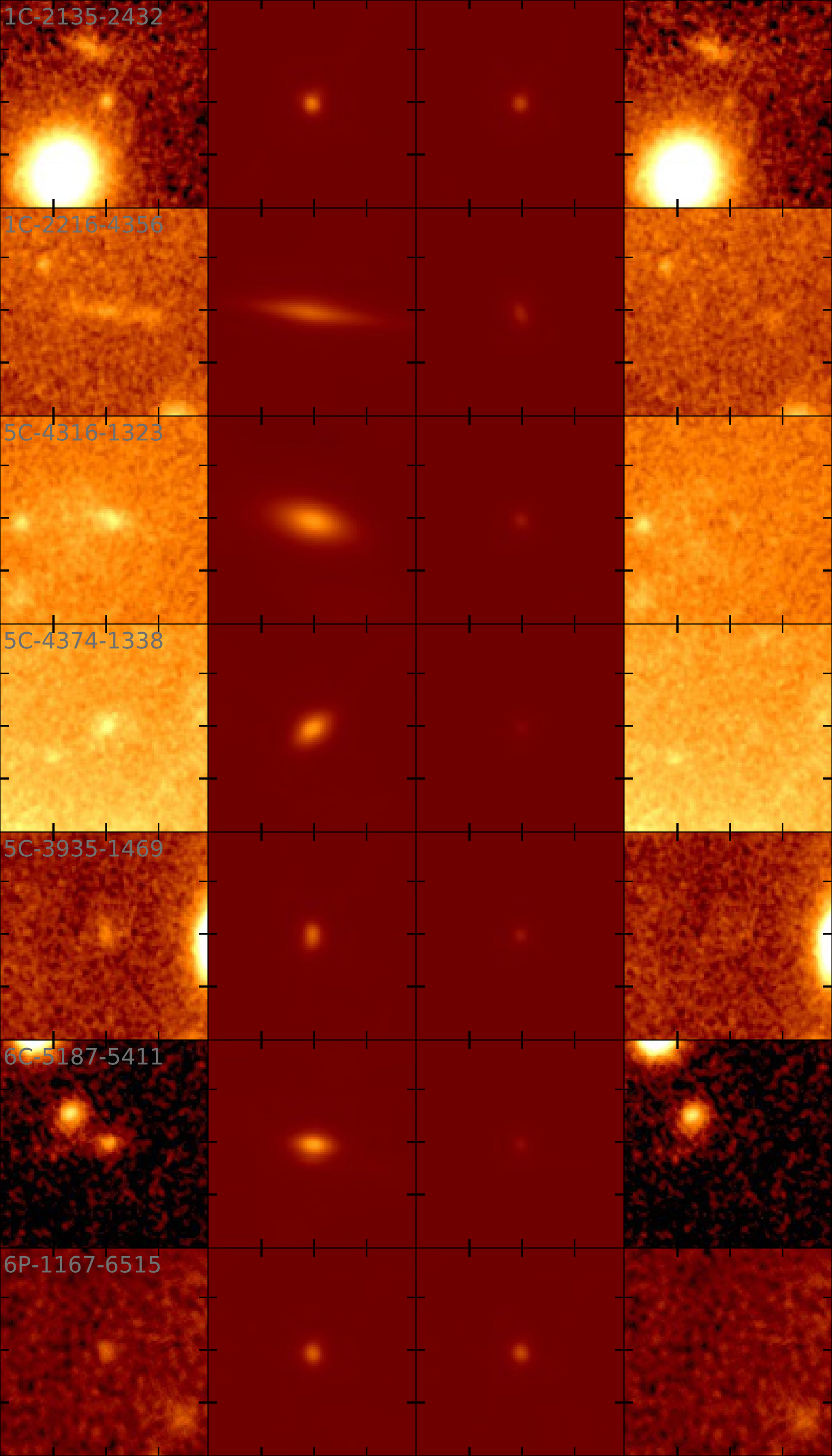}
      \includegraphics[width=0.45\linewidth, trim=0 0 0 0, clip]{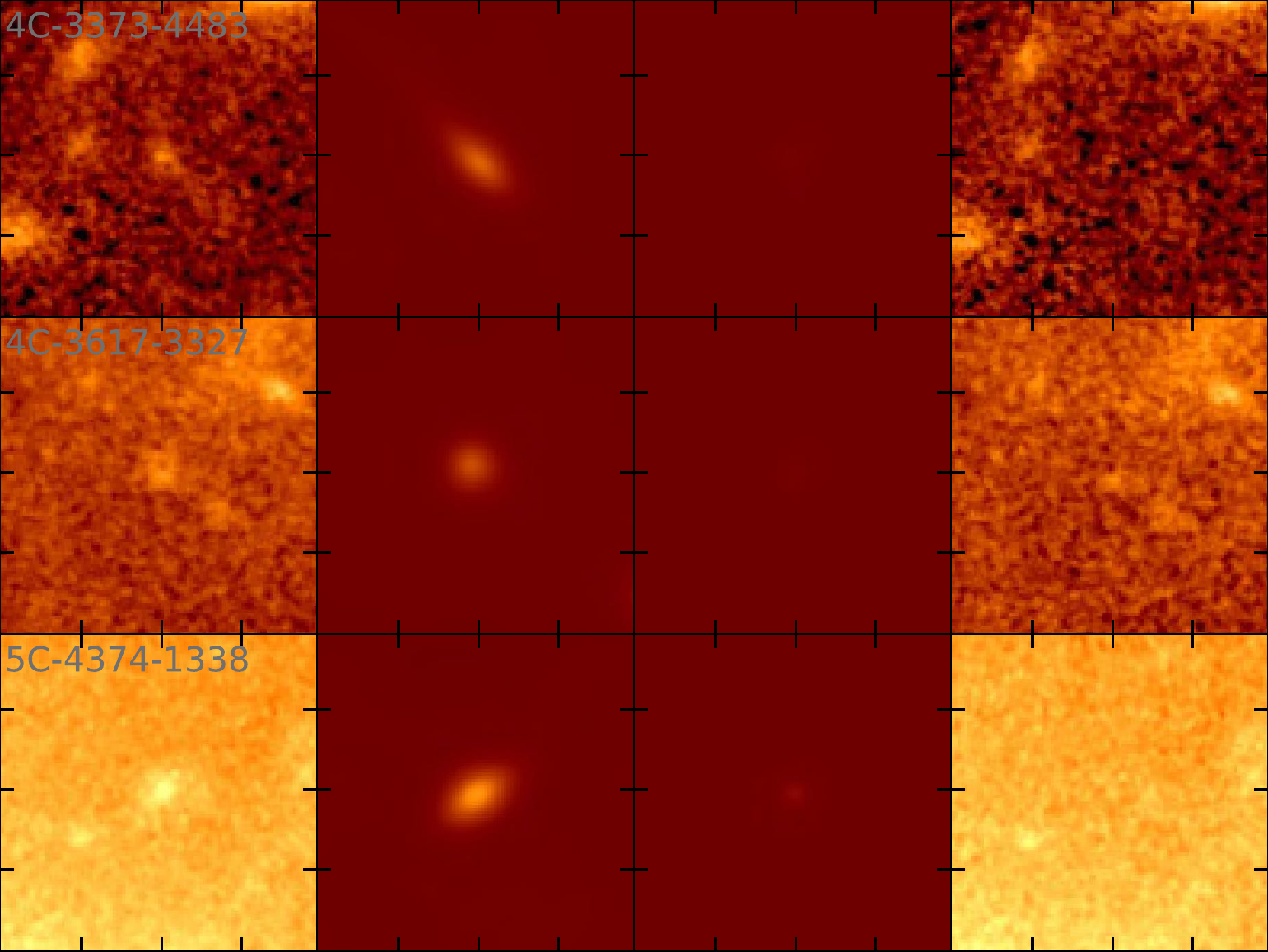}
  \caption{Same as Figure~\ref{fig:faceimages_z7_1} but for $z\sim8$ (\textit{left}) and $z\sim9$ (\textit{right}).
   }
  \label{fig:faceimages_z89}
\end{figure}

\end{appendix}

\end{document}